\documentclass[fleqn,usenatbib]{mnras}

 \usepackage{newtxtext,newtxmath}
 \usepackage{pdflscape}
 
\usepackage[T1]{fontenc}
\usepackage{ae,aecompl}

\usepackage{caption}
\usepackage{subcaption}
\usepackage{graphicx}	% Including figure files
\usepackage{amsmath}	% Advanced maths commands
\usepackage{color}

\newcommand\hi{\mbox{H\,{\sc i}}\ }

\title[\hi absorption against MRC 1-Jy radio sources]{The FLASH pilot survey: an \hi absorption search against MRC 1-Jy radio sources }

\author[ ]{J. N. H. S. Aditya$^{1,2}$\thanks{ Email: aditya.jammalamadaka@sydney.edu.au}, Hyein Yoon$^{1,2}$, James R. Allison$^{2,3}$, Tao An$^{9,10}$, Rajan Chhetri$^{4,5}$, 
\newauthor 
Stephen J. Curran$^7$, Jeremy Darling$^{12}$, Kimberly L. Emig$^{11}$, Marcin Glowacki$^6$, 
\newauthor
Emily Kerrison$^{1,2,8}$, 
B{\"a}rbel S. Koribalski$^{8,15}$,
Elizabeth K. Mahony$^8$, Vanessa A. Moss$^8$, 
\newauthor
John Morgan$^5$, 
Elaine M. Sadler$^{1,2,8}$, 
Roberto Soria$^{13,14,1}$,
Renzhi Su$^{8,9,10}$, Simon Weng$^{1,2}$,
\newauthor
 Matthew Whiting$^8$ 
 \\
$^{1}$Sydney Institute for Astronomy, School of Physics A28, University of Sydney, NSW 2006, Australia \\
$^{2}$ARC Centre of Excellence for All Sky Astrophysics in 3 Dimensions (ASTRO 3D) \\
$^{3}$First Light Fusion Ltd., Unit 9/10 Oxford Pioneer Park, Mead Road, Yarnton, Kidlington OX5 1QU, UK \\
$^{4}$ATNF, CSIRO Space and Astronomy, P.O. Box 1130, Bentley, WA 6102, Australia \\
$^{5}$International Centre for Radio Astronomy Research, Curtin University, GPO Box U1987, Perth, WA 6845, Australia \\
$^{6}$International Centre for Radio Astronomy Research (ICRAR), Curtin University, Bentley, WA 6102, Australia \\
$^{7}$School of Chemical and Physical Sciences, Victoria University of Wellington, PO Box 600, Wellington 6140, New Zealand \\
$^{8}$ATNF, CSIRO Space and Astronomy, PO Box 76, Epping, NSW 1710, Australia \\
$^{9}$ Shanghai Astronomical Observatory, 80 Nandan Road, Shanghai 200030, China \\
$^{10}$ University of Chinese Academy of Sciences, 19A Yuquan Road, Beijing 100049, China \\
$^{11}$ National Radio Astronomy Observatory, 520 Edgemont Road, Charlottesville, VA 22903, USA \\
$^{12}$ Center for Astrophysics and Space Astronomy, Department of Astrophysical and Planetary Sciences, University of Colorado, 389 UCB, Boulder, CO 80309-0389, USA \\
$^{13}$ INAF-Osservatorio Astrofisico di Torino, Strada Osservatorio 20, I-10025 Pino Torinese, Italy\\
$^{14}$ College of Astronomy and Space Sciences, University of the Chinese Academy of Sciences, Beijing 100049, China \\
$^{15}$ School of Science, Western Sydney University, Locked Bag 1797, Penrith, NSW 2751, Australia \\
}

\date{Accepted XXX. Received YYY; in original form ZZZ}

\pubyear{2023}

\begin{document}
\label{firstpage}
\pagerange{\pageref{firstpage}--\pageref{lastpage}}
\maketitle

\begin{abstract}
We report an ASKAP search for associated \hi 21-cm absorption against bright radio sources from the Molonglo Reference Catalogue (MRC) 1-Jy sample. The search uses pilot survey data from the ASKAP First Large Absorption Survey in \hi (FLASH) covering the redshift range $0.42 < z < 1.00$. From a sample of 62 MRC 1-Jy radio galaxies and quasars 
in this redshift range we report three new detections of associated \hi 21-cm absorption, yielding an overall detection fraction of $1.8\%^{+4.0\%}_{-1.5\%}$. 
The detected systems comprise two radio galaxies (MRC\,2216$-$281 at $z=0.657$ and MRC\,0531$-$237 at $z=0.851$) and one quasar (MRC\,2156$-$245 at $z=0.862$). 
The MRC\,0531$-$237 absorption system is the strongest found to date, with a velocity integrated optical depth of $\rm 143.8 \pm 0.4 \ km \ s^{-1}$. All three objects with detected \hi 21-cm absorption are  peaked-spectrum or compact steep-spectrum (CSS) radio sources, classified based on our SED fits to the spectra. Two of them show strong interplanetary scintillation at 162\,MHz, 
implying that the radio continuum source is smaller than 1 arcsec in size even at low frequencies. 
Among the class of peaked-spectrum and compact steep-spectrum radio sources,
the \hi detection fraction is $23\%^{+22\%}_{-13\%}$. This is consistent within $1\sigma$ with a detection fraction of $\approx 42\%^{+21\%}_{-15\%}$ in earlier reported GPS and CSS samples at intermediate redshifts ($0.4 < z < 1.0$). All three detections have a high 1.4 GHz radio luminosity, with MRC 0531$-$237 and MRC 2216$-$281 having the highest values in the sample, $\rm > 27.5 \ W \ Hz^{-1}$. 
The preponderance of extended radio sources in our sample could partially explain the overall low detection fraction, while the effects of a redshift evolution in gas properties and AGN UV luminosity on the neutral gas absorption still need to be investigated. 
\end{abstract}


%%%%%%%%%%%%%%%%%%%%%%%%%%%%%%%%%%%%%%%%%%%%%%%%%%

%%%%%%%%%%%%%%%%% BODY OF PAPER %%%%%%%%%%%%%%%%%%
\section{INTRODUCTION}\label{intro}

Neutral hydrogen (\hi) is the most abundant gaseous element in the interstellar medium of galaxies, and acts as a key reservoir for star formation, particularly in its cold phase. \citep[e.g.][]{krumholz2009}. Probing the distribution of \hi is hence critical for understanding the evolution of galaxies. \hi can be traced using the Lyman $\alpha$ (n=2-1 transition; Ly$\alpha$) and the hyperfine \hi 21-cm transitions in the hydrogen atom. The Ly$\alpha$ transition traces both the cold and warm phases of the neutral gas, however, in general the line is not a good indicator of gas kinematics because of the complex radiative transfer mechanism involved \citep[see e.g.][]{cantalupo2005}. Most observations of this line are conducted for high redshift ($z > 1.7$) galaxies, as the rest-frame wavelength of the transition, 1215.6 \AA, redshifts to the optical bands, which in turn can be observed using ground based optical telescopes \citep[][]{noterdaeme2012, bird2017}. 
The \hi 21-cm hyperfine transition is an excellent alternative to trace neutral gas in galaxies from low to high redshifts \citep[][]{kanekar2004}, and the neutral gas can be detected through both \hi 21-cm emission and absorption. In the case of \hi 21-cm emission, the signal strength is solely dependent on the \hi column density, and, given the sensitivity limitations of the currently available telescopes, the \hi 21-cm emission from individual galaxies has been detected in systems up to a redshift of only $z \approx 0.376$ \citep[][]{fernandez2016}. An exception to this is a recent detection at $z \approx 1.3$, where the emission is strongly boosted by a gravitational lens \citep[][]{chakraborty2023}. On the other hand, \hi can also be detected in absorption arising against a background radio source which could be powered by an active galactic nucleus (AGN). Here, the signal strength is dependent on both the \hi column density and the strength of the background radio source, and hence, against bright radio sources the absorption strength is independent of redshift.
\subsection{Associated \hi\ absorption in radio AGN}
`Associated' \hi 21-cm absorption (defined to be within $\rm 3000 \ km \ s^{-1}$ of the AGN redshift \citep[see e.g.][]{gupta2021} is thought to arise from within the AGN host galaxy, whereas for intervening absorbers the gas is located somewhere along the line of sight towards a background source. The observed absorption line will be a summation of the features arising from individual clouds that aintercept our line of sight towards the radio source, and the combined signal strength is sensitive to the cold component of the gas \citep[see e.g.][]{kanekar2004}. 

In the case of associated systems, the absorption profile can be used to understand the overall motion of the gas relative to the AGN. Absorption that is redshifted with respect to the AGN systemic redshift could indicate gas inflows, whereas blueshifted features could indicate outflowing gas pushed out either by starbursts or the expanding AGN jets \citep[e.g.][]{vermeulen2003, morganti2013, maccagni2017}.
However, we note that such features could also represent  gas clouds (particularly in the narrow-line regions) or disk-like structures rotating around the nucleus. The shape and the width of the absorption can be used to infer the structure and location of the absorbing gas in some cases. For example, features with either narrow (Full Width at Half Maximum, FWHM $\rm \lesssim few \times 100 \ km \ s^{-1}$) and deep absorption, and\slash or with a double-horned profile would indicate that the gas could be located in a disk-like rotating structure around the nucleus \citep[see e.g.][]{murthy2021}. Whereas, wide and shallow profiles, with signatures of disturbed kinematic features would indicate jet-gas interactions that could be perturbing the surrounding gas \citep[][]{morganti2013, aditya2018a}. The key advantages of the associated \hi 21-cm absorption studies are namely, the strength of absorption (arising from an isolated cloud against a background radio source) being independent of redshift, absorption being a good tracer of cold gas that is believed to be the precursor for star formation, and the line being sensitive to the gas kinematics. These advantages make these studies suitable for understanding the physical conditions of neutral gas in the host galaxies of AGN at various redshifts. 

\subsection{Targeted searches for associated \hi 21-cm absorption }
Over a thousand radio sources have been targeted in searches for associated \hi 21-cm absorption at redshifts $0.1 < z < 5$ \citep[e.g.][]{vermeulen2003, pihlstrom2003, gupta2006, carilli2007, curran2010, curran2016, allison2015, maccagni2017, aditya2018b, grasha2019, murthy2022}. 

Most of the detections ($\approx 190$) made in these searches are associated with low-redshift ($z \lesssim 0.4$) systems, and only eleven detections have been reported at $z > 1$ \citep[see][and references therein]{chowdhury2020}. The detection fraction in systems at $z \gtrsim 1$ was found to be significantly lower ($\approx 3\sigma$) in the combined sample of all literature searches for associated \hi 21-cm absorption \citep[][]{curran2019}. 
The \hi 21-cm absorption strength was also found to be significantly lower ($\approx 3.5\sigma$) at $z > 1$ in a sample of compact flat- spectrum radio sources studied by \citet[][]{aditya2018b}. \citet[][]{grasha2019} deduced similar results in their sample of 89 compact radio sources distributed over redshifts $0 < z < 3.8$. 

The main explanation for the low detection fraction is that the relatively higher ultraviolet (UV) luminosities of the high-redshift optically selected AGNs would ionise the surrounding neutral gas, thereby reducing the \hi optical depth \citep[][]{curran2008, curran2019}. However, it has long been known that the proximity of neutral gas to a strong radio source could alter the hyperfine populations, and raise the spin temperature of the gas \citep[][]{field1959}, and as well reduce the optical depth of the gas. And so, the higher radio luminosities of the optically-selected high-z AGNs could be an alternative explanation for the lower detection rate at high redshifts \citep[e.g.][]{aditya2018b}. Further, \citet[][]{kanekar2014} found that the absorption-selected damped Ly$\alpha$ systems at high redshifts typically have higher spin temperatures, probably due to inefficient cooling therein, which could as well be the case in AGN host galaxies (i.e. for `associated' \hi absorption systems) \citep[see e.g.][]{aditya2018b, murthy2022}. Despite the efforts of various studies mentioned above, it is not yet clear to what level of significance the above factors affect the \hi 21-cm absorption strength at high redshifts, and, if any of these factors is dominant compared to others.

We note that nearly all earlier studies used targeted searches where the samples were selected based on specific radio spectral or optical properties. 
 
The number density of the searches is unevenly distributed across redshift intervals, with the intermediate redshifts $0.4 < z < 1.0$ having only $\approx15\%$ of the total \citep[see][]{curran2019} and being dominated by compact radio sources \citep[see e.g.][]{grasha2019}. Such differences in the samples across redshift could introduce biases in the interpretations. Indeed, it is already known that the detection fraction in compact radio sources is relatively high compared to extended sources, due to a high gas covering factor in compact sources \citep[][]{gupta2006, maccagni2017, curran2013b}. Thus an unbiased estimate of the detection fraction and the overall absorption strength at intermediate redshifts is currently not available. Although the study of a small sample of 11 targets (at $0.7 < z < 1.0$) analysed by \citet[][]{aditya2019} resulted in 4 new detections, indicating that the detection fraction could be as high as $30\%$, similar to that at low redshifts, larger samples of 89 and 24 sources studied by \citet[][]{grasha2019} and \citet[][]{murthy2022}, respectively, yielded no new detections at these redshifts. It should, however, be pointed out that while the sample size of \citet[][]{aditya2019} is quite small to obtain reliable statistics, the samples chosen by \citet[][]{grasha2019} and \citet[][]{murthy2022} consist of strictly compact and extended radio sources respectively. The observed trend of decreasing detection rate of the \hi 21-cm absorption line at higher redshifts, if confirmed, has significant implications for our understanding of galactic evolution and its gaseous components. The causes for this trend are not yet clear; the biases in previous studies due to restricted range of source luminosities and redshifts, need to be addressed. Therefore, further studies should be conducted by including a more diverse and balanced sample distribution across different redshifts and radio source types.

\subsection{This paper}
The First Large Absorption Survey in \hi (FLASH; \citealp[][]{allison22}) is a large-area survey to search for \hi 21-cm absorption at intermediate redshifts, $0.42 < z < 1.00$ (covering a  frequency range of $\sim 700-1000$ MHz), using the ASKAP radio telescope (\citealp[][]{hotan2021}). The survey is planned to cover almost the entire southern sky (excluding $\rm |b| < 8.5 \ deg$ ) and two pilot surveys have recently been completed. While the pilot surveys have covered a total of $\sim 3000$\,deg$^2$ of the sky, the full survey will span $\sim 25000$\,deg$^2$ (see Yoon, H. et al.,in prep.).  We note that the sources in the survey have no prior selection criterion. In this paper we report the results of FLASH observations towards a sample of bright radio sources from the Molonglo Radio Catalogue 1\,Jy (MRC 1-Jy; \citealp[][]{mccarthy1996, kapahi1998a, kapahi1998b}). 

In the below sections, we describe the properties of the MRC sample, FLASH pilot observations towards the MRC sources and data reduction, the radio properties of the MRC sample used in this paper, the new detections of \hi 21-cm from our observations, and finally the deductions from our results. We adopt a $\Lambda$CDM cosmology with $H_{0}=70\,$km$\,$s$^{-1}\,$Mpc$^{-1}$,  $\Omega_{m}$ = 0.3, and $\Omega_{\Lambda}$ = 0.7 for our estimates.

 \section{The Molonglo Reference Catalogue and the MRC 1-Jy sample}\label{mrc_desc}
 
 The Molonglo Reference Catalogue \citep[MRC;][]{large1981} is a catalogue of over 12,000 radio sources with a flux density above 0.7\,Jy at 408\,MHz, derived from a survey of 7.85\,sr of sky carried out by the Molonglo Radio Telescope between 1968 and 1978. The catalogue is substantially complete above a 408\,MHz flux density of 1.0\,Jy. \citet{large1981} note that the reliability of the catalogue, in terms of completeness, is $>99.9$\,per cent. 
 
 The MRC 1-Jy sample \citep[][]{mccarthy1996, kapahi1998a, kapahi1998b} is a complete sample of 560 radio sources with $\rm S_{408 \ MHz} > 0.95$ Jy within a 0.59\,sr region of sky defined as follows: \\
 
 \begin{tabular}{ll}
 RA (B1950) & between 09h20m and 14h04m, \\ 
            & or between 20h20m and 06h14m; \\
Dec (B1950) & between $-20{^\circ}$ and $-30^\circ$ \\
 & \\
\end{tabular}
  
  The MRC 1-Jy sample is relatively unbiased, in the sense that only the radio flux density was used to select the sample members. The sample contains a mixture of radio galaxies \citep{mccarthy1996} and quasars \citep{kapahi1998a, baker1999}. 
  
  Detailed information about the radio structure on arcsec scales, obtained through VLA observations at either 4.8 GHz (C band) or 1.4 GHz (L band), is available for most of the MRC 1-Jy sources. Optical and or infrared identifications, astrometry, and r-band magnitudes are also available for a large fraction of the sample \citep[][]{kapahi1998b}.

  Importantly, 393 sources ($\approx 70\%$) of the MRC 1-Jy sample have spectroscopic redshifts available either from the optical follow-up observations by the authors, or from later studies in the literature \citep[e.g.][]{pocock84, mccarthy1996, kapahi1998a, kapahi1998b, brown2001}.  Among the 393 sources with redshifts, 217 have $0.4 < z < 1.0$; these 217 high redshift sources are suitable for searches for \hi 21-cm absorption in the FLASH survey (see Section \ref{pilot1}). The remaining 167 sources in the MRC 1-Jy catalogue do not have spectroscopic redshifts available presently. Most targets among these are not covered in existing optical spectroscopic surveys like SDSS.   
 
 The three main characteristics of the sample, namely, the unbiased selection of radio sources, the flux densities being $\gtrsim 1$ Jy at low radio frequencies, and the availability of optical redshifts for a large fraction of the sample, makes it highly suitable for \hi 21-cm absorption studies. The high radio flux densities allow us to achieve stringent optical depth sensitivity in a reasonable integration time, while the optical redshifts allow us to assess whether an \hi absorber is associated with the source, and if so, to assess the kinematics of the gas relative to the AGN.
 
For this reason, the fields observed in the two FLASH pilot surveys (Yoon, H. et al., in preparation) were selected to cover a substantial fraction of the MRC 1-Jy survey area. 

 %Table 1
  \begin{table*}
  \caption{Summary of observations of the 23 FLASH Pilot Survey fields overlapping with the MRC 1-Jy survey area. The columns are: (1) the field name, (2) the unique CASDA scheduling block identification (SBID) assigned to each observation, (3) and (4) the RA and DEC of the pointing centres, (5) the date of observation, (6) the width of beam-forming interval used in the observations, (7) the RMS noise on the continuum, for region within 3.2 deg of the phase centre, (8) the spectral root mean square (RMS) noise per 18.5\,kHz channel in the central part of the frequency band, and for the sources within 3.2 deg of the phase centre, (9) and (10) the sizes of the synthesized beams in continuum and spectral cubes.
  SBIDs in the range 10849 to 15212 were observed during the first FLASH Pilot Survey, while those from 34547 to 37797 were observed during the second pilot survey. The observing time for the below mentioned SBIDs was 2 hrs each. } \label{obs_summary}
\begin{tabular}{lccccccccclll}
\hline
Name & CASDA & \multicolumn{2}{c}{Pointing Centre} & Obs. date &  BF  & $\sigma_{\rm cont}$ &  $\sigma_{\rm line}$ & \multicolumn{2}{l}{Synthesized beam (arcsec)} \\
%& Notes \\
& SBID & RA\,[J2000] & Dec.\,[J2000] &   UTC &  [MHz]  & [$\mu$Jy\,beam$^{-1}$] &[mJy\,beam$^{-1}$ & Continuum & Line \\ 
&&&&&&& channel$^{-1}$] & \\
(1) & (2) & (3) & (4) & (5) & (6) &  (7) & [8] & [9] & [10] \\
\hline

FLASH\_302P & 10850 & 00:00:00.15 & -25:07:52.2  & 17-12-2019 & 5 & 102 & 5.03 & $20.0 \times 14.6 $ & $30.4 \times 28.6 $  \\
FLASH\_303P & 11053 & 00:27:10.19 & -25:07:47.2  & 05-01-2020 & 5 & 123 & 4.83 & $19.7 \times 14.3 $ & $29.9 \times 26.9 $ \\
FLASH\_304P & 13291 & 00:54:20.38 & -25:07:47.2  & 19-04-2020 & 9 & 93 & 4.62 & $18.3 \times 11.2 $ & $30.9 \times 22.7 $  \\

FLASH\_305P & 13372 & 01:21:30.50 & -25:07:47.2  & 22-04-2020 & 9 & 90 & 4.84 &  $16.9 \times 11.0$ & $30.3 \times 22.7$\\

FLASH\_305P & 37449 & 01:21:30.57 & -25:07:47.2 & 18-02-2021 & 5 & 90 & 4.65 & $15.1 \times 13.2 $ & $24.5 \times 21.8 $  \\

FLASH\_306P & 37475 & 01:48:40.70 & -25:07:47.2 & 19-02-2022 & 5 & 116 & 4.65 & $21.4 \times 12.1 $ & $31.7 \times 18.8 $  \\

FLASH\_306P & 13281 & 01:48:40.75 & -25:07:47.2  & 18-04-2020 & 9 & 84 & 4.51 & $13.7 \times 11.4 $ & $27.0 \times 22.9 $ \\

FLASH\_255P & 35939 & 01:52:56.47 & -31:23:14.7 & 15-01-2021 & 5 & 81 & 4.40 & $12.7\times11.9$ & $22.2\times20.1$ \\

FLASH\_255P & 41226 & 01:52:56.63 & -31:23:20.8 & 03-06-2022 & 2 & 85 & 4.65 & $14.5 \times 11.3$ & $24.6 \times 19.6$ \\

FLASH\_307P & 13268 & 02:15:50.94 & -25:07:47.2  & 17-04-2020 & 9 & 85 & 4.56 & $14.6 \times 11.3 $ & $27.9 \times 22.8 $ \\

FLASH\_308P & 15212 & 02:43:01.13 & -25:07:47.2  & 04-07-2020 & 9 & 91 & 4.57 & $13.0 \times 11.8 $ & $22.8 \times 20.1 $ \\

FLASH\_257P & 37451 & 02:49:24.70 &    -31:23:14.7 & 18-02-2022  & 5 & 87 & 4.84 & $17.4 \times 12.4 $ & $27.4 \times 21.2 $ \\

FLASH\_309P & 13269 & 03:10:11.32  & -25:07:47.2  & 17-04-2020 & 9 & 93 & 4.56 & $18.6 \times 11.0 $ & $31.2 \times 22.7 $  \\

FLASH\_258P & 37452 & 03:17:38.82 & -31:23:14.7 & 18-02-2021 & 5 & 79 & 4.58 & $13.8 \times 11.6 $ &  $22.5 \times 19.4 $  \\
%&& \\

FLASH\_310P & 37432 & 03:37:21.51 & -25:07:47.2 & 17-02-2021 & 5 & 110 & 4.75 & $21.8 \times 11.5 $ & $32.7 \times 18.7 $ \\

% FLASH\_311P & 37453 & 04:04:31.70 &   -25:07:47.2 & 18-02-2022 & 5 & 85 & 4.59 & $17.0 \times 11.6 $ & $27.0 \times 18.8 $\\

FLASH\_312P & 37797 & 04:31:41.89 & -25:07:47.2 & 02-03-2022 & 5 & 107 &  5.13 & $22.6 \times 11.5 $ & $33.4 \times 18.7 $ \\
FLASH\_313P & 34547 & 04:58:52.08  &  -25:07:47.2 & 17-12-2021 & 5 & 107 & 4.60 & $19.2 \times 11.6 $ & $30.6 \times 19.3 $ \\
FLASH\_314P & 34570 & 05:26:02.26 & -25:07:47.2 & 19-12-2021 & 5 & 104 & 4.52 & $18.2 \times 11.6 $ & $29.3 \times 19.3 $ \\

FLASH\_314P & 41061 & 05:26:02.41 & -25:07:52.2 & 29-05-2022 & 5 & 78 & 4.53 & $13.5 \times 11.9$ & $22.3 \times 20.5$ \\

FLASH\_377P & 34571 & 09:36:00.00 & -18:51:45.4 & 19-12-2021 & 5 & 92 & 4.81 & $13.1 \times 12.1 $ & $22.7 \times 20.0 $ \\
FLASH\_378P & 34561 & 10:02:10.90 & -18:51:45.4  & 18-12-2021 & 5 & 101 & 4.79 & $22.2 \times 11.8 $ & $34.3 \times 19.1 $ \\
FLASH\_351P & 10849 & 22:11:19.25 & -25:07:47.2  & 17-12-2019 & 5 & 101 & 4.80 & $19.7 \times 14.3 $ & $30.1 \times 28.0 $ \\
FLASH\_352P & 11051 & 22:38:29.43 & -25:07:47.2  & 05-01-2020 & 5 & 92 & 5.07 & $16.2 \times 11.7 $ & $25.2 \times 22.8 $  \\
FLASH\_353P & 11052 & 23:05:39.62 & -25:07:47.2  & 05-01-2020 & 5 & 92 & 4.98 & $17.9 \times 12.2 $ & $27.0 \times 23.4 $  \\
FLASH\_354P & 13279 & 23:32:49.81 & -25:07:47.2  & 18-04-2020 & 9 & 87 & 4.90 & $13.2 \times 11.6 $ & $26.4 \times 22.4 $ \\

\hline
  \end{tabular}
\end{table*}
 
 \section{FLASH Pilot Survey observations in the MRC 1-Jy region} 
 \subsection{FLASH Pilot Surveys} \label{pilot1}
 The ASKAP radio interferometer is located at Inyarrimanha Ilgari Bundara, CSIRO's 
 Murchison Radio-Astronomy Observatory, and consists of 36 antennas, each 12\,m in diameter, with a maximum baseline of 6 km \citep[][]{deboer2009, hotan2021}.  The observing frequency range is 0.7 - 1.8 GHz, and has an instantaneous bandwidth of 288\,MHz that can be split into 18.5\,kHz channels. These features are suitable for the studies of \hi 21-cm absorbers from the local Universe out to redshift $z \approx 1.0$. 
 
 Based on the ASKAP sensitivity characteristics, a frequency range of 711.5 to 999.5\,MHz was chosen for the FLASH survey in order to optimise the discovery potential and detection yield of \hi 21-cm absorption systems \citep{allison22}.  
 For a system at redshift $z\approx1$, the 18.5\,kHz channel width  corresponds to a velocity width of $\rm \approx 7.6 \ km \ s^{-1}$. This is sufficient to resolve the typical narrow absorption lines of $\rm few \times 10 \ km \ s^{-1}$, discovered in the literature \citep[][]{maccagni2017}. 
 
  A total observing time of 200 hrs was allocated to the two pilot surveys of FLASH. The observations for the first pilot survey were conducted from December 2019 to September 2020. 
  Each ASKAP field, with a size of $\rm 6.4 \times 6.4 \ deg^{2}$, was observed for two hours, and the typical noise characteristics are described in Table \ref{obs_summary}. Eleven fields of this survey have a complete overlap, and one field has a partial overlap, with the regions of sky covered by the MRC 1-Jy sample.  
 
 The second pilot survey was carried out from November 2021 to August 2022. Seven fields have a complete overlap and six fields have partial overlap with the MRC 1-Jy region of sky. Among the above, one field was also observed in the first pilot survey. A few of the fields were repeated in both first and second surveys for multiple reasons; to test the data quality, to re-observe specific fields where data were corrupted, and to confirm any tentative detection.      
 Standard observing techniques as described by \citet[][]{hotan2021} and \citet{mcconnell2020} were used in both of the pilot surveys. The bright radio source PKS B1934-638 was used for primary calibration, and also to set the flux density scale. 
 
  \subsection{Data reduction}
  The standard data reduction pipeline for ASKAP, described in \citet[][]{whiting2020} and \citet[][]{hotan2021} was used for initial data processing. The data reduction steps include calibration, continuum imaging, spectral cube imaging, continuum subtraction, production of continuum source catalogues, spectra extraction, and measurement of a range of quality control parameters. The median RMS noise acheived on the continuum for the pilot survey fields, for a region within 3.2 deg of the phase centre is $\rm 93 \mu \ Jy \ beam^{-1}$. And, the median RMS noise per 18.5 kHz channel in the central part of the frequency band, and for the source within 3.2 deg of the phase centre is 4.75 mJy beam$^{-1}$. The median resolution of the synthesised beam on the continuum images is $18.2\arcsec \times 11.7\arcsec$. The synthesised beam sizes for the fields can be seen in Table~\ref{flash_results}. A complete description of these aspects and the data-reduction steps, specific to FLASH survey, will be given in our forthcoming data release paper (Yoon, H. et al. in preparation). 
  
  The reduced pipeline (ASKAPsoft) data products for the pilot surveys 1 and 2 were released through the CSIRO ASKAP Science Data Archive (CASDA) \url{https://research.csiro.au/casda/}. The data products include continuum catalogues, continuum images and cubes, spectral-line cubes, individual spectra at the location of the continuum sources, and validation reports. 

The pipeline does two rounds of continuum subtraction; first in the visibility domain by subtracting the components from the self-calibrated adta, and then, an image-based continuum subtraction to clean up any residual continuum. The spectral-line data of pilot 1 observations, released through CASDA, were processed using 1 MHz intervals while smoothing the bandpass solutions, and during continuum subtraction using the spectral cube. 
Due to this, broad absorption lines with widths larger than $\rm \approx 300 \ km \ s^{-1}$ were subtracted out during processing, and were not detected in the final spectra (that are available in CASDA). To overcome this issue, a few post-processing steps were carried out by the FLASH team.
Here, we undid the image-based continuum subtraction (as this used 1 MHz intervals), and did a second-round of continuum subtraction by fitting a second-order polynomial to the spectra, over 5 or 9 MHz  intervals. We note that a method of subtracting the residual continuum features on the scale of the beam-forming frequency intervals, was developed for ASKAP commissioning work by \citet[][]{allison2015, allison2017}. Further details on these procedures, and details on the beam-forming frequency intervals, will be described in detail in Yoon, H. et al.(in preparation). In this paper we use these `post-processed' spectra for our analysis, and these results will also be released through CASDA in future, as part of a `value-added' FLASH data release. For pilot 2 observations the image-based continuum subtraction was done over a 5 MHz beam-forming intervals, by the ASKAPsoft pipeline. Thus post-processing was not required.
  
\subsection{FLASH observations of MRC 1-Jy sources}
 
  We cross-matched the sky coordinates of the radio sources in the MRC 1-Jy catalogue with the source components from the FLASH pilot survey data catalogues using a search radius of $\approx45$\,arcsec.
  The beam size in the MRC observations is $\approx 2$ \,arcmin \citep[see][]{kapahi1998a}.
  We used the \textbf{TOPCAT} \citep[][]{taylor2005} software for cross-matching the source coordinates. Out of the 217 sources in the MRC catalogue with redshifts in the range $0.4 < z < 1.0$ (please see Section \ref{mrc_desc}), a total of 64 have counterparts in our pilot survey.

The redshifted \hi 21-cm line for the targets MRC 0337-216 and MRC 0055-258 ($z=0.414$ for both) falls at the edge of the ASKAP band, so we removed these two sources from our sample. Further, for the sources MRC 0930-200, MRC 1002-215, and MRC 1002-216 the band is corrupted by ripples in the band; we conservatively use a peak-to-peak value on the spectrum, instead of the RMS, for estimating the upper limits on the \hi integrated optical depths. The source MRC 0201-214 lies at the edge of the FLASH beam, and a part of the continuum was chopped-out in the image. However, the peak of the continuum lies within the beam and we could extract the spectrum at the location. Our final sample has 62 targets, which are the main focus of this paper.

Our sample includes 40 galaxies and 22 quasars. The quasar fraction of 35\% is higher than in the full MRC-1Jy catalogue (20\% quasars). This reflects the higher redshift completeness of the MRC-1Jy sample for objects with strong optical emission lines. We note that \cite{mccarthy1996} quote a spectroscopic completeness of $\sim60$\% for galaxies, while the completeness for quasars is $\sim 100$\% in the MRC sample. 

In Figure~\ref{flash_results}, we show examples of ASKAP continuum images, FLASH \hi 21-cm spectra covering $\rm \sim 3000 \ km \ s^{-1}$ on each side of the AGN redshift (except for those which fall at the edge of ASKAP band), and the radio SEDs.
  In Table \ref{sample1} we list the compact sources that have a single continuum component, with peak flux density $> 40$mJy, while Table \ref{sample2} lists sources with multiple components that have peak flux density $> 40$mJy.

\section{Radio properties of the MRC 1-Jy sample}

\subsection{ASKAP and VLA-5GHz radio sizes}

In our sample of 62 radio sources, 25 are single-component sources in our ASKAP continuum images, while the remaining 37 sources are best fitted by multiple components (see Tables~\ref{sample1}, \ref{sample2}, respectively). The typical resolution in our ASKAP images is $\approx 15$\,arcsec at 856 MHz, which is insufficient to ascertain the radio morphology on kiloparsec scales. As a  result, the single-component sources in Table \ref{sample1} could still have an extended morphology on scales of a few kpc to tens of kpc. For these single-component sources, we used the 5-GHz radio images published by \citet[][]{kapahi1998a, kapahi1998b} to classify the sources as either 
compact or extended (lobe-dominated, core+lobes, core+hotspots and hotspot with no core). On this basis, 14 of the single-component sources are core-dominated at 5\,GHz and classified as compact in Table~\ref{sample1}, while the remaining 48 have extended morphology. We note that in the SUMSS radio survey \citep[][]{mauch2003} conducted at 843 MHz, among the sources with flux densities $\rm S_{843MHz} > 50 \ mJy$, the number density of steep-spectrum radio sources (with $\alpha^{843}_{1400} < -0.5$) is atleast three times larger than the flat-spectrum radio sources ($|\alpha^{843}_{1400}| < 0.5$). Compact radio sources tend to have flatter radio spectra compared to extended sources \citep[e.g.][]{wang1997, healey2007}. The SUMSS survey results thus imply a preponderance of extended sources among bright radio sources ($\rm S > 50 \ mJy$) surveyed at $\sim 850$ MHz. We can, thus, expect a high fraction of extended radio sources in our `full' FLASH survey, similar to the number fraction in the current MRC sample.

\subsection{Radio SEDs}
Further, we used the low radio frequency spectral-energy distribution (SED) as an indicator of compactness of the radio source.
For this purpose, we collected the flux densities at frequencies $\rm \approx 70 \ MHz - 5 \ GHz$ of all sources from the literature \citep[][]{wayth2015, intema2017, douglas1996, kapahi1998a, kapahi1998b, mauch2003, becker1995, condon1998}, and fit 
SEDs based on the following models to the data:
\newline
1. power-law (PL) and re-triggered (power-law embedded with a peaked-spectrum profile) 
\newline
2. peaked-spectrum (PS; see Equation~\ref{snellen})
\newline
3. a high-frequency turnover (HFT)
\newline
The SED fitting code is availabe at \url{https://doi.org/10.5281/zenodo.8336847}.
A typical radio source with an extended morphology is expected to have a power-law spectrum, with the radio flux density decreasing rapidly with frequency due to synchrotron energy losses. On the other hand, a peaked-spectrum source is expected to be compact, since here, the radio source would be embedded in a dense medium and the radiation would undergo synchrotron self-absorption or free-free absorption, leading to a negative turnover in the spectrum towards low radio frequencies \citep[e.g.][]{odea1991, odea1998}. We use this peaked-spectrum classification to refer collectively to the GPS (gigahertz-peaked spectrum), CSS (compact steep-spectrum), HFP (high-frequency peakers) and MPS (megahertz-peaked spectrum) sources. 

\begin{figure*}
\caption[]{Top row: Examples of ASKAP continuum images of three sources; MRC 0035-231, MRC 0424-268 and MRC 2236-264. MRC 0035-231 and MRC 0424-268 have only a single Gaussian component with peak flux $> 40$ mJy, while MRC 2236-264 has four components with peak flux $> 40$ mJy. Second and third rows: The FLASH pilot spectra for the three sources, extracted at the location of peak flux density. For MRC 2236-264, four spectra at the locations of the four peaks, with corresponding alphabetical labels are shown. To note: The \hi 21-cm line of MRC 2236-264 falls near the edge of ASKAP band. So, velocities at $\rm < -2000 \ km \ s^{-1}$ are not shown. Fourth row: The radio SEDs for the three example sources. The fainter red lines are iterations of SED fits to the data. And, the faint markers correspond to TGSS \citep[][]{intema2017}, Texas \citep[][]{douglas1996} and Molonglo \citep[][]{kapahi1998a, kapahi1998b} surveys, which have relatively larger beams. The same description applies to all other SED fits in this paper. }\label{flash_results}
\begin{tabular}{ccc}
\includegraphics[width = 5.3cm]{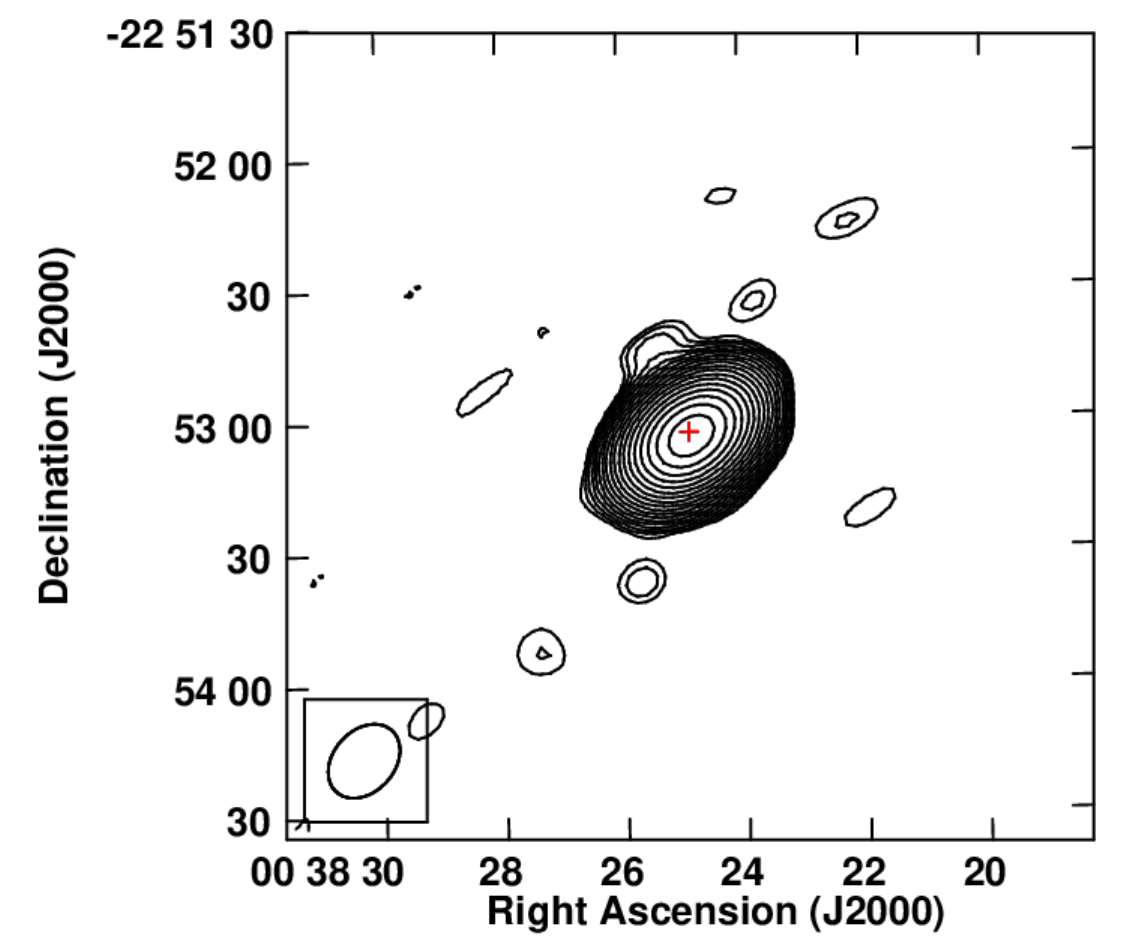} & \includegraphics[width = 5.3cm]{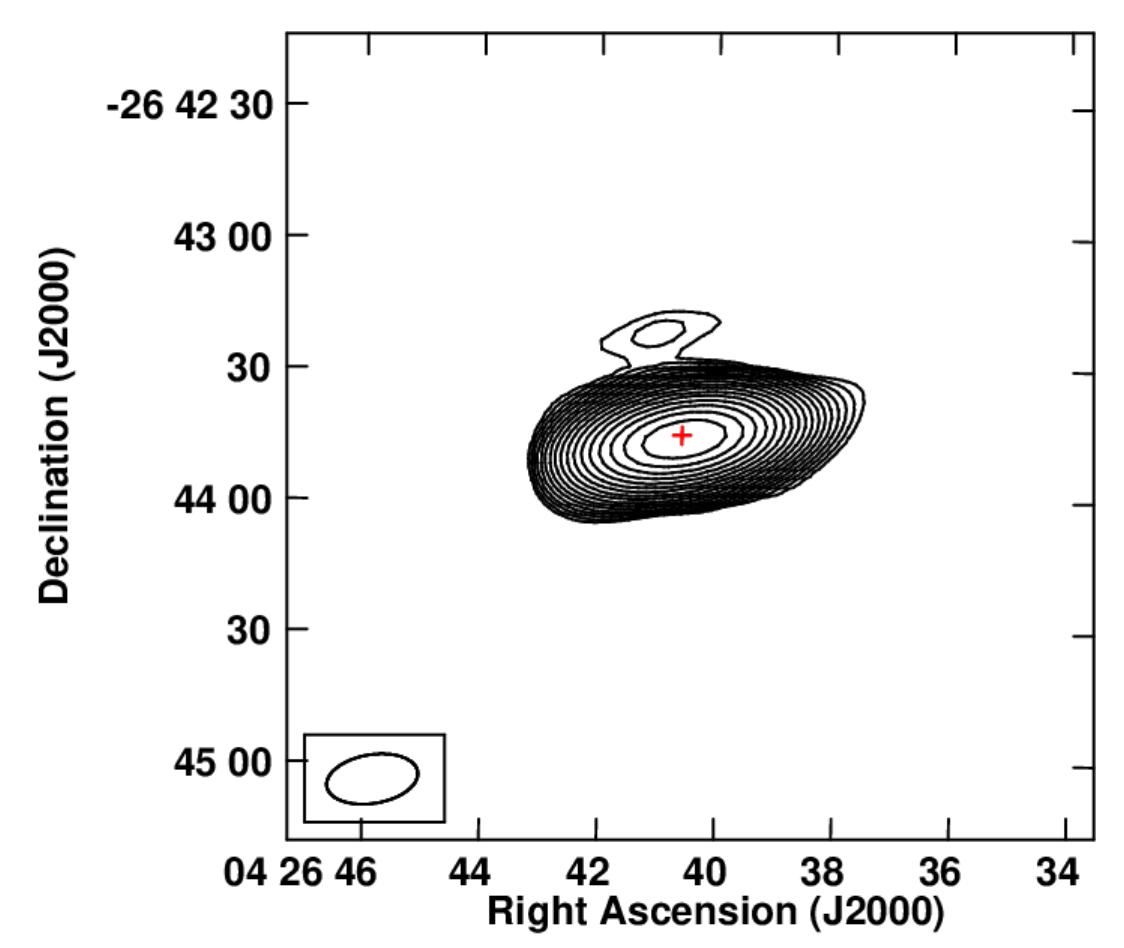} &  \includegraphics[width = 5.3cm]{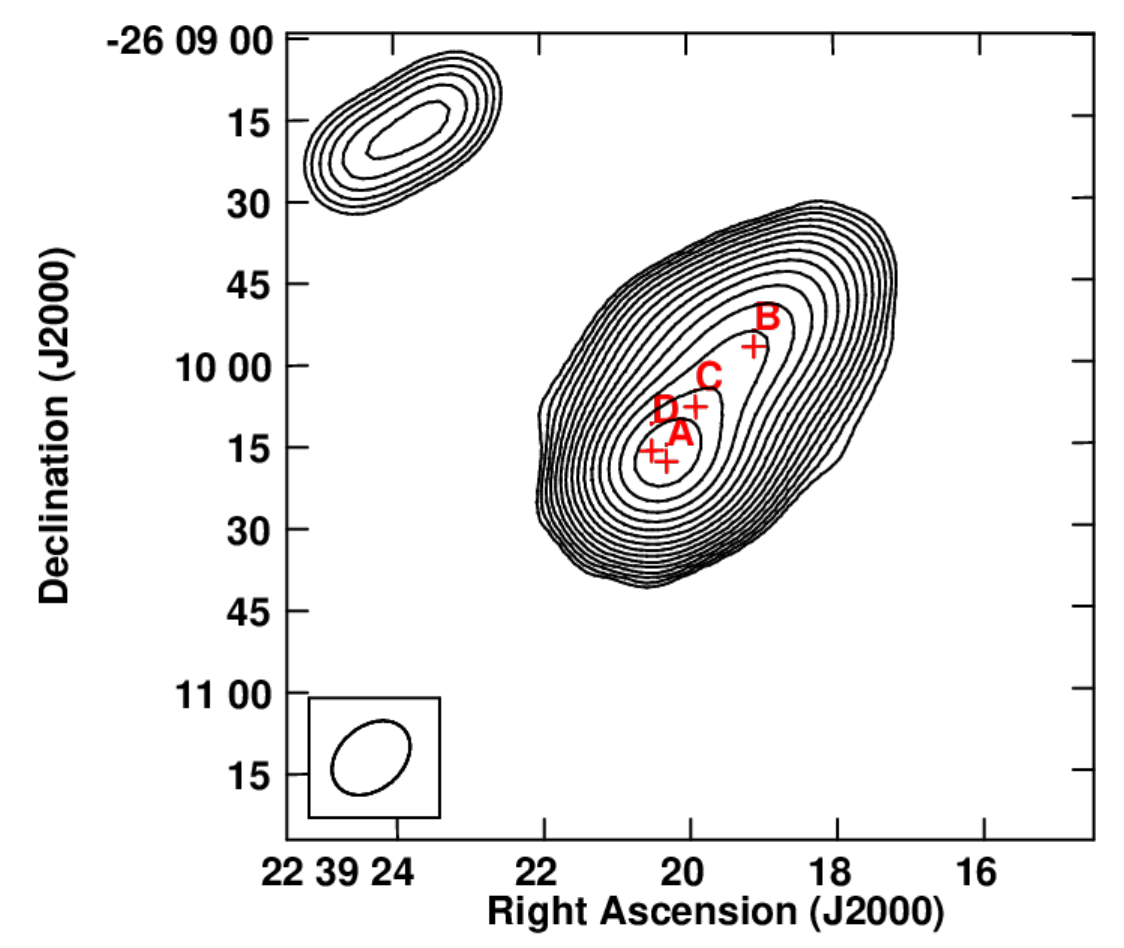}\\

\includegraphics[width = 5.7cm]{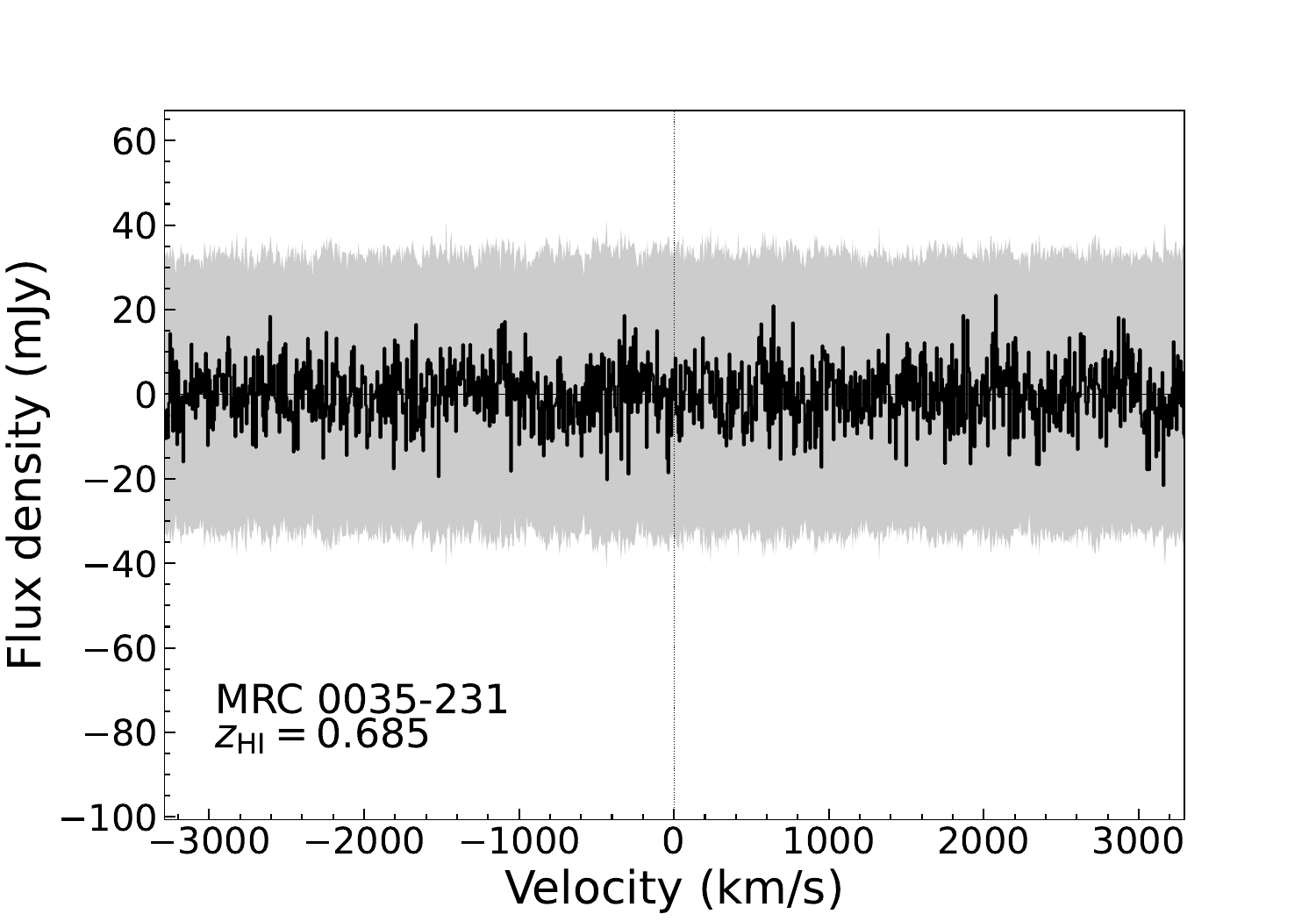} & \includegraphics[width = 5.7cm]{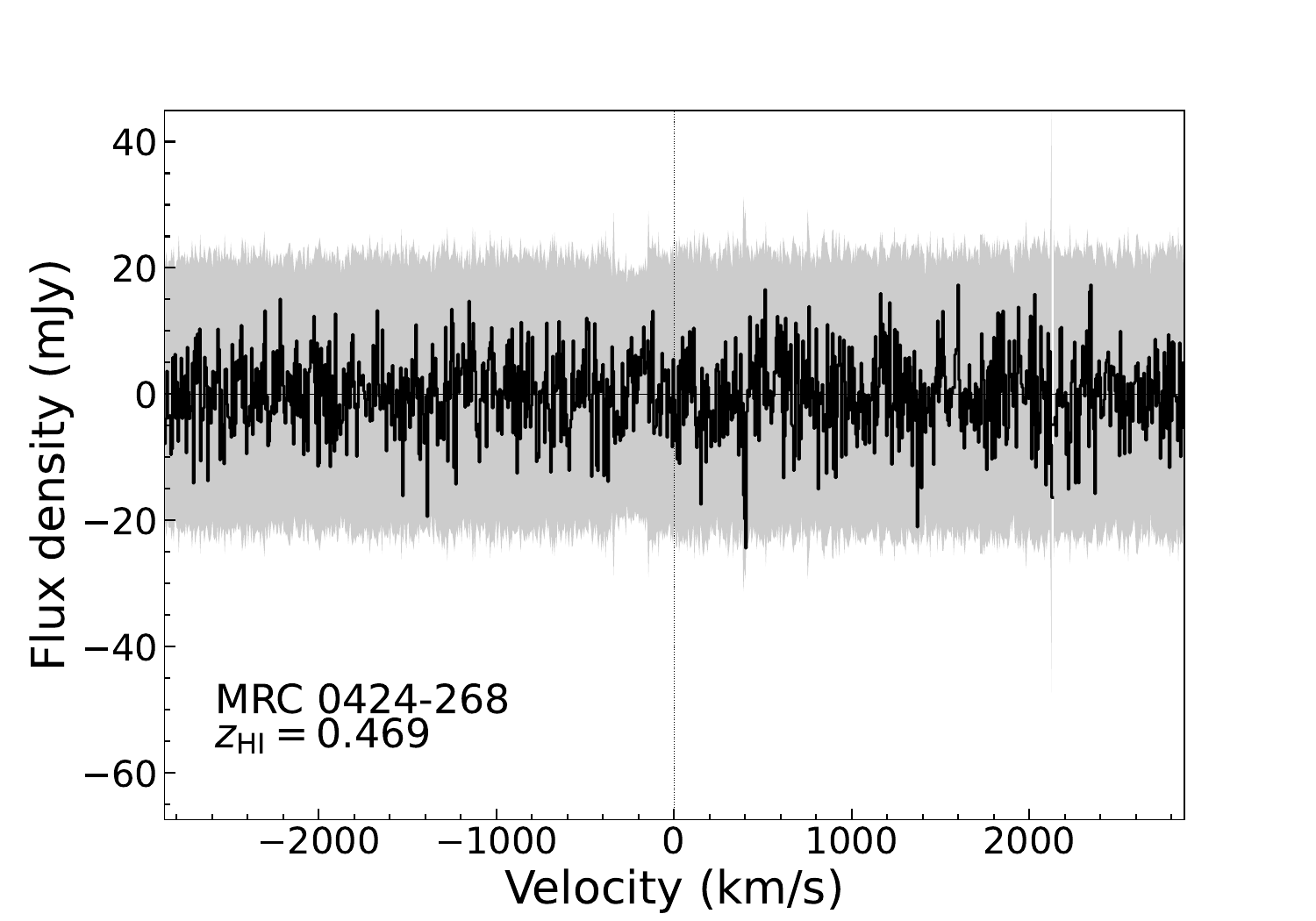} &  \includegraphics[width = 5.7cm]{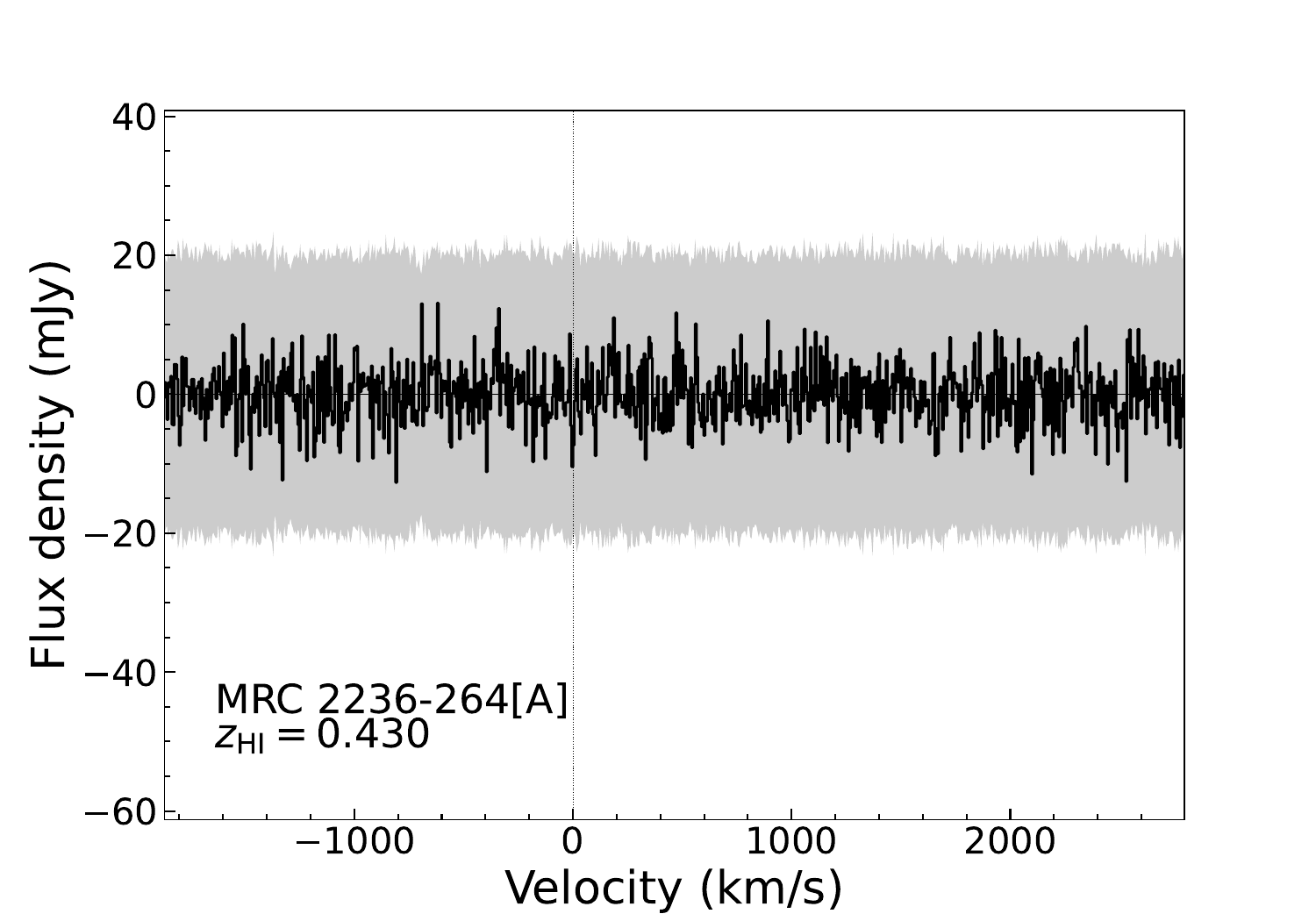}\\

\includegraphics[width = 5.7cm]{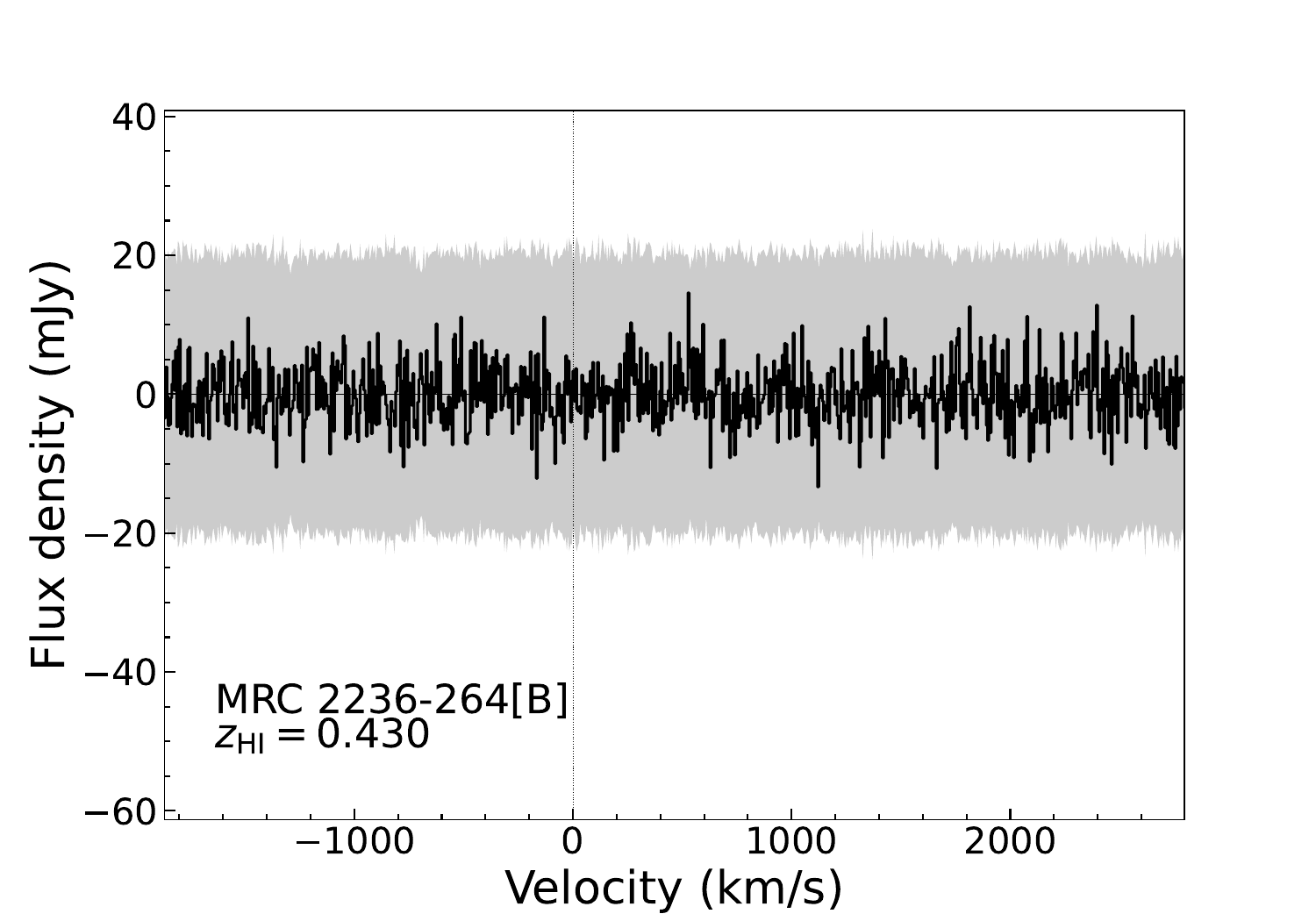} & \includegraphics[width = 5.7cm]{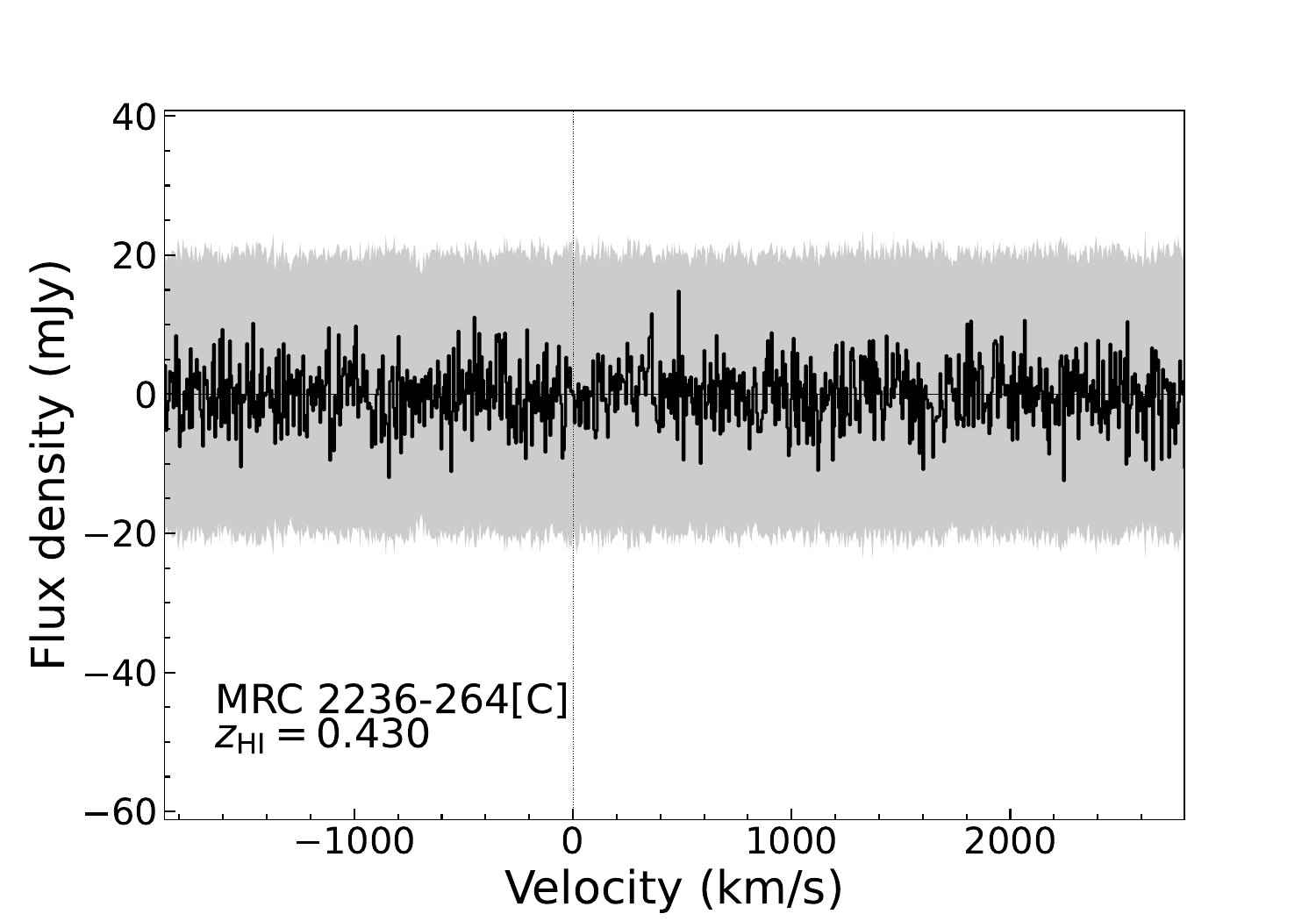} &  \includegraphics[width = 5.7cm]{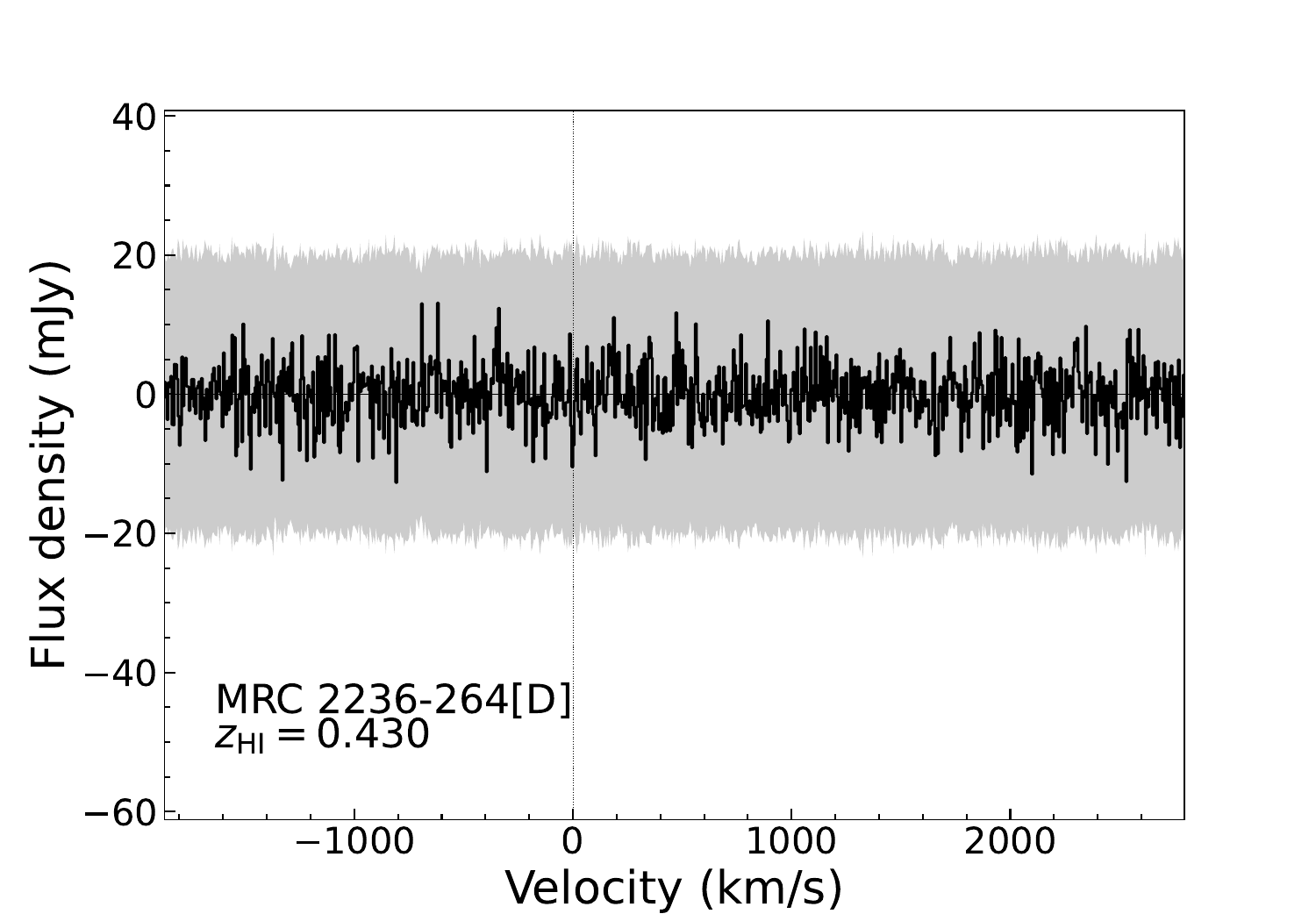}\\

\includegraphics[width = 5.3cm]{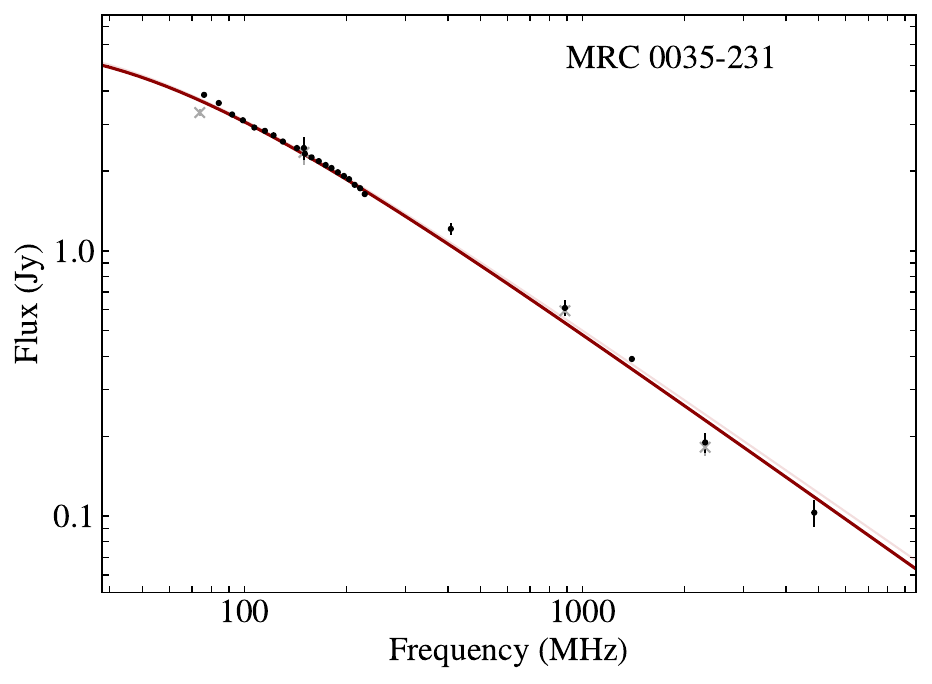} & \includegraphics[width = 5.3cm]{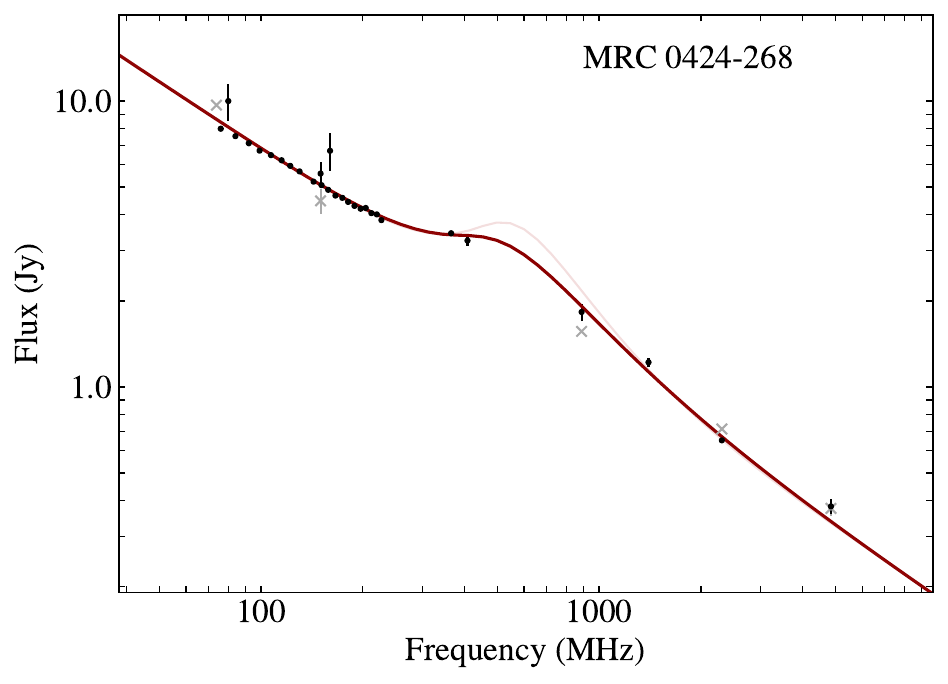} &  \includegraphics[width = 5.3cm]{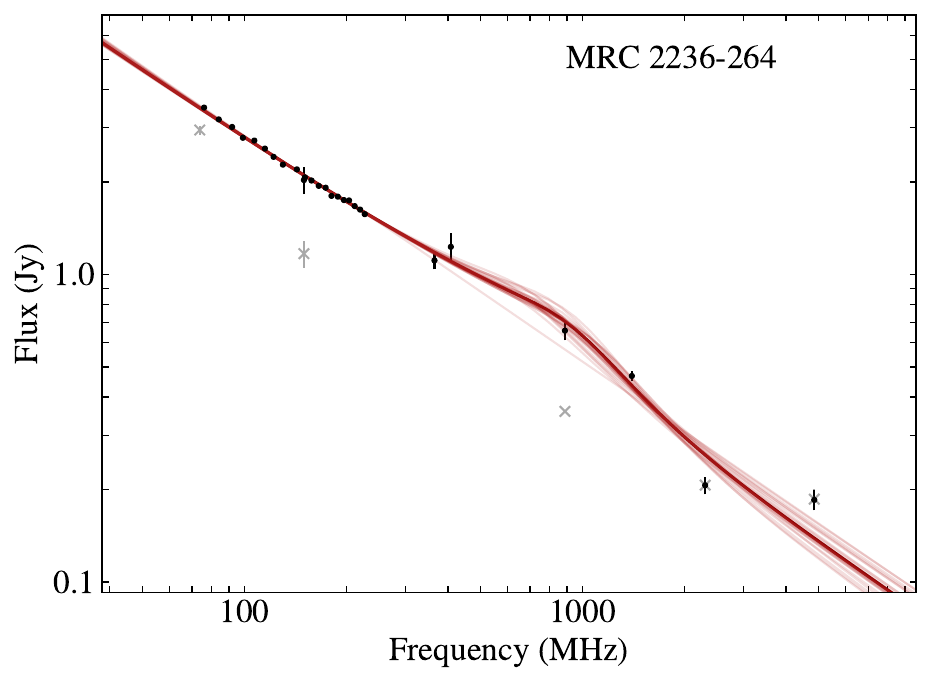}\\

\end{tabular}

\end{figure*}

A re-triggered model represents the case where the AGN has undergone multiple episodes of nuclear activity \citep[e.g.][]{callingham2017}. The source would have one or few radio blobs, along with a compact component, yielding a SED which is a combination of a 
steep-spectrum power-law and a peaked-spectrum component. We note, however, that some of the literature flux density measurements were obtained using instruments with relatively larger observing beams (e.g. the Texas survey at 365 MHz and Molonglo survey at 408 MHz), which could have systematically higher flux estimates. Hence, for the targets for which the significance of the SED curvature could not be assessed, we classified them being consistent with a power-law profile. For targets classified as HFT, the SEDs show a positive turn from a steep power law, at frequencies of $\rm few \times GHz$. These could be cases where a recent flare-up could be happening in the immediate vicinities of the core. Examples of the SED fits are shown in Figure~\ref{flash_results}; here we show the plots for six sources, and the full set for the 59 targets (for three targets there are not enough data points available in the literature for obtaining a reliable SED) is available along with the online version of the paper.

Among the 59 SEDs, the fits of 4 sources are consistent with a peaked-spectrum profile, 6 are consistent with HFT, and the remaining 49 are consistent with PL (or steep) profiles. 
We note that of the four peaked-spectrum sources, 3 are compact in the 5-GHz images. And for one of them, MRC 0233-245, while the source is unresolved in 5-GHz maps, two components are detected in the ASKAP image, suggesting that it could have extended structure at low radio frequencies (see also, notes in APPENDIX). Further, among the 49 sources with PL profiles, 9 are compact in the 5-GHz images, thus classified as compact steep-spectrum (CSS) sources. We listed the peaked-spectrum and compact steep-spectrum sources from our MRC sample in the Table~\ref{table_peak}.

%Figure 5

\subsection{Interplanetary scintillation}

Interplanetary scintillation (IPS) is the phenomenon in which the flux density of a radio source fluctuates on $\sim1$-second timescales due to scintillation induced by turbulent solar wind plasma \citep[][]{clarke1964}. Only compact sources with sizes $<2$\,arcsec show rapid scintillation, so this technique can hence be used to identify radio sources that have sub-arcsec components without the need to use Very Large Baseline Interferometry \citep[e.g.][]{hewish1964}. 

\citet[][]{morgan2018} have demonstrated that synthesis imaging techniques with the Murchison Widefield Array (MWA) can be used to detect the IPS signal from a large number of radio sources simultaneously. Observations were made putting the Sun in the null of the primary beam. An image cube with two spatial axes and one time axis, with a time resolution of $0.5$s, was constructed to estimate the temporal fluctuation in the source flux densities.
A normalised scintillation index (NSI) was calculated from source variability for each source removing the effects of solar elongations  \citep[][]{chhetri2018a}.

Radio sources dominated by a sub-arcsec compact component should have a high NSI, $\approx 1$, with angular size comparable to, or smaller than, the Fresnel scale ($0.3 \arcsec$ for MWA IPS observations) of the scintillation inducing medium \citep[][]{little1966}.

\begin{figure}
\includegraphics[width = 9.5cm]{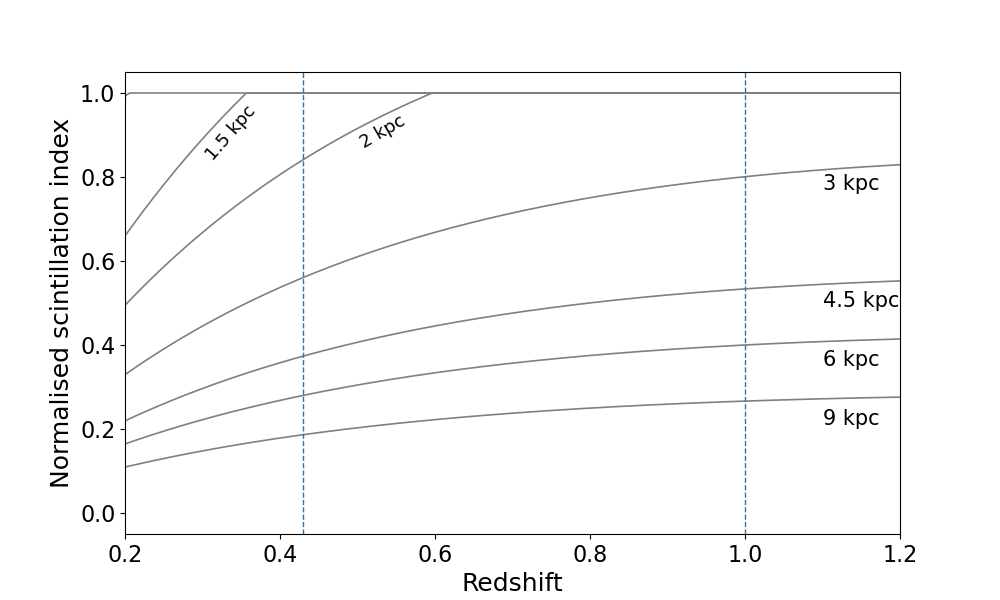} \\
\caption[]{The predicted normalised scintillation index (NSI) for sources of different linear sizes as a function of redshift, similar to \cite{chhetri2018a}. Vertical lines indicate the redshift range considered in this study. The plotted values assume an observing frequency of 150\,MHz and an effective scattering distance $D=1\,$AU.} \label{ips_size}
\end{figure}

Figure \ref{ips_size} shows the expected relationship between source size and NSI at different redshifts. For the redshift range considered in this paper ($0.42<z<1$), the normalised scintillation index is expected to be close to 1 for AGN with linear sizes $\lesssim$0.5 kpc. 

%Figure 1

We cross-matched the source coordinates of our sample of 62 targets with an updated catalogue of IPS sources compiled by two of the authors (JM and RC). 
%\citet[][]{chhetri2018a, chhetri2018b}, and found 23 
We found counterparts for 55 of our 62 sources, and the NSI estimates for these sources are listed in Table~\ref{table_ips}.

For most of our sources, there is good agreement between the angular size at 5\,GHz and the NSI values derived at 162\,MHz. In particular, all five of the sources with NSI $>0.75$ (implying source sizes smaller than 2-3\,kpc at low frequency) are also unresolved on arcsec scales at 5\,GHz. The advantage of the NSI measurements is that they probe structure on smaller scales than the 5\,GHz images, and can also reveal the existence of extended low-frequency emission that may not be apparent in high-frequency surveys. 
 
\section{FLASH Detections of \hi 21-cm absorption}
  
  Among the 62 targets, we detected \hi 21-cm absorption towards three sources: 
  MRC 0531-237 at $z = 0.851$,  MRC 2156-245 at $z = 0.862$, and MRC 2216-281 at $z=0.626$. The detections were identified initially using an automated line-finding script that uses a Bayesian analysis to identify the absorption lines \citep[][]{allison2012}, and further confirmed by visual inspection. The parameters of the absorption features are given in Table~\ref{detections}.

\begin{table*}
         \caption{The parameters of the absorption features in the three detections. The columns are (1) the MRC source name, (2) the source redshift, (3) the peak flux density in ASKAP observations, (4) the integrated flux density, (5) the width of the line at nulls, (6) the peak optical depth, (7) the velocity integrated optical depth, and (8) the \hi column density assuming $\rm T_{sp} = 100 \ K$ and a covering factor of unity. }\label{detections}
%    \centering
    \begin{tabular}{lcllllll}
    \hline
    MRC name & $z$ & $\rm F_{peak}$ & $\rm F_{int}$ & $\Delta V$ & $\tau _{\rm peak}$  &  $\int \tau dV$ & $\rm N_{HI}$ \\  
      &   & [mJy]  & [mJy] &  [$\rm km \  s^{-1}$] & & [$\rm km \  s^{-1}$] &  [$\rm \times 10^{20} \ cm^{-2}$]\\
         (1) & (2) & (3) & (4) & (5) & (6) & (7) & (8) \\
\hline
MRC 0531-237 & 0.851 & 1841.1 & 1913.6 & 662.9 & 0.59 & $143.80 \pm 0.35$ & $262.15 \pm 0.64$ \\
MRC 2156-245 & 0.862 & 805.7 & 827.7 &  155.5 & 0.04 & $3.12 \pm 0.33$  & $5.69 \pm 0.60$\\
MRC 2216-281 & 0.626 & 3652.8 & 3706.3 & 40.7 & 0.01 & $0.26 \pm 0.04$  & $0.47 \pm 0.01$\\

    \hline
    \end{tabular}
\end{table*}

  Below we describe the characteristics of the radio source, and \hi 21-cm absorption in these three targets. 
  
  \subsection{MRC 0531-237}

  %Figure 2
    \begin{figure*}
    \centering
    \begin{subfigure}[t]{0.45\textwidth}
        \centering
             \includegraphics[width = 8.2cm]
{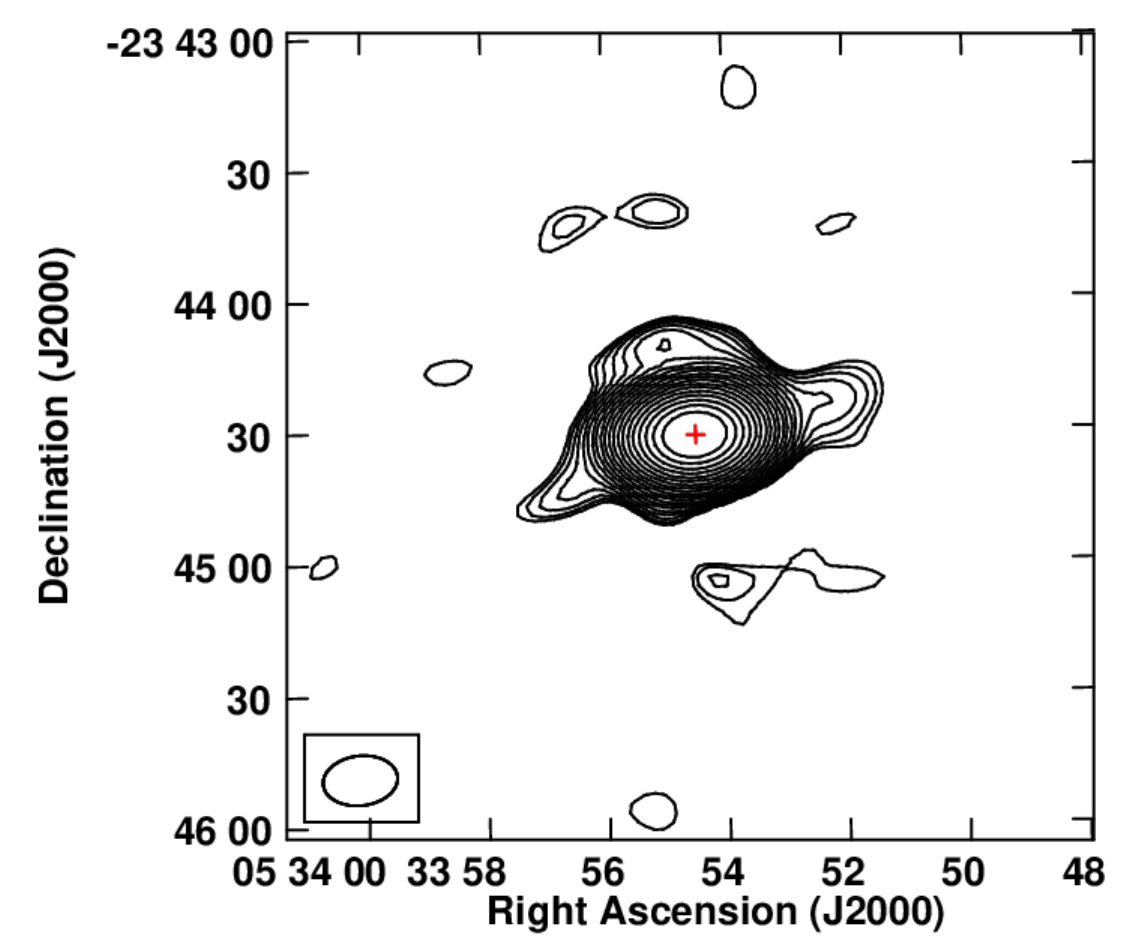} 
        \caption{The ASKAP continuum image at 856 MHz. The contours are at 1.05 mJy $\times$ (1, $\sqrt{2}$, 2, 2$\sqrt{2}$, 4, 4$\sqrt{2}$, 8, 8$\sqrt{2}$, 16, 16$\sqrt{2}$, 32, 32$\sqrt{2}$, 64, 64$\sqrt{2}$, 128, 128$\sqrt{2}$, 256).} \label{mrc0531-237_cont}
    \end{subfigure}
    \hfill
    \begin{subfigure}[t]{0.5\textwidth}
        \centering
   \includegraphics[width = 8.6cm]{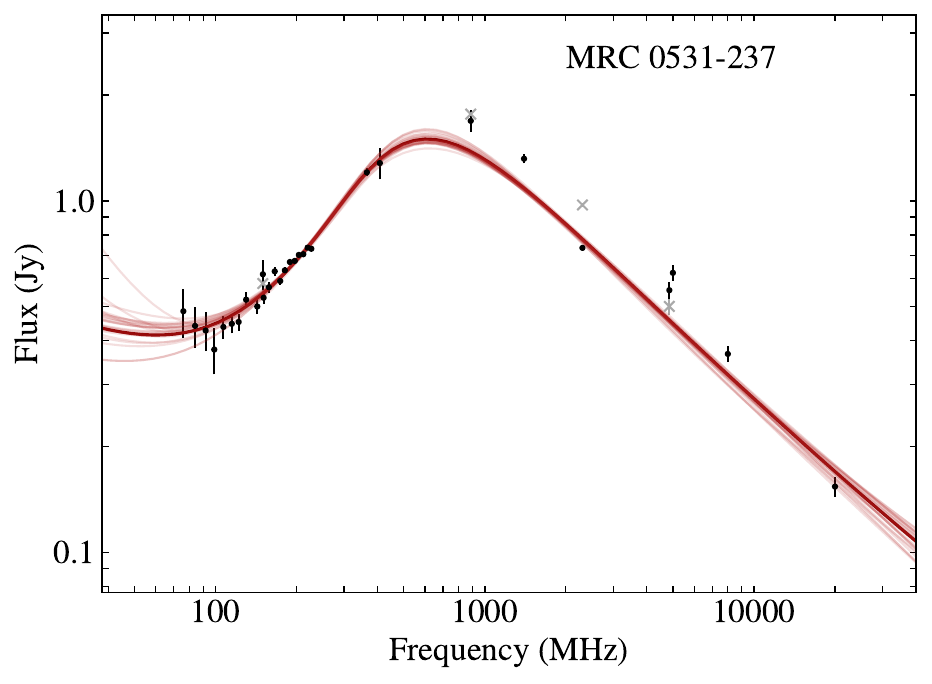} 
        \caption{SED fit for the radio flux densities of MRC 0531-237.} \label{mrc0531-237_sed}
    \end{subfigure}

    \vspace{0.5cm}
    \begin{subfigure}[t]{0.45\textwidth}
    \centering
    \includegraphics[ width = 9.0cm]
        {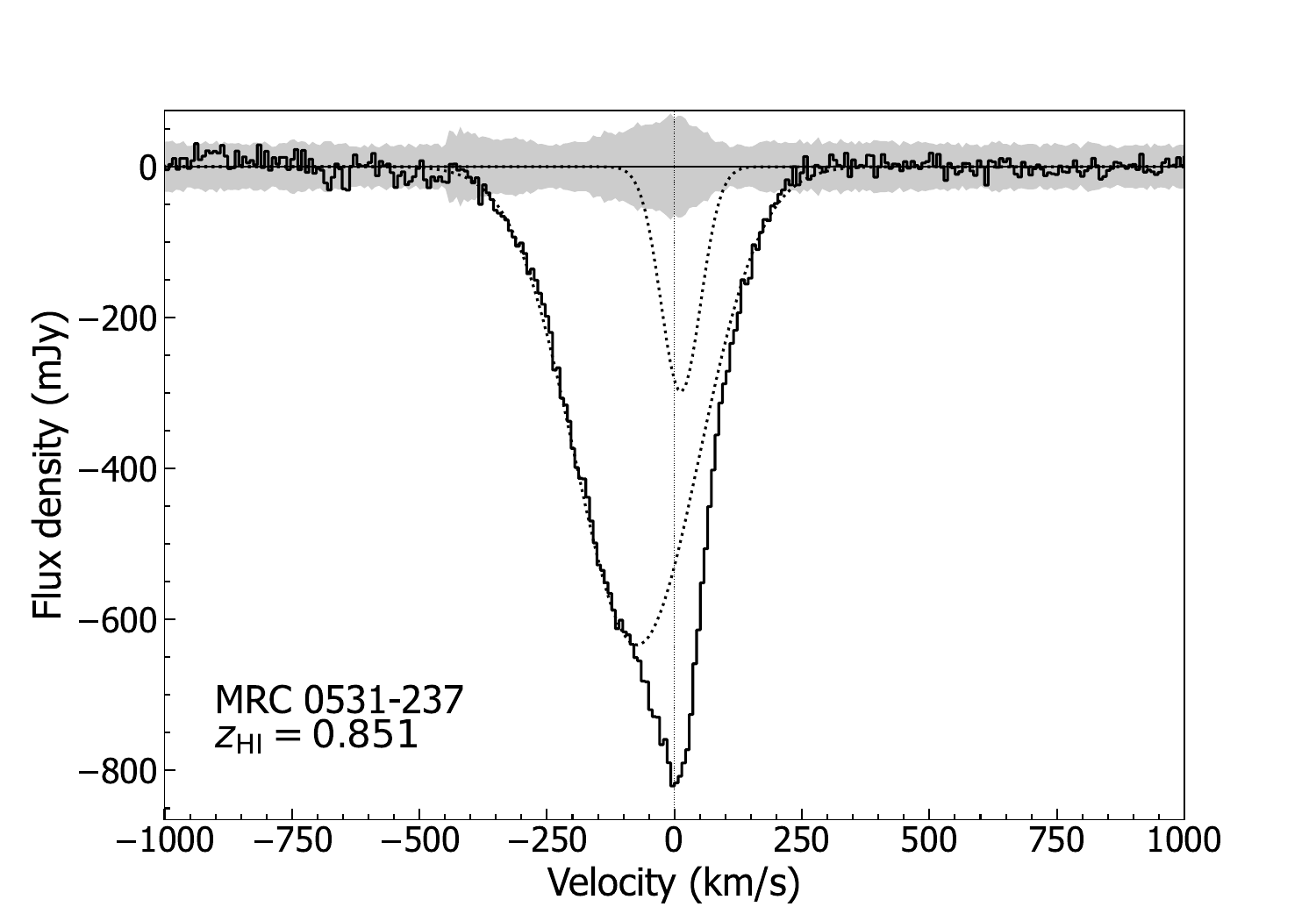} 
        \caption{The \hi 21-cm absorption spectrum. The dotted lines are Gaussian fits to the absorption. The vertical line represents the AGN redshift, $z=0.851$. And, the grey shaded region represents $5\sigma$ noise level. The higher noise level in channels sowing absorption is due to presence of low-level instrumental data corruption.} \label{mrc0531-237_hi}
    \end{subfigure}
\hfill
    \begin{subfigure}[t]{0.5\textwidth}
        \centering

               \includegraphics[width = 9.0cm]
        {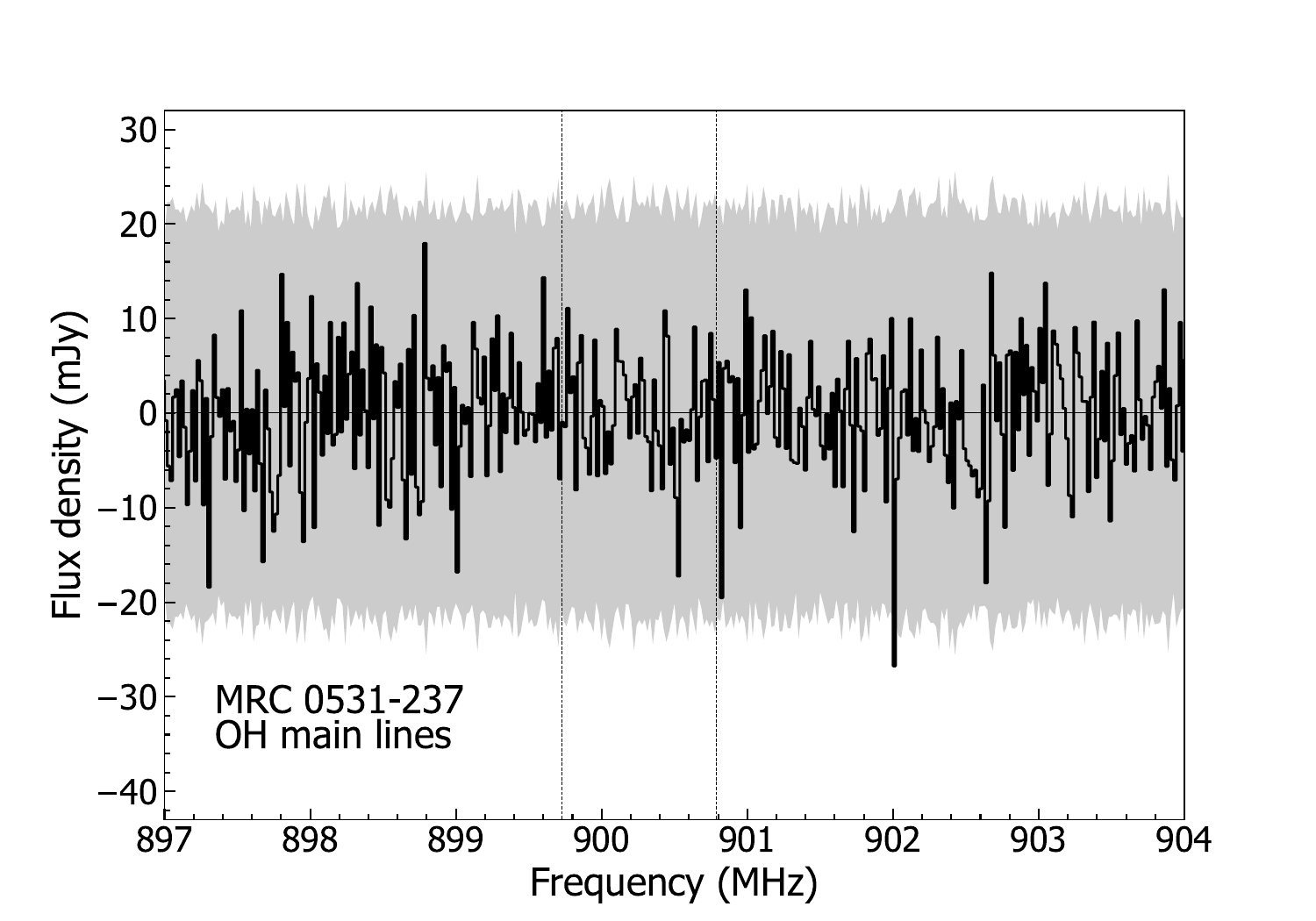} 
        \caption{OH spectrum towards MRC 0531-237. The vertical dashed lines represent the redshifted frequencies (w.r.t. $z=0.851$) of OH main lines at 1665.402 MHz and 1667.359 MHz, respectively. The grey shaded region represents $5\sigma$ noise level.} \label{mrc0531-237_oh}
    \end{subfigure}

       \caption{ASKAP radio continuum, radio SED, and \hi 21-cm absorption spectrum of MRC 0531-237} 
\end{figure*}
  
 While detailed optical studies are not available for this source in the literature, \citet[][]{mccarthy1996} reported an optical spectroscopic redshift of $z = 0.851$ (no uncertainty is reported).

 The radio source in the ASKAP image at 856 MHz (see left panel of Fig. \ref{mrc0531-237_cont}) is bright with a peak flux density of 1841.1 mJy, and has a dominant core. The milli arcsecond scale VLBI image at 8.4 GHz, archived in the ASTROGEO VLBI data base (see \href{http://astrogeo.org/vlbi_images/}{astrogeo}), shows an elongated resolved structure extending from North-East to South-West, with a scale of $\rm \approx 465 \ pc \ \times \ 97 \ pc$. The North-Eastern half of the emission looks relatively diffuse while the south-west half is collimated, though both have similar projected lengths. This suggests probably that the core lies in the central region and the radio lobes are propagating nearly perpendicular to the line of sight. Alternatively, the core could correspond to the South-Western node at the end, and the emission could represent a one-sided jet extending to North-East. In both cases, the North-East lobe is likely interacting with a dense ambient medium. 
  
  We used the measurements of flux densities at multiple radio frequencies from the literature, to fit a radio SED  to the data using the expression: 
  
 \begin{equation}\label{equation:radio_sed_fit_equ}
S_{\nu}=\frac{S_{\rm max}}{1-e^{-1}}\times\left(\frac{\nu}{\nu_{\rm max}}\right)^{\alpha_{\rm thick}}\times [(1-e^{-(\frac{\nu}{\nu_{\rm max}})^{\alpha_{\rm thin}-\alpha_{\rm thick}}}]
\end{equation}\label{snellen}
  
where $S_{\rm max}$, $\nu_{\rm max}$, $\alpha_{\rm thick}$, and $\alpha_{\rm thin}$ are peak flux density, peak frequency, optically thick index, and optically thin index, respectively \citep{snellen1998}. 
The fit has a clear peaked-spectrum profile (Figure~\ref{mrc0531-237_sed}), with a peak at $\approx 560.2$ MHz, and spectral indices of $\alpha_{thick}=2.2$ and $\alpha_{thin}=-0.8$. The steep spectrum on the optically-thin side is consistent with the extended nature of the radio emission at parsec scales (in the 8.4 GHz VLBI image), and, the turn-over of the spectrum at low frequencies, likely due to synchrotron self-absorption or to free-free absorption \citep[e.g.][]{odea1991}, indicates that the source is likely embedded in a dense ionised medium.

We detected a strong associated \hi 21-cm absorption towards MRC 0531-237, with a peak absorption fraction of 0.35, and a velocity-integrated optical depth of $\rm (143.80 \pm 0.35) \ km \ s^{-1}$.
The velocity integrated optical depth was estimated using the expression 
\begin{equation*}
    \rm \int \tau (V) dV = - \int ln[1 + \frac{\Delta S}{f * S_{c}}] \ dV
\end{equation*}
Here, $\tau $ is the \hi 21-cm optical depth, $\rm f$ is the gas covering factor, which assumed to be unity.
This is the strongest known absorber till date, with the highest integrated optical depth; the earlier highest detected value being $\rm \int \tau \ dV = 118.4 \ km \ s^{-1}$, towards SDSS J090331+010847 \citep[][]{su2022}. We estimate a high column density of $\rm (2.62 \pm 0.01) \times 10^{22} \ cm^{-2}$,
\begin{equation*}
  \rm N_{HI} = 1.823 \times 10^{18} \times T_{s} \times \int \tau \ dV \  
\end{equation*}
\citep[e.g.][]{morganti2018}, 
assuming a spin temperature ($\rm T_{s}$) of 100 K. We could fit two Gaussian components to the absorption profile, which is the minimum number with which after subtracting the fits, the residual was consistent with noise. Among these, the peak of the stronger component is blueshifted by $\rm \approx 73 \ km \ s^{-1}$,
and the weaker component is marginally redshifted by $\rm \approx 12.6 \ km \ s^{-1}$ w.r.t the AGN redshift of $z = 0.851$. While the absorption could be arising either from a few, or all, of the compact and diffuse components visible in the 8.4 GHz VLBI image, the multiple Gaussian components indicate absorption against different radio emission components.
The optical depths of various absorption components could be added up in velocity domain, giving rise to the observed absorption profile. Among associated \hi 21-cm absorbers the detection fraction is known to be relatively higher in low redshift compact peaked-spectrum sources \citep[][]{gupta2006, aditya2018a}, a reason for which could be higher gas covering factor due to compact background emission in these systems. Given that the total extension of emission at 8.4 GHz is only $\approx 465$ pc, and that the typical size of the \hi cloud could be $\approx 100$ pc \citep[see e.g.][]{braun2012} which accounts to a significant fraction of the size of the emission, it is likely that the complete radio emission is obscured by neutral gas.  

The blueshifted absorption wing in Figure~\ref{mrc0531-237_hi} (bottom-left panel) extends up to $\rm \approx 431 \ km \ s^{-1}$ at the null, indicating fast gas outflows, and the absorption has a total width of $\rm \approx 662.9 \ km \ s^{-1}$ at the nulls. While a majority of narrow absorption features with widths $\rm \lesssim 200 \ km \ s^{-1}$ are known to be arising from rotating disk-like gaseous structures around the nucleus \citep[][]{maccagni2017, murthy2021}, wider features are thought to created by shocks and turbulences resulting form strong jet-gas  interactions. The likely scenario of the radio emission being embedded in a dense medium is consistent with a  situation where strong interactions between the ambient gas and the expanding jets could be taking place.

\subsubsection{\hi gas density and OH upper limits}

The physical conditions of the atomic and molecular gases in AGN vicinities could be drastically different from those in the Galactic clouds. Strong accretion to the nuclear regions, and, the kinematic feedback from the AGN jets could be significantly affecting the gas densities. 
Beyond the local Universe ($z>1.0$), currently there are limited number of studies that have probed the \hi gas densities, estimated through \hi 21-cm absorption at high spatial resolution \citep[e.g.][]{araya2010, morganti2013}. \citet[][]{morganti2013} have reported 
densities of $\rm 150-300 \ cm^{-3}$ in the nuclear regions of 4C12.50 (at $z=0.12174$), while \citep[][]{araya2010} have estimated densities of $\rm 10^{2} - 10^{3} \ cm^{-3}$ in B2352+495 (at $z=0.2379$). 

We estimate a gas density of $\rm \approx 85 \ cm^{-3}$, based on $\rm N_{HI} = (2.62 \pm 0.01) \times 10^{22} \ cm^{-2}$, assuming $\rm T_{s} = 100 \ K$. Further, we conservatively assume that the \hi gas is uniformly distributed across the radio emission of size $\rm \approx 97 \ pc \ \times \ 465 \ pc$ (in the 8.4 GHz VLBI image, and assume a scale of $\rm 100 \ pc$ \citep[e.g.][]{braun2012} for the gas clouds, which is similar to the width of the projected radio emission along the line of sight. We note that a spin temperature of $\rm T_{s} = 100 \ K$ is only a fiducial value, used for a comparison of the column density with literature estimates. For the gas present in close vicinities of the radio AGN, as in the present case, the values could be as high as 1000 - 8000 K \citep[e.g.][]{maloney1996, araya2010}. Or, the temperature could be lower than 100 K, in dense isolated clouds. The density would vary significantly depending on the assumed spin temperature.
Also, for a localised and clumpy distribution of \hi gas, and higher spin temperatures, the gas density could be higher than our estimate, even by an order of magnitude. The fiducial estimate of gas density appears to be higher than the densities of cold \hi clouds ($\rm \approx 30 \ cm^{-3}$) found in our Galaxy, and, is comparable to the densities of diffuse molecular gas clouds ($\rm \approx 100 \ cm^{-3}$), where abundant $\rm H_{2}$ molecules could be present in the cloud interiors \citep[][]{draine2011}.   

We did not detect any OH 1665 MHz or 1667 MHz lines in our ASKAP spectrum (see Figure~\ref{mrc0531-237_oh}). By using the expression 

\begin{equation*}
  \rm N_{OH} = T_{ex} \times C_{o} \ \times \int \tau _{1667} dv \  
\end{equation*} 

where $\rm C_{o} = 2.2 \times 10^{14} \ cm^{-2} \ K^{-1} [km \ s^{-1}]^{-1}$ for the 1667 MHz line, and $\rm T_{ex} = 8 \ K$ (upper limit for excitation temperature in Galactic diffuse clouds; \citealp[][]{dickey1981}), we estimate a $3\sigma$ upper limit of $\rm N_{OH} < 1.6 \times 10^{14} \ cm^{-2}$ for the OH column density. 
The OH column densities in the Galactic diffuse clouds are known to be typically of the order of $\rm N_{OH} \sim 10^{14} \ cm^{-2}$, while in the dense molecular clouds they are $\rm N_{OH} \sim 10^{16} \ cm^{-2}$ \citep[e.g.][]{dickey1981, gupta2018}. 
Our upper limit is stringent enough to rule out the presence of any dense molecular clouds, while it is comparable to the column density of diffuse clouds. However, we note that while the OH excitation temperature is weakly dependent on the gas kinetic temperature, and thus, on the \hi spin temperature, proximity of the gas to the radio jets could raise the excitation temperature significantly \citep[][]{dickey1981}, yielding a low optical depth for a given column density. Also, the gas covering factor could be playing a critical role here; the localised distribution of molecular gas within the \hi cloud would mean that it would have a small covering factor against the background radio emission, making it difficult to detect in absorption.

\subsection{MRC 2156-245}

%Figure 3
\begin{figure*}
    \centering
    \begin{subfigure}[t]{0.45\textwidth}
        \centering
        \includegraphics[width = 8.2cm]{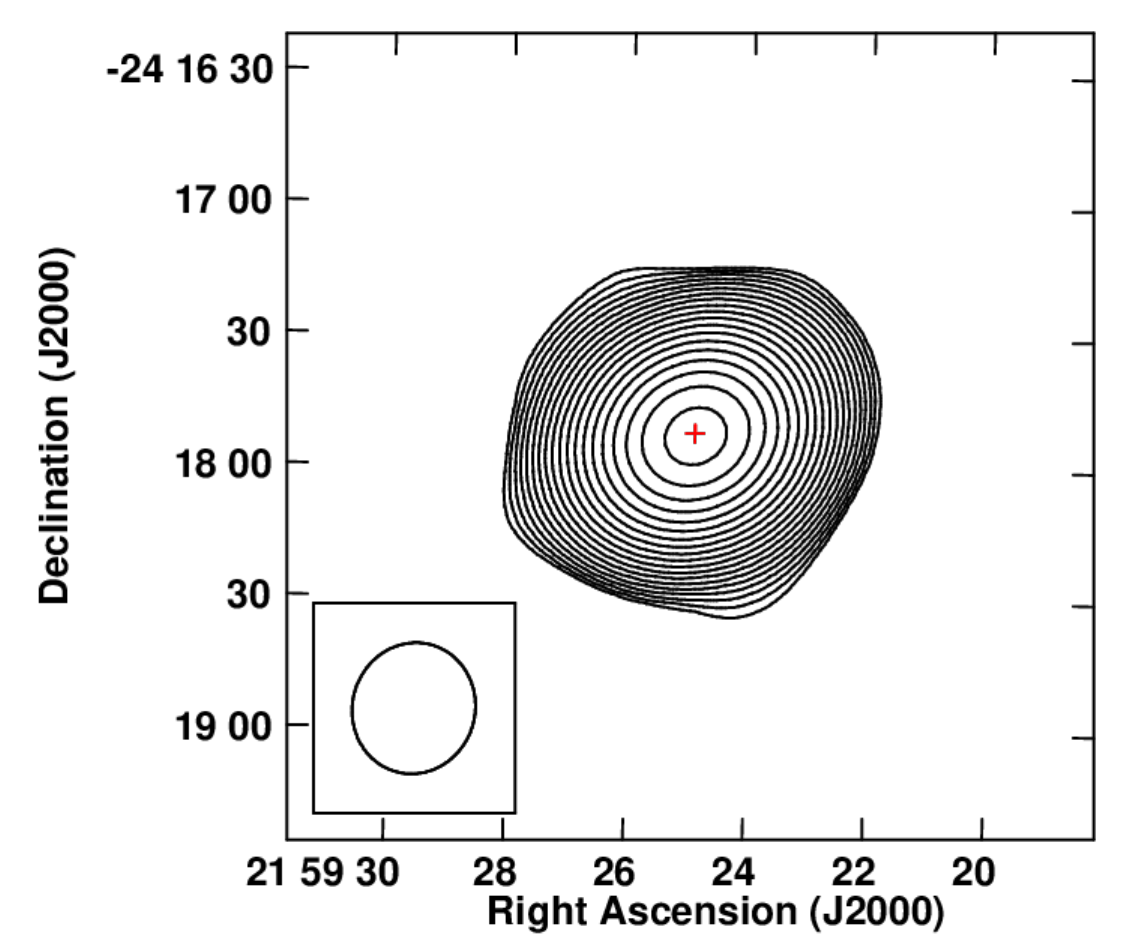} 
        \caption{The ASKAP continuum image at 856 MHz. The contours are at 0.95 mJy $\times$ (1, $\sqrt{2}$, 2, 2$\sqrt{2}$, 4, 4$\sqrt{2}$, 8, 8$\sqrt{2}$, 16, 16$\sqrt{2}$, 32, 32$\sqrt{2}$, 64, 64$\sqrt{2}$, 128, 128$\sqrt{2}$, 256).} \label{mrc2156-245_cont}
    \end{subfigure}
    \hfill
    \begin{subfigure}[t]{0.45\textwidth}
        \centering
        \includegraphics[width = 8.2cm]{ 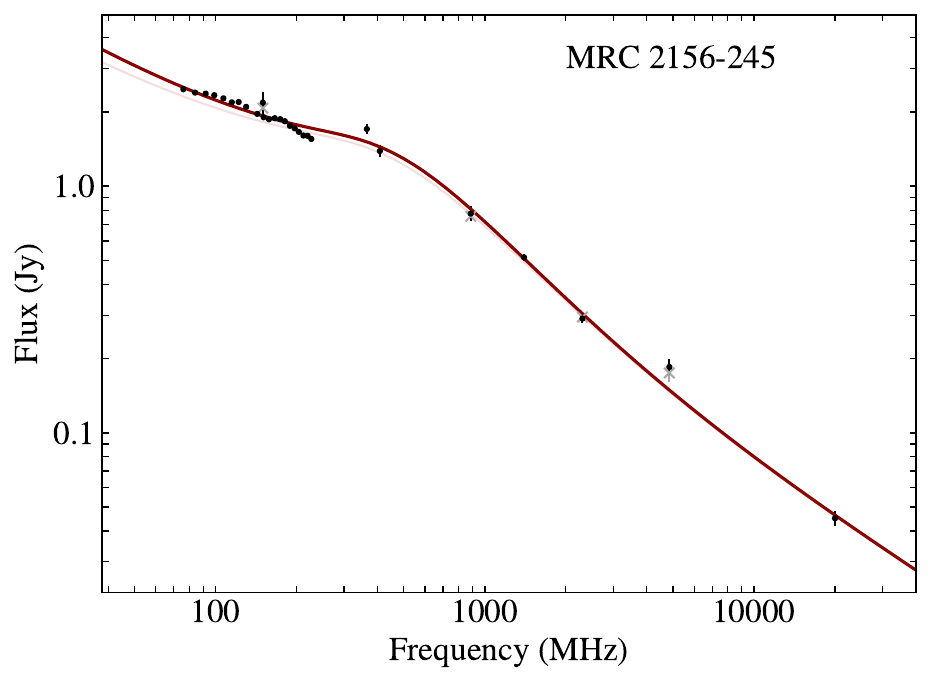} 
        \caption{SED fit for the radio flux densities of MRC 2156-245.} \label{mrc2156-245_sed}
    \end{subfigure}

    \vspace{0.5cm}
    \begin{subfigure}[t]{\textwidth}
    \centering
        \includegraphics[ width = 17.5cm]
%        , height = 9.0cm]
        {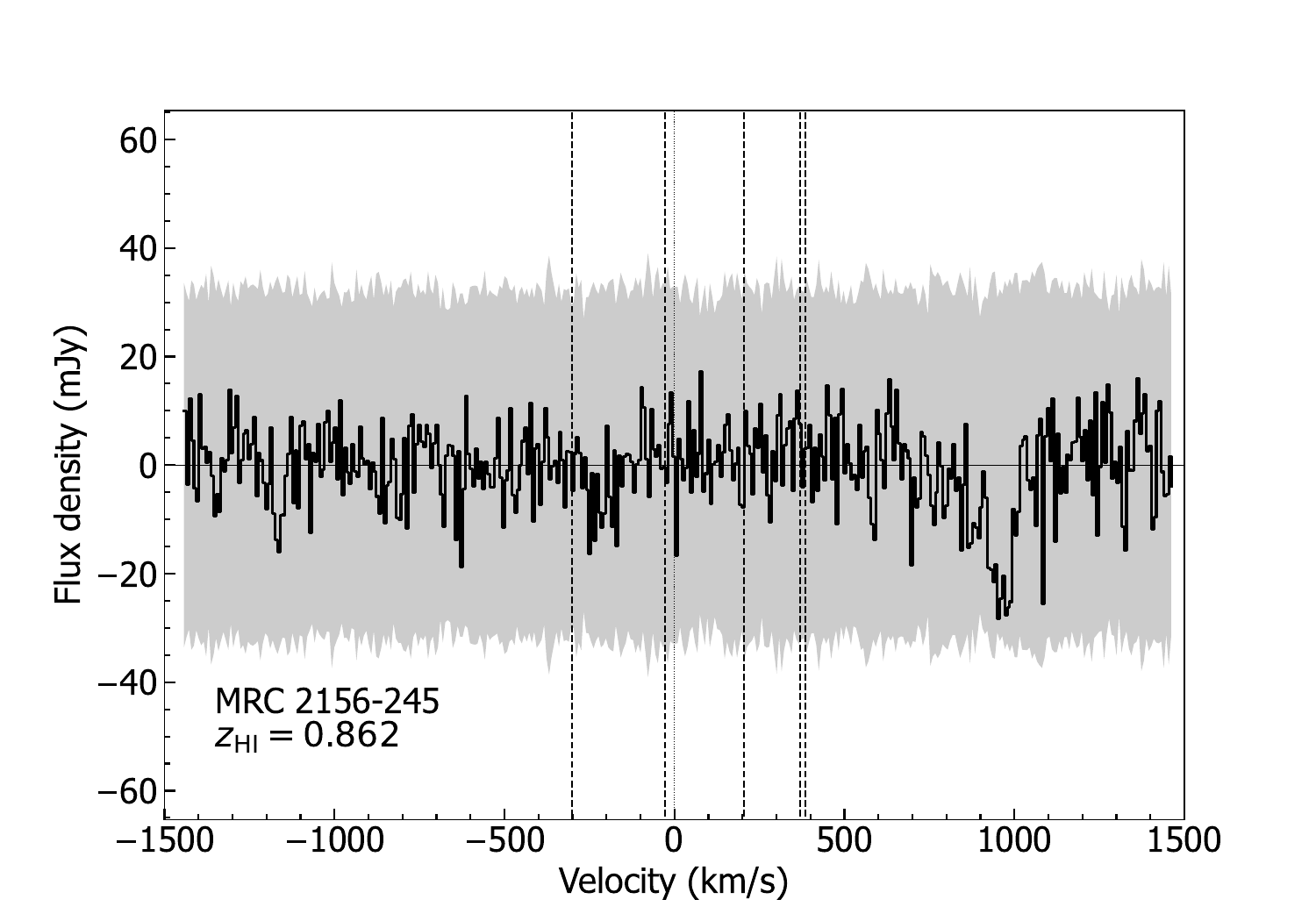} 
        \caption{The \hi 21-cm absorption spectrum. The vertical dashed lines represent, from left to right, the positions of [Ne V], H$\beta$, Mg II, [O II] and H$\gamma$ respectively. The vertical dotted line at $\rm 0 \ km \ s^{-1}$ represents the reference for the average redshift, $z=0.862$. The absorption feature can be seen near the velocity of $\rm \approx 1000 \ km \ s^{-1}$. } \label{mrc2156-245_hi}
    \end{subfigure}
    \caption{ASKAP radio continuum, radio SED, and \hi 21-cm absorption spectrum of MRC 2156-245}
\end{figure*}

\citet[][]{baker1999} reported detections of [Ne V], H $\beta$, Mg II, [O II] and H$\gamma$ lines in the optical observations towards the source. They reported a redshift of $z = 0.862$ for the source, which is the average of the estimates for individual lines.

The radio source is unresolved in the 
%$30\arcsec \times 28 \arcsec$ 
ASKAP continuum image at 856 MHz, with a peak flux density of 805.7 mJy (see Figure~\ref{mrc2156-245_cont}). The SED fit to radio flux measurements at frequencies ~70 MHz to 11 GHz is shown in Figure~\ref{mrc2156-245_sed}, which has an overall steep shape. The steep spectral shape due to synchrotron energy losses at high frequencies, indicates that the radio sources could have an extended structure at smaller scales.

 We have detected \hi 21-cm absorption in our ASKAP observations, against the peak of the radio source (Figure~\ref{mrc2156-245_hi}). The absorption has a velocity-integrated optical depth of $\rm 3.12 \pm 0.33 \ km \ s^{-1}$, implying a column density of $\rm (5.69 \pm 0.60) \times 10^{20} \ cm^{-2}$, assuming a spin temperature of 100 K, and a unity covering factor. The peak of the absorption is redshifted from the source redshift ($z=0.862$) by $\rm \approx 955 \ km \ s^{-1}$.

\subsection{MRC 2216-281}

  %Figure 4
  \begin{figure*}
    \centering
    \begin{subfigure}[t]{0.45\textwidth}
        \centering
        \includegraphics[width = 8.2cm]{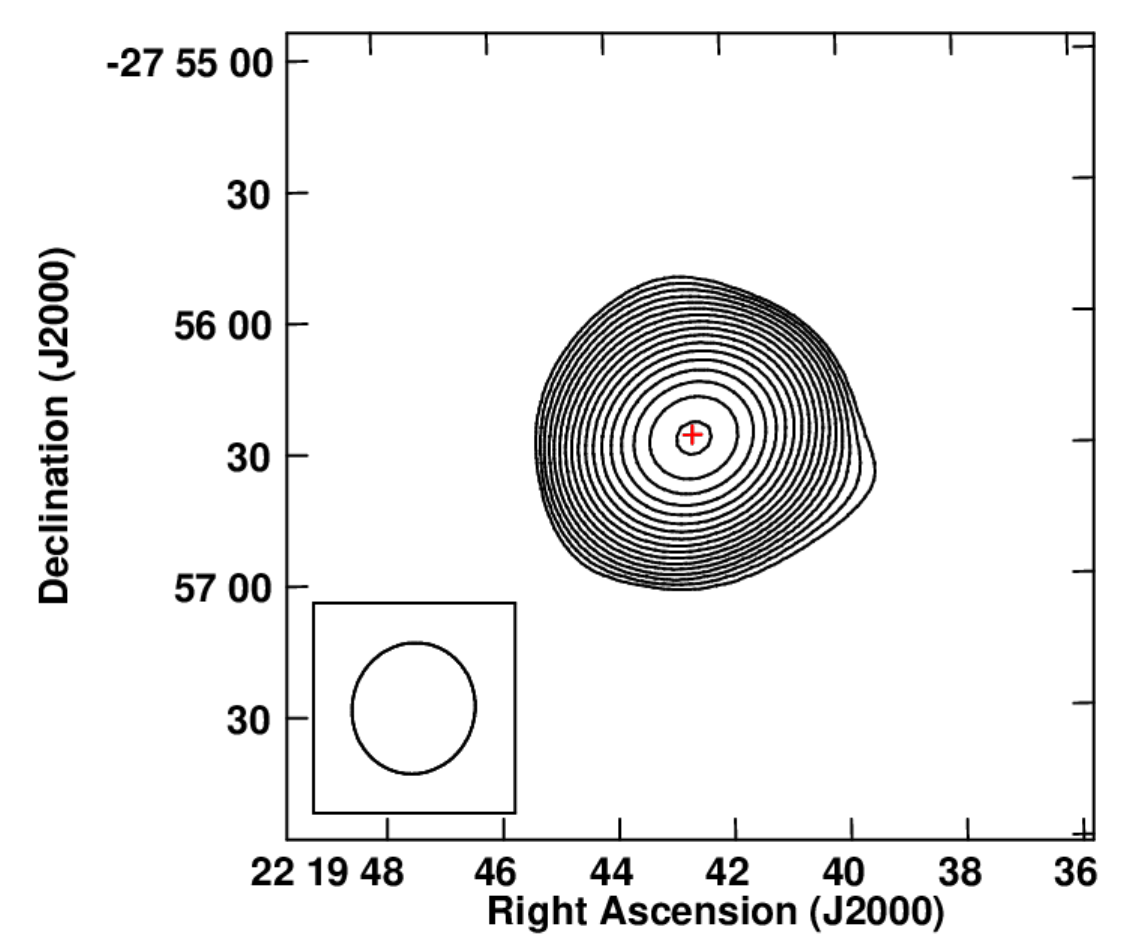} 
        \caption{The ASKAP continuum image at 856 MHz. The contours are at 5.7 mJy $\times$ (1, $\sqrt{2}$, 2, 2$\sqrt{2}$, 4, 4$\sqrt{2}$, 8, 8$\sqrt{2}$, 16, 16$\sqrt{2}$, 32, 32$\sqrt{2}$, 64, 64$\sqrt{2}$, 128, 128$\sqrt{2}$, 256).} \label{mrc2216-281_cont}
    \end{subfigure}
    \hfill
    \begin{subfigure}[t]{0.45\textwidth}
        \centering
        \includegraphics[width = 8.2cm]{ 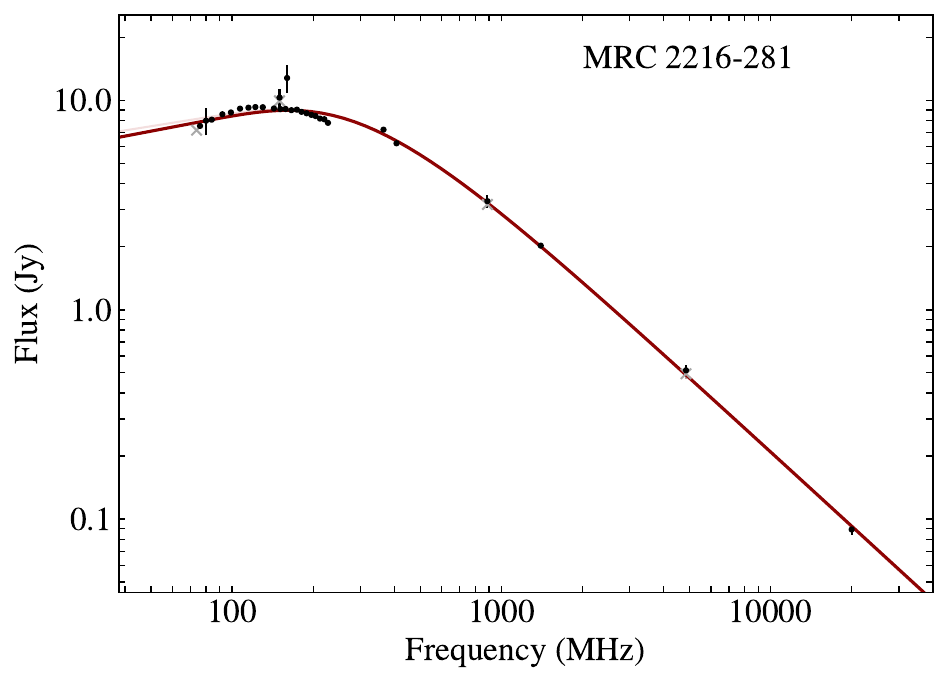} 
        \caption{SED fit for the radio flux densities of MRC 2216-281.} \label{mrc2216-281_sed}
    \end{subfigure}

    \vspace{0.5cm}
    \begin{subfigure}[t]{\textwidth}
    \centering
        \includegraphics[ width = 17.5cm]
%        , height = 9.0cm]
        {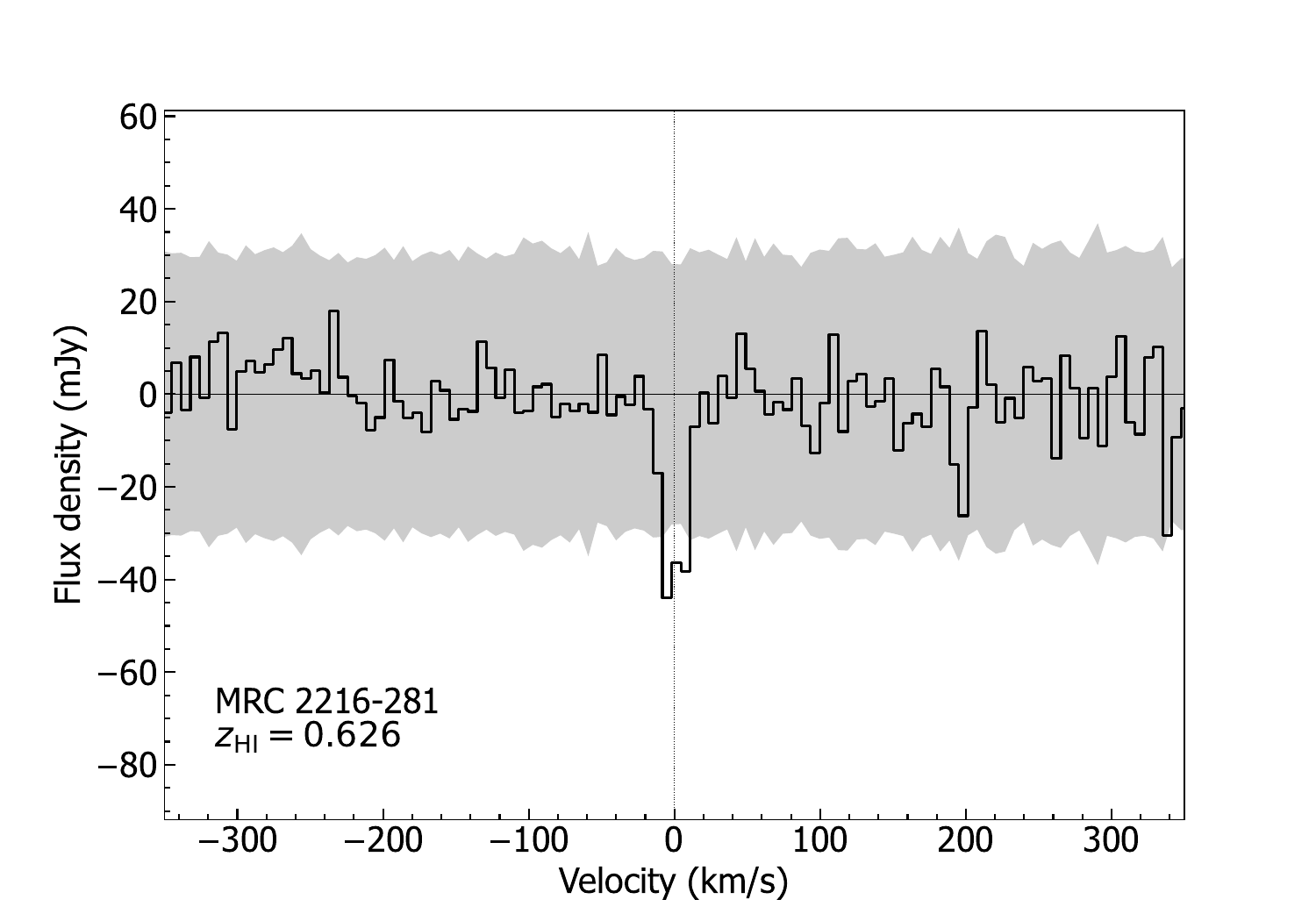} 
        \caption{The \hi 21-cm absorption spectrum.
        The velocities are relative to $z=0.626$. To note: the AGN redshift reported in the literature is $z=0.657 \pm 0.050$ \citep[][]{mccarthy1996}} \label{mrc2216-281_hi}
    \end{subfigure}
    \caption{ASKAP radio continuum, radio SED, and \hi 21-cm absorption spectrum of MRC 2216-245}
\end{figure*}

  \citet[][]{mccarthy1996} have reported an optical spectroscopic redshift of $z = 0.657 \pm 0.050$ for the source. To investigate the large uncertainty on the redshift, we downloaded the archival data of the optical observations conducted using Anglo-Australian Telescope (AAT) during 1989-1993. However, we find that the quality of the data is relatively poor. Our efforts on calibration and subsequent data reduction, to identify the optical emission/absorption lines, were not successful.
  
  The radio source is unresolved in the ASKAP image at $\approx 856$ MHz, and the peak flux density of the radio source is 3652.8 mJy.
  The radio frequency SED has a slight turn-over at $\sim 110$ MHz as shown in Figure~\ref{mrc2216-281_sed} (see also Section~\ref{radioprop}).
  
  We have detected a narrow \hi 21-cm absorption line towards the source as shown in Figure~\ref{mrc2216-281_hi}, where the line peak has a redshift of $z=0.6260$. The large uncertainty ($\rm \approx 14000 \ km \ s^{-1}$ in velocity) on the redshift implies that it is not possible to assess whether the absorption is associated with the source, or if it is arising from a lower-redshift gas cloud that is intervening our line of sight. The line has a width of $\rm 40.7 \ km \ s^{-1}$ at the nulls, and a velocity-integrated optical depth of $\rm 0.26 \pm 0.04 \ km \ s^{-1}$, implying a column density of  $\rm (4.74 \pm 0.07) \times 10^{19} \ cm^{-2}$. It is interesting to note a recent detection at $z \approx 0.7$ by \citet[][]{mahony2022}, which also has a similarly large difference between the \hi peak and the AGN redshift. Here, the \hi is detected against the radio lobe of PKS 0409-75, which is not associated with the source but arises from gas in the same galaxy group.    
   
  MRC 2216-281 is the only one of our three HI-detected sources with a published X-ray detection \citep[J221942.6$-$275627 in the 4XMM-DR13 catalogue; ][]{webb2020}. Assuming a suitable K correction for a power-law index $\Gamma=1.7$ gives a rest-frame X-ray luminosity L(2--10\,keV) = 3.5 ($\pm1.0$) $\times 10^{43}$\,erg s$^{-1}$.

\section{DISCUSSION}

\begin{figure}

\includegraphics[width = 9.5cm]{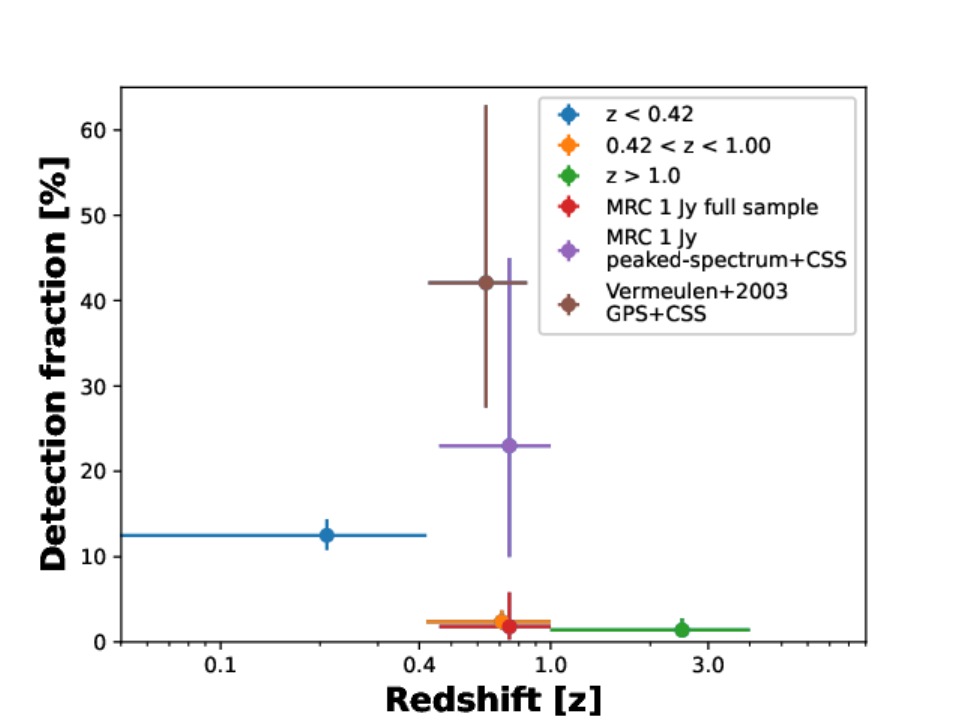} 
\\

\caption[]{The detection fraction of associated \hi 21-cm absorbers in the literature, at various redshift intervals. 
The detections fractions for various markers are: $12.5^{+1.9}_{-1.7}\%$ (Blue), $2.4^{+1.3}_{-0.9}\%$ (Orange), $1.4^{+1.4}_{-0.8}\%$ (Green), $1.8^{+4.0}_{-1.5}\%$ (Red), $23.0^{+22.1}_{-13.3}\%$ (Violet) and $42.1^{20.8}_{14.6}\%$ (Brown).
}\label{det_frac}

\end{figure}

\subsection{Detection fraction}

As discussed above, our sample is unbiased in terms of radio source properties, and covers the intermediate redshift range of $0.42 < z < 1.00 $. To compare the detection fraction in our sample with other studies in the literature, we smoothed the spectra of non-detections to a velocity resolution of $\rm 100 \ km \ s^{-1}$ and estimated the $3\sigma$ upper limits to the velocity integrated optical depth, $\int \tau dV$, assuming a line width of $\rm 100 \ km \ s^{-1}$. Here, the assumed line width of $\rm 100 \ km \ s^{-1}$ is used as a conventional value to compare the optical depth sensitivity in various literature samples. A number of earlier studies have used this velocity \citep[e.g.][]{vermeulen2003, aditya2016, aditya2018b} for the assumed \hi 21-cm line width in estimating the optical depth sensitivity; therefore it is reasonable to compare the optical depth sensitivities at $\rm 100 \ km \ s^{-1}$ in various samples.  

The median $3\sigma$ limit on the integrated optical depth in our sample, for the radio components that are closest to their WISE counterparts, is $\rm < 2.07 \ km \ s^{-1}$. Five sources in our sample have shallow upper limits (due to poor spectra with large noise) in the range of $\rm 12 - 30 \ km \ s^{-1}$, which are $\approx 6 - 15 $ times the median value, making them outliers in the sample; we hence exclude them in the detection fraction estimate. Among the remaining 57 sources, the shallowest optical depth upper limit is $\rm < 9.08 \ km \ s^{-1}$, which means that the spectrum has an optical depth sensitivity of 0.09 per $\rm 100 \ km \ s^{-1}$ channel. We consider only those detections that have an integrated optical depth $\rm > 9.08 \ km \ s^{-1}$ in our estimates. In our sample, only one detection (towards MRC 0531-237) has an integrated optical depth higher than this limit; one detection among 57 sources implies a detection fraction of $1.8\%^{+4.0\%}_{-1.5\%}$.  
The uncertainties here and elsewhere in the paper correspond to $1\sigma$, which are estimated using small-number Poisson statistics \citep[][]{gehrels1986}.

We compile a sample of all sources at redshifts $0 < z < 0.42$, $0.42 < z < 1.00$ and $z > 1.00$, from the literature that have a sensitivity same or better than the above \footnote{The references for the data are: \citet[][]{dewaard1985, vermeulen2003, yan2016, renzhi2022, srianand2015, salter2010, ostorero2017, murthy2021, murthy2022, moss2017, mirabel1989, maccagni2017, ishwar2003, gupta2006, grasha2019, glowacki2017, emonts2010, dutta2018, debreuck2003, curran2016, curran2016b, curran2017, curran2013, curran2013a, curran2008, curran2011a, curran2011b, curran2006, chandola2010, chandola2011, chandola2012, chandola2013, carilli2007, allison2012, aditya2016, aditya2017, aditya2018a, aditya2018b, aditya2019, aditya2021, chowdhury2020, uson1991, mhaskey2020}.}. Here, for optical depth sensitivities which are reported for velocity resolutions smaller than $\rm 100 \ km \ s^{-1}$, the upper limits on optical depth limits are divided by a factor of $\rm \sqrt{100.0 / v_{rep} } $, where $\rm v_{rep}$ is the reported velocity resolution. In the redshift range $0 < z < 0.42$ there are 54 detections and 378 non detections (with upper limits) implying a detection fraction of $12.5\%^{+1.9\%}_{-1.7\%}$ (see Figure~\ref{det_frac}). In the range $0.42 < z < 1.00$, there are 7 detections and 283 non-detections, yielding a detection fraction of $2.4\%^{+1.3\%}_{-0.9\%}$. Finally, at redshifts $z > 1.00$ there are 3 detections and 213 non-detections, implying a detection fraction of $1.4\%^{+1.4\%}_{-0.8\%}$.

The detection fraction in our sample is consistent (within $1\sigma$) with that in the literature sources at $0.42 < z < 1.0$. In Figure~\ref{det_frac}, a significant decline between the detection fractions of sample at $z < 0.42$, and the samples at $0.42< z < 1.00$ and $z > 1.0$ can be seen, although a difference in the detection fractions between $0.42 < z < 1.0$ and $z > 1.0$ intervals is not clear. Similar trends were reported earlier by various studies by \citet[][]{curran2008, curran2013, aditya2018b}. Also, recent FLASH pilot observations towards the Galaxy And Mass Assembly (GAMA) survey fields \citep[e.g.][]{liske2015}, reported by \citet[][]{su2022}, have yielded a detection fraction of just $2.9\%^{+9.7\%}_{-2.6\%}$.  

The detectability of AGN associated \hi 21-cm could depend on multiple factors, namely, the AGN UV and radio luminosities, the gas spin temperature, dust and cold gas content, and the gas covering factor. In the following sections, we will examine the effects of these factors in our sample.

\subsection{WISE colours}\label{wise_prop}

\begin{figure*}

\begin{tabular}{cc}

\includegraphics[width = 8.5cm]{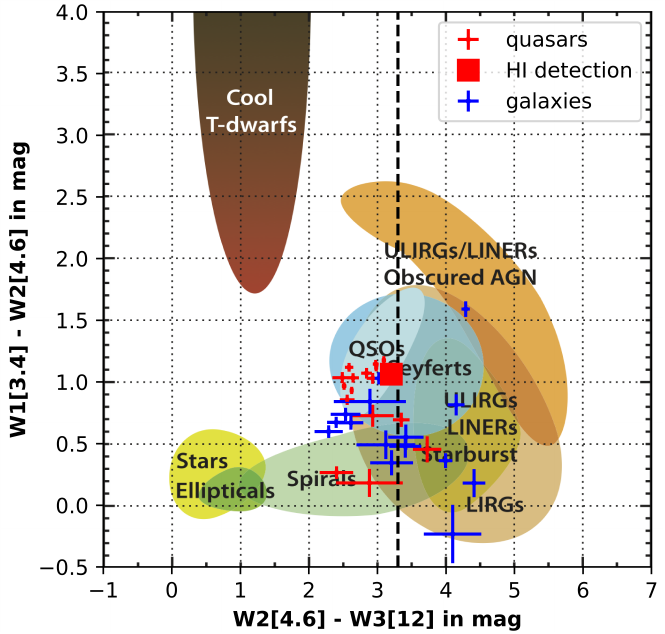} & \includegraphics[width = 8.5cm]{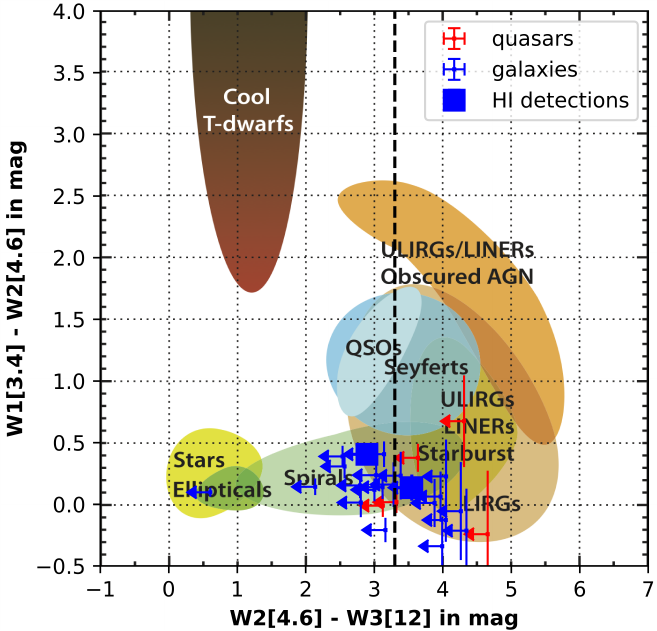}\\

\end{tabular}
\caption[]{WISE colour-colour plot for 60 sources in our sample, plotted over the source classification obtained from \citet[][]{wright2010}. For two sources, MRC 0430-235 \& MRC 2348-235 the emission is detected only in W1 band. The left panel shows the WISE colours for 33 sources which are detected in W1, W2 \& W3 bands. The panel at right shows 27 sources that are detected in only W1 \& W2 bands, and have an upper limit for the W2-W3 mag. The vertical dashed line at W2-W3 = 3.3 mag represents the cut-off reported by \citet[][]{hickox2017} to segregate dust-obscured sources.    }\label{wise_plot}

\end{figure*}

The mid- to far-infrared spectral energy distribution (i.e. 6 - 100$\mu m$ SED) of dust-obscured AGNs is
known to rise steeply with wavelength, when compared to unobscured sources \citep[][]{mullaney2011}. This is mainly due to the excess mid-infrared emission at longer wavelengths arising from the heated dust present along the line-of-sight. The Wide-field Infrared Survey Explorer (WISE) observes at 3.4 $\mu m$ (W1), 4.6 $\mu m$ (W2), 12.0 $\mu m$ (W3), 24.0 $\mu m$ (W4) wavelengths \citep[][]{wright2010}, and is particularly sensitive to the infrared luminous, dust-obscured galaxies. The obscured AGNs will have redder WISE W2-W3 mag colour, compared to unobscured ones. Based on this, \citet[][]{wright2010} have classified various galaxies using the WISE colours, as shown in Figure~\ref{wise_plot}. 

We cross-matched the source co-ordinates with AllWISE catalogue \citet[][]{cutri2003}, again using the \textbf{TOPCAT} application. Out of 62 sources, 33 sources have detections ($SNR\geq 5$) in W1, W2 \& W3 bands, 27 have detections in only W1 \& W2 bands, while 2 have detections in only W1 band.
We have plotted the WISE W2-W3 mag and W1-W2 mag colours for the 60 sources with detections in W1 \& W2 bands (and upper limit in W3 band) in Figure~\ref{wise_plot}. Here, the red markers represent quasars while the blue markers represent galaxies, as classified in the Tables~\ref{sample1} \&~\ref{sample2}, and the solid square markers represent detections of associated \hi 21-cm absorption. The sources appear to be majorly distributed in the regions of QSOs, spirals, seyferts, starburts and LIRGs, and overall there is no significant clustering of the sources around any of the background classifications. Only one source (MRC 2232-232) is lying in the region of ULIRGs/obscured AGN. Further, \citet[][]{hickox2017} used a cut-off colour of $\rm W2-W3 > 3.3$ mag to segregate obscured sources from unobscured ones; the vertical dashed line in Figure~\ref{wise_plot} represents this cut-off. In the Figure~\ref{wise_plot} only 9 out of 33 sources with measurements of W2-W3 colour have W2-W3 $>$ 3.3 mag, and, 13 out of 27 sources with upper limits on W2-W3 colour have limits higher than W2-W3 $=$ 3.3 mag (i.e. lying on the right side of the vertical dashed line). It is interesting to note that among the three detections of \hi 21-cm absorption in our sample, only one target, MRC 2216-281 at $z=0.657$, has a W2-W3 colour of 3.5 mag, marginally higher than the cut-off of 3.3 mag, while the remaining two targets (MRC 2156-245 at $z = 0.862$ and MRC 0531-237 at $z=0.851$) have $\rm W2-W3 < 3.3$ mag and do not show any strong signature for dust-obscuration \citep[e.g.][]{wright2010, hickox2017}. This indicates that associated \hi 21-cm absorption could have little relation to the presence of dust in the line-of sight, in our sample. It is interesting to note that \citet[][]{glowacki2019} found null detections in their sample with $\rm W2-W3 > 3.5$ mag. However, their \hi sensitivities were significantly poorer than the current survey; just 6 ASKAP antennas were used the observations. \citet[][]{curran2019} found a strong correlation between \hi 21-cm absorption strength and V-K mag colour which indicates dust-reddening. Also, we note that \citet[][]{carilli1998} found 4 detections in a sample of 5, where all the sources have $\rm W2-W3 < 3.5$ mag. Overall, given the small sample sizes, a further investigation using a larger sample would be needed to investigate the relation between \hi 21-cm absorption and the WISE colours. Finally, a single source in our sample, MRC 2313-277, has a very low W2-W3 colour; MRC 2313-277 has a flat near-infrared spectrum, with W1 and W2  $\approx 12$ mag, and may be representing a dust-poor elliptical galaxy.

\subsection{Radio properties}\label{radioprop}

Earlier studies at $z < 0.4$ have consistently reported a lower detection fraction of \hi 21-cm absorption in extended radio sources ($\approx 10-15\%$) compared to compact sources with linear sizes smaller than a few kpc \citep[$\approx 30\%$; GPS and CSS sources; see e.g.][]{gupta2006, maccagni2017, murthy2021}.
We note that earlier ASKAP searches towards GPS/CSS sources at $z < 0.1$, conducted by \citet[][]{allison2012} and \citet[][]{glowacki2017}, yielded 7 detections in a sample of 66 sources ($11\%^{+6\%}_{-4\%}$). However, these observations had relatively very low spectral-sensitivity, where just 6 ASKAP antennas (now 36) were used. In the case of extended radio sources, the foreground absorbing gas, which is expected to be concentrated in the nuclear regions, could be only partially covering the background radio source.
This will yield a low covering factor \citep[see][]{curran2013b}, causing a low apparent \hi optical depth and detection fraction. In addition to a low covering factor, the expanding jets could ionise and expel the ambient gas away from the jet axis. In such a scenario, while the 
nuclear regions in larger sources would be
cleared of gas and dust, young and compact sources would still be embedded in a dense medium (see more discussion in Section~\ref{dust_prop}). 

At higher redshifts, a redshift evolution in the gas properties and effects due to high AGN luminosities are also expected to become prominent, leading to a further reduction in the \hi absorption strength and detection fraction \citep[e.g.][]{curran2008, aditya2018b}. However, as noted in Section~\ref{intro}, at $z>1$, the available individual searches are small in sample size, and are mostly targeted for compact sources. While at lower redshifts, the searches are heterogeneous with a mixture of extended and compact sources,  
which inhibits rigorous testing of various scenarios.

In our sample, there are 4 peaked-spectrum sources and 9 compact steep-spectrum sources (see Table~\ref{table_peak}). Two sources with detections of \hi 21-cm absorption are classified as peaked-spectrum sources (MRC 0531-237 and MRC 2216-281), while the remaining one is classified as a CSS source (MRC 2156-245).
Overall, the detection fraction among peaked-spectrum and CSS class sources is $23\%^{+22\%}_{-13\%}$ (3 out of 13).
For comparison, \citet[][]{vermeulen2003} detected 8 absorbers in a sample of 19 GPS and CSS sources at redshifts $0.42 < z < 0.84$ (similar to our range), yielding a detection fraction of $42.1\%^{+20.8\%}_{-14.6\%}$ (see Figure~\ref{det_frac}). Albeit with large uncertainties, our estimates are consistent with those of \citet[][]{vermeulen2003} within $1\sigma$. We find no detection in the remaining 48 targets, implying a detection fraction of $0\%^{+4\%}$. The preponderance of such sources in our sample, which have an extended radio morphology, would partially explain the overall low detection fraction in our sample (as discussed above), although the effects of a redshift evolution in gas properties and the AGN luminosity, would still need to be investigated.

\begin{figure*}

\begin{tabular}{cc}
\includegraphics[width = 8.7cm]{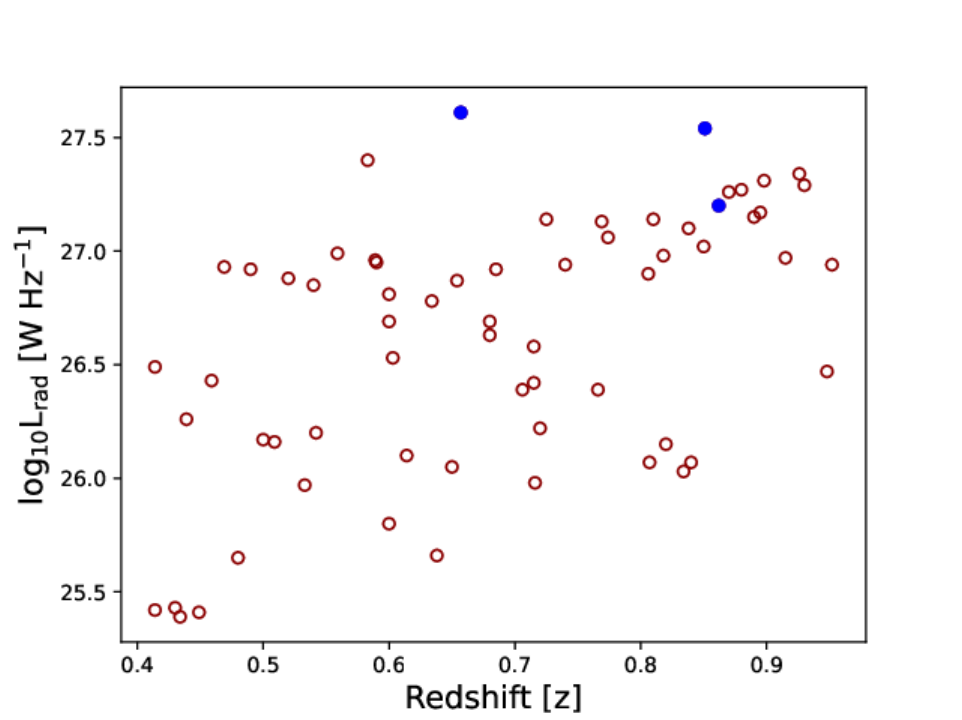} & \includegraphics[width = 9.3cm]{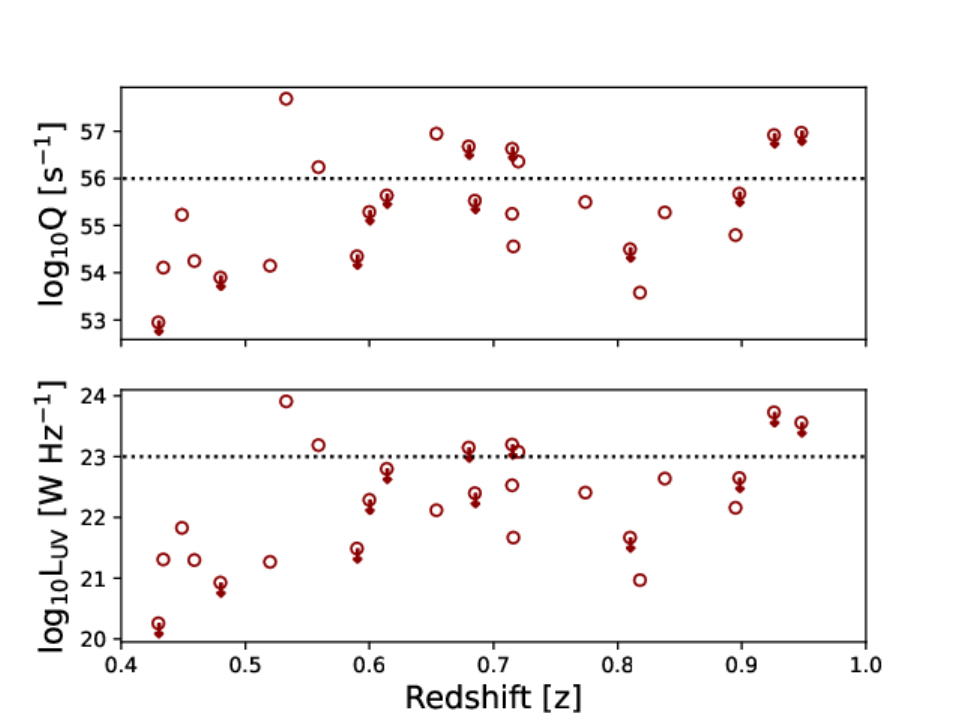} \\

\end{tabular}
\caption[]{Left panel: Rest-frame 1.4\,GHz radio luminosity as a function of redshift for all 62 targets. The solid blue markers represent detections of \hi 21-cm absorption, while the hollow red markers represent non detections. Upper right panel: the ionising photon rate as a function of redshift. Bottom right panel: ultraviolet luminosity, at 1216\AA, as a function of redshift. In both these panels the red hollow markers represent non-detections of \hi 21-cm absorption for 26 targets with available data. Among these, 14 are measurements (of ionising photon rate and UV luminosity), while remaining 12 are upper limits represented by `down' arrows. We particularly note that the three detections of \hi 21-cm absorption do not have UV data. The horizontal dotted lines represent the cut-offs suggested by \citet[][]{curran2008, curran2019}. However, we note that the cut-offs are to be treated as approximate limits. Complete ionisation would result in cases of significant departure of the luminosity from these limits.}\label{luminosity}

\end{figure*}

\subsection{\hi absorption in sources with interplanetary scintillation}

The two targets MRC 2156-245, at $z = 0.862$, and MRC 2216-281, at $z = 0.657$, that have been detected with associated \hi 21-cm absorption, are among the ones with the highest NSI in the sample, 0.79 and 0.82 (see Table~\ref{table_ips}), respectively. For the remaining sample the median detected NSI is 0.36. While the ASKAP observations have a spatial resolution of $\approx 15$ \,arcsec, which cannot reveal sub-arcsec structure, the high NSI for the two radio sources with associated \hi 21-cm absorption indicates a compact sub-arcsec structure, or, presence of compact components.  

The detection of \hi 21-cm absorption reiterates the radio source compactness, and demonstrates the effectiveness of the IPS technique, because the neutral gas covering factor is expected to be higher ($\approx 1$ \citealp[see e.g.][]{curran2013b}) for compact sources, yielding high apparent optical depth and detection chances of \hi 21-cm absorption. Indeed, the detection fraction of associated \hi 21-cm absorption is known to be higher in low-redshift gigahertz peaked-spectrum (GPS) and compact steep-spectrum (CSS) sources compared to extended radio sources \citep[e.g.][]{ pihlstrom2003, gupta2006, maccagni2017}. Given the complexities involved in conducting VLBI surveys, like the calibration and slow survey rate, the IPS technique could be an efficient alternative to identify compact radio sources, and precursors for \hi 21-cm absorption surveys.  We also emphasise the need to further test the high \hi 21-cm detection fraction in sources with high NSI.

\subsection{AGN UV and radio luminosities}
Here, we examine the effect of radio and UV luminosities of the AGN on the detections of \hi 21-cm absorption in our sample. We plot the rest-frame 1.4 GHz radio luminosity ($\rm L_{rad}$) for all our targets in the left panel of Figure~\ref{luminosity}. A gradual increase in the source luminosities as a function of redshift can be seen, which is due to a Malmquist selection bias (in a brightness-limited survey) where the high-redshift sources would have higher intrinsic luminosities.
The radio luminosities span over $\rm 25.4 - 27.7 \ W \ Hz^{-1}$. We note that the two sources with detection of \hi 21-cm absorption, MRC 0531-237 and MRC 2216-281, have the highest radio luminosities, and the third detection, MRC 2156-245, has $\rm L_{rad} > 27 \ W \ Hz^{-1}$ which is higher than a majority of the sources in the sample (see left panel of Figure~\ref{luminosity}). This indicates that the 1.4 GHz radio luminosity is likely less effective in raising the gas spin temperature, and thus, reducing the \hi absorption strength. And thus, it is probably not a dominant factor for the low detection rate in our sample.

In the upper-right and the lower-right panels of Figure~\ref{luminosity}, we plot the ultraviolet ionising photon rate ($\rm Q$; in $\rm s^{-1}$), and 1216 \AA~UV luminosity ($\rm L_{UV}$; in $\rm W \ Hz^{-1}$) respectively, for 26 targets of our sample. For the remaining sources there were not enough optical or UV photometric data points available in the literature, either to estimate the 1216 \AA~luminosity (and ionising photon rate) or to estimate an upper limit for the same. 
We note that among the 26 targets, 12 have only upper limits for the 1216 \AA~luminosity and photon rates.
Unfortunately, the three detections of \hi 21-cm absorption in our sample do not have enough literature data, and are thus not plotted in these figures. While it is difficult to draw conclusions from the limited data, we note that among the 26 sources a majority (18) have $\rm Q < 10^{56} \ s^{-1}$ and $\rm L_{UV} < 10^{23} \ W \ Hz^{-1}$. A 1216 \AA~UV photon rate of $\rm \approx 10^{56} \ s^{-1}$ is proposed by \citet[][]{curran2019} as the cut-off value above which all the neutral gas in the AGN surroundings is expected to be completely ionised, while $\rm 10^{23} \ W \ Hz^{-1}$ is proposed as an indicative AGN luminosity in excess of which most neutral gas is expected to be ionised.
We note that the cut-offs are to be treated as approximate limits. Complete ionisation would result in cases of significant departure of the luminosity from these limits. We also note the cases of TXS 1954+513 (at $z\sim1.2$; \citealp[][]{aditya2017}) and TXS 0604+728 (at $z\sim 3.5$; \citealp[][]{aditya2021}) that are examples of systems that have higher ionising photon rate and luminosity than the proposed cut-offs, but have been detected with associated neutral hydrogen gas.
Assuming a similar distribution of luminosities and ionising photon rates for the full sample of 62, only 19 sources would have 1216 \AA~luminosity and photon rate higher than the cut-offs, while the remaining 43 sources would have lower values, suggesting that the 1216 \AA~radiation from the AGN may not effectively ionise all the neutral gas in the AGN surroundings for a majority of the sources in our sample. However, we strongly note that this effect needs to be further investigated using a larger, statistically significant sample.         

\subsection{Dust \& C IV properties}\label{dust_prop}

Metal absorption lines associated with quasars provide valuable information about the nuclear environments, probing the gas along various sight-lines towards the quasar. \citet[][]{baker2002} conducted observations with HST, Anglo-Australian Telescope (AAT), the ESO 3.6m telescope at La Silla,
Chile, and VLT at Paranal, Chile,
to probe the CIV 1548 \AA \ \& 1550 \AA \ absorption and emission lines associated with the quasars in the MRC sample. Eight sources from the current sample (including MRC 2156-245, detected with \hi 21-cm absorption) are a part of this study. While both C IV 1550 \AA \ emission and 1548 \AA \ absorption were detected towards six of these sources, only emission was detected towards one source, and, MRC 2156-245 was too faint in the $\rm b_{J}$ band (20.2 mag; \citealp[see][]{baker2002}) to obtain an optical spectrum. However, MRC 2156-245 has a very red optical spectral index, $\rm \alpha_{opt} = 2$, measured between 3500 and 10000 \AA, which is the second highest in the sample studied by \citet[][]{baker2002}. Both the faintness and red spectral properties indicate presence of significant amount of dust along the line of sight. We estimate a gas to dust ratio, $\rm N_{HI} \ / \ E(B-V)$, of $\rm 5.6 \times 10^{21} \ cm^{-2} \ mag^{-1}$ \citep[see][for B and V magnitudes]{baker2002, neeleman16}, which is close to the mean Galactic ratio of $\rm 5.2 \times 10^{21} \ cm^{-2} \ mag^{-1}$ \citep[][]{shull1985}, again implying high dust content \citep[see also][]{curran2019}. However, we note here that the $\rm N_{HI}$ is estimated using the \hi 21-cm absorption line which probes only the cold component of the gas, while the total \hi column density could be larger than this. We also note that the WISE colours for the source does not indicate any strong signature for dust-obscuration, as discussed in Section~\ref{wise_prop} above.

As discussed in Section~\ref{radioprop}, earlier studies reported a relatively high detection fraction of \hi 21-cm absorption towards compact radio sources. The three detections in the current sample also have inverted radio SED shapes, and two of them have high NSI, suggesting the compact nature of the radio sources. A high detection fraction of \hi 21-cm absorption among compact radio sources is interpreted as the result of a high gas covering factor against the background radio source, yielding a high apparent optical depth \citep[e.g.][]{pihlstrom2003, curran2013b}, whereas, the covering factor would be low in the case of extended radio sources. However, it is currently not clear from these studies whether the apparently low absorption strength in extended radio sources is solely due to a low gas covering factor, or, if these systems indeed harbour lower gas densities. 

\citet[][]{baker2002} find a correlation between the C IV equivalent width and the optical spectral index, $\alpha_{opt}$, for the sample of MRC quasars in their study. The authors argue that a high optical spectral index is an indication of presence of dust, and the absorbing clouds comprise of both dust and highly-ionised gas.
They also find a strong inverse correlation between the C IV 1548 \AA \ absorption strength and the projected linear size of the radio sources; CSS and GPS sources, with sizes $\rm \lesssim few \times \ kpc$, show the strongest absorption.
Stronger absorption against the cores of compact radio sources indicate that the immediate surroundings of young radio sources could be hosting excess dust and gas, when compared to extended radio sources. Compact sources are believed to be the `younger' or undeveloped stages of an AGN \citep[e.g.][]{odea1991, odea1998}, when they are engulfed in a cocoon of gas and dust. The ambient dust would either be destroyed or pushed away from the jet axis by the expanding radio lobes, over a time scale of $\rm \sim 10^{4-6}\ years$ \citep[e.g.][]{deyoung1998}. Strong absorption lines thus arise in the opportune stages of the AGN when the lobes are still embedded in the cocoon. The high \hi 21-cm detection fraction in GPS and CSS sources could then be a result of both high gas covering factor and neutral gas density.

\section{SUMMARY}

The First Large Absorption Survey in \hi (FLASH) is the first large-area survey with no prior selection of radio sources, 
%covering all declination below $\delta < 40 \deg$, 
to search for \hi 21-cm absorption at intermediate redshifts ($0.42 < z < 1.00$). Here, we report a pilot search for associated \hi 21-cm absorption towards bright radio sources from the Molonglo Reference Catalogue (MRC) / 1 Jy sample. 
We achieved a median $3\sigma$ upper limit on integrated optical depth of $\rm 2.07 \ km \ s^{-1}$, for our sample of 62 MRC sources.
We discover three new associated \hi 21-cm absorbers from the search; towards MRC 0531-237 at $z=0.851$, MRC 2156-245 at $z=0.862$ and MRC 2216-281 at $z=0.626$. %The absorbers have line widths of $\rm \approx 30 \ km \ s^{-1}$, $\rm \approx 200 \ km \ s^{-1}$  and $\rm \approx 300 \ km \ s^{-1}$, and velocity integrated optical depths of $\rm 143.80 \pm 0.35 \ km \ s^{-1}$, $\rm 3.12 \pm 0.33 \ km \ s^{-1}$ and $\rm 0.26 \pm 0.04 \ km \ s^{-1}$, respectively. 
The overall detection fraction is $5\%^{+5\%}_{-3\%}$, which is relatively low compared to the detection fraction at low redshifts ($z < 0.4$), reported in earlier studies. The absorber towards 
MRC 0531-237 has the strongest \hi 21-cm optical depth ($\rm 143.80 \pm 0.35 \ km \ s^{-1}$) so far reported. 
We find that the continuum sources of all the three detections are intrinsically compact, showing either an inverted SEDs and/or high NSI. We do not find any significant relation between the \hi 21-cm absorption and dust obscuration as inferred from the WISE colours. Also, we do not find that a high 1.4 GHz radio luminosities of the sources has a significant effect on \hi 21-cm absorption strength and detection fraction.
The preponderance of extended radio sources in our sample could be a dominant cause for an overall low detection fraction in our sample, although the effects of high UV luminosities of AGNs and a redshift evolution in gas properties would still need to be investigated using a larger sample.

\section*{ACKNOWLEDGEMENTS}
%** Acknowledgements go here...
This scientific work uses data obtained from Inyarrimanha Ilgari Bundara / the Murchison Radio-astronomy Observatory. We acknowledge the Wajarri Yamaji People as the Traditional Owners and native title holders of the Observatory site. CSIRO’s ASKAP radio telescope is part of the Australia Telescope National Facility (\url{https://ror.org/05qajvd42}). Operation of ASKAP is funded by the Australian Government with support from the National Collaborative Research Infrastructure Strategy. ASKAP uses the resources of the Pawsey Supercomputing Research Centre. Establishment of ASKAP, Inyarrimanha Ilgari Bundara, the CSIRO Murchison Radio-astronomy Observatory, and the Pawsey Supercomputing Research Centre are initiatives of the Australian Government, with support from the Government of Western Australia and the Science and Industry Endowment Fund.
This research was supported by the Australian
Research Council Centre of Excellence for All Sky Astrophysics in 3 Dimensions (ASTRO 3D), through project number CE170100013.
 Support for the operation of the MWA is provided by the Australian Government (NCRIS), under a contract to Curtin University administered by Astronomy Australia Limited. MG acknowledges support from IDIA and was partially supported by the Australian Government through the Australian Research Council's Discovery Projects funding scheme (DP210102103).

\section*{DATA AVAILABILITY} 
All data from FLASH Pilot 1 and 2 observations are available online at the CSIRO ASKAP Science Data Archive (CASDA; \url{https://research.csiro.au/casda/}), with the project code AS109.      
\bibliographystyle{mnras}
\bibliography{ms}

\newpage

\begin {landscape}
\begin{table}

\caption{Single-component MRC sources (with peak flux density $> 40$ mJy) observed in our FLASH pilot surveys. The columns are: (1) the source name, (2) the FLASH component ID, (3) the source right ascension, (4) the source declination, (5) the source classification; QSO represents a quasar and G represents a galaxy (6) the source redshift, z, (7) the peak flux density at redshifted \hi 21-cm frequency, from our ASKAP observations, (8) the integrated flux density, (9) rms noise on the continuum image in units of mJy per beam, (10) rms noise on the spectrum in mJy per beam per channel, (11) velocity integrated optical depth, $\int \tau dv$, and the 3$\sigma$ upper limits to optical depth for non-detections, assuming a line width of $\rm 100 \ km \ s^{-1}$, (12) the largest angular size (LAS) of the source, measured at 5 GHz by \citep{kapahi1998a} for the galaxies and \citep{kapahi1998b} for the quasars, (13) notes on the radio morphology, based on the 5 GHz images presented by \citep{kapahi1998a} and \citep{kapahi1998b}. }\label{sample1}

\begin{tabular}{|l|l|l|l|c|r|r|r|r|r|r|r|l}
\hline
\hline
  \multicolumn{1}{|c|}{Source name} &
  \multicolumn{1}{c|}{Component ID} &
  \multicolumn{1}{c|}{RA} &
  \multicolumn{1}{c|}{DEC} &
  \multicolumn{1}{l|}{Source$^1$} &
  \multicolumn{1}{c|}{Redshift$^2$} &
  \multicolumn{1}{c|}{$\rm F_{\rm peak}$} &
  \multicolumn{1}{c|}{$\rm F_{\rm int}$} &
  \multicolumn{1}{c|}{$\sigma _ {image}$} &
  \multicolumn{1}{c|}{$\sigma _ {channel}$} &
  \multicolumn{1}{c|}{$\tau$} &
  \multicolumn{1}{l|}{LAS$\_5$\,GHz} &
   \multicolumn{1}{c|}{Notes}  \\
 
\multicolumn{1}{|c|}{ } &
  \multicolumn{1}{c|}{[FLASH]} &
  \multicolumn{1}{c|}{ } &
  \multicolumn{1}{c|}{ } &
  \multicolumn{1}{l|}{type} &
  \multicolumn{1}{c|}{[z] } &
  \multicolumn{1}{c|}{ [mJy/beam] } &
  \multicolumn{1}{c|}{[mJy] } &
  \multicolumn{1}{c|}{ [mJy/beam] } &
  \multicolumn{1}{c|}{ [mJy/beam] } &
  \multicolumn{1}{c|}{[$\rm km \ s^{-1}$] } & 
  \multicolumn{1}{l|}{[arcsec]} & \\
      \multicolumn{1}{|c|}{(1) } &
  \multicolumn{1}{c|}{(2)} &
  \multicolumn{1}{c|}{(3) } &
  \multicolumn{1}{c|}{(4) } &
  \multicolumn{1}{c|}{(5) } &
  \multicolumn{1}{c|}{(6) } &
  \multicolumn{1}{c|}{ (7) } &
  \multicolumn{1}{c|}{(8) } &
  \multicolumn{1}{c|}{ (9) } &
  \multicolumn{1}{c|}{ (10) } &
  \multicolumn{1}{c|}{ (11)} &
  \multicolumn{1}{c|}{ (12)} &
  \multicolumn{1}{c|}{ (13)} \\  
  \hline
\hline
MRC 0030-220 & SB11053\_component\_10a & 00:32:44.6 & -21:44:21 & QSO & 0.806 & 472.5 & 490.3 & 0.21 & 9.5 & $<$2.3 & 3.9 & lobe-dominated \\
MRC 0035-231 & SB11053\_component\_6a & 00:38:24.9 & -22:53:03 & G & 0.685 & 682.3 & 711.5 & 0.20 & 7.2 & $<$1.1 & 2.3 & compact \\
MRC 0050-222 & SB13291\_component\_7a & 00:52:42.8 & -21:55:47 & G & 0.654 & 660.6 & 689.7 & 0.14 & 6.7 & $<$0.9 & $<2$ & compact \\
MRC 0106-233 & SB13291\_component\_9a & 01:09:02.9 & -23:07:30 & QSO & 0.818 & 545.4 & 575.4 & 0.20 & 11.6 & $<$1.9 & 2.5 & lobe-dominated \\
MRC 0118-272 & SB37449\_component\_5a & 01:20:31.5 & -27:01:23 & QSO & $>$0.557 & 1220.1 & 1332.4 & 0.27 & 5.4 & $<$0.4 & $<2$ & compact, BL Lac \\
&& \\
MRC 0144-227 & SB13281\_component\_6a & 01:47:09.2 & -22:32:41 & G & 0.600 & 699.3 & 720.3 & 0.11 & 4.9 & $<$0.6  & $<2$ & compact \\
MRC 0201-214 & SB13268\_component\_15a & 02:03:32.1 & -21:13:46 & G & 0.915 & 421.5 & 466.5 & 0.28 & 26.2 & $<$5.5 & $<2$ & compact \\ 
MRC 0209-237 & SB13268\_component\_12a & 02:11:28.3 & -23:28:23 & QSO & 0.680 & 403.5 & 869.4 & 0.08 & 4.7 & $<$1.1 & 18 & core+hotspots \\
MRC 0225-241 & SB13268\_component\_4a & 02:27:33.5 & -23:54:55 & G & 0.520 & 1116.5 & 1317.4 & 0.41 & 5.1 & $<$0.6 & 5.2 & compact triple$^a$  \\
MRC 0230-245 & SB15212\_component\_5a & 02:32:30.4 & -24:22:05 & G & 0.880 & 922.4 & 1284.6 & 0.57 & 6.7 & $<$0.8 & 11.3 & lobe-dominated  \\
&& \\
MRC 0233-290 & SB37451\_component\_1a & 02:35:56.2 & -28:50:46 & G & 0.725 & 922.4 & 1000.3 & 0.37 & 10.6 & $<1.0$ & $<5$ & compact  \\
MRC 0319-298 & SB37452\_component\_1a & 03:21:28.8 & -29:40:45 & G & 0.583 & 2858.3 & 2990.2 & 0.89 & 6.5 & $<$0.3 & $<2$ & compact \\ 
%MRC 0337-216 & SB37432\_component\_9a & 03:39:39.1 & -21:29:26 & G & 0.414 & 708.9 & 752.1 & 0.65 & 17.3 & $<$2.070 \\ %%% Remove, z<0.42 
MRC 0338-259 & SB37432\_component\_25a & 03:40:39.4 & -25:47:32 & QSO & 0.439 & 371.1 & 590.4 & 0.18 & 5.4 & $<$1.1 & 19.7 & core+lobes \\
MRC 0418-288 & SB37797\_component\_12a & 04:20:36.4 & -28:41:15 & QSO & 0.850 & 556.5 & 575.9 & 0.49 & 22.6 & $<$3.2 & $<2$ & compact \\    
MRC 0424-268 & SB37797\_component\_3a & 04:26:40.6 & -26:43:45 & G & 0.469 & 1529.3 & 2044.5 & 0.58 & 6.3 & $<$0.4 & 22.5 & jets+hotspots  \\
&& \\
MRC 0450-221 & SB34547\_component\_6a & 04:52:44.7 & -22:01:18 & QSO & 0.898 & 974.3 & 1508.0 & 0.23 & 13.2 & $<$1.8 & 14.3 & core+hotspots \\
MRC 0531-237 & SB41061\_component\_1a & 05:33:54.5 & -23:44:31 & G & 0.851 & 1841.1 & 1913.6 & 0.21 & 6.7 & $143.8 \pm 0.4$ & $<2$ & compact  \\
MRC 0930-200 & SB34571\_component\_1a & 09:32:45.9 & -20:17:53 & G & 0.769 & 870.4 & 1992.2 & 0.21 & 60.0 & $<$5.7 & 20.2 & hotspots, no core \\
MRC 1002-216 & SB34561\_component\_1a & 10:05:08.3 & -21:52:25 & G & 0.490 & 1364.5 & 1391.6 & 0.31 & 61.5 & $<$3.7 & $<2$ & compact \\
MRC 2024-217 & SB13372\_component\_4a & 20:27:04.2 & -21:36:19 & QSO & 0.459 & 507.4 & 1266.6 & 0.20 & 10.7 & $<$1.8 & 31 & core+hotspots \\
&& \\
MRC 2156-245 & SB10849\_component\_8a & 21:59:24.9 & -24:17:52 & QSO & 0.862 & 805.7 & 827.7 & 0.19 & 7.5 & $3.1 \pm 0.3$ & $<2$ & compact \\
MRC 2216-281 & SB10849\_component\_1a & 22:19:42.6 & -27:56:27 & G & 0.657 & 3652.8 & 3706.3 & 1.91 & 8.2 & $0.3 \pm 0.1$ & $<2$ & compact \\
MRC 2255-282 & SB11052\_component\_3a & 22:58:06.0 & -27:58:21 & QSO & 0.926 & 976.3 & 1051.9 & 0.18 & 8.7 & $<$0.8 & $<1$ & compact \\  
MRC 2303-253 & SB11052\_component\_9a & 23:06:26.9 & -25:06:54 & G & 0.740 & 614.7 & 1285.0 & 0.15 & 4.3 & $<$0.7 & 18.6 & core+hotspots \\
MRC 2341-244 & SB13279\_component\_3a & 23:44:12.2 & -24:07:41 & G & 0.590 & 979.7 & 1025.7 & 1.21 & 5.4 & $<$0.5 & 2.1 & compact \\
\hline
\hline
\end{tabular}

Notes $^1$ $^2$: Reference for source classification: \citet[][]{mccarthy1996, kapahi1998a, kapahi1998b}. References for redshifts are \citet[][]{mccarthy1996, kapahi1998a}. 
Compact triple$^a$: A source with a relatively bright compact core, with faint and nearly symmetric extended lobes. \cite[e.g.][]{cseh2008}

\end{table}
 
\end{landscape} 

%Table 3
\begin {landscape}
\begin{table}
%\label{sample2}
\caption{ The MRC 1-Jy sources observed in the FLASH pilot surveys that have multiple Gaussian components (with peak flux density $> 40$ mJy) in the ASKAP continuum image. The columns are same as those of Table~\ref{sample1}. The Gaussian components in various sources are labelled with alphabets [a], [b], [c] etc., appended with the MRC source name.
} \label{sample2} %\begin{tabular}{llrrrrclllllllll}

\begin{tabular}{|l|l|l|l|l|l|l|l|l|l|l|l}
\hline
\hline
  \multicolumn{1}{|c|}{Source name} &
  \multicolumn{1}{c|}{Component ID} &
  \multicolumn{1}{c|}{RA} &
  \multicolumn{1}{c|}{DEC} &
  \multicolumn{1}{l|}{Source$^1$} &
  \multicolumn{1}{c|}{Redshift$^2$} &
  \multicolumn{1}{c|}{$\rm F_{peak}$} &
  \multicolumn{1}{c|}{$\rm F_{int}$} &
  \multicolumn{1}{c|}{$\sigma_{image}$} &
  \multicolumn{1}{c|}{$\sigma_{channel}$} &
  \multicolumn{1}{c|}{$\tau$} & \\
% \multicolumn{1}{c|}{Notes} \\
\multicolumn{1}{|c|}{ } &
  \multicolumn{1}{c|}{[FLASH]} &
  \multicolumn{1}{c|}{ } &
  \multicolumn{1}{c|}{ } &
  \multicolumn{1}{l|}{type } &
  \multicolumn{1}{c|}{[z] } &
  \multicolumn{1}{c|}{ [mJy] } &
  \multicolumn{1}{c|}{[mJy] } &
  \multicolumn{1}{c|}{ [mJy] } &
  \multicolumn{1}{c|}{ [mJy] } &
  \multicolumn{1}{c|}{[$\rm km \ s^{-1}$] } \\
  
      \multicolumn{1}{|c|}{(1) } &
  \multicolumn{1}{c|}{(2)} &
  \multicolumn{1}{c|}{(3) } &
  \multicolumn{1}{c|}{(4) } &
  \multicolumn{1}{c|}{(5) } &
  \multicolumn{1}{c|}{(6) } &
  \multicolumn{1}{c|}{ (7) } &
  \multicolumn{1}{c|}{(8) } &
  \multicolumn{1}{c|}{ (9) } &
  \multicolumn{1}{c|}{ (10) } &
  \multicolumn{1}{c|}{ (11)} & \\
%    \multicolumn{1}{c|}{ (12)} \\

  \hline
\hline

  MRC 0042-248[a] & SB13291\_component\_12a & 00:45:05.0 & -24:34:10 & QSO & 0.807$^a$ & 416.7 & 465.8 & 0.13 & 5.6 & $<$1.4 \\ %& lobe \\
  MRC 0042-248[b] & SB13291\_component\_12b & 00:45:00.7 & -24:34:43 & QSO & 0.807$^a$ & 145.4 & 187.4 & 0.13 & 5.9 & $<$3.9 \\ %& lobe \\
  MRC 0042-248[c] & SB13291\_component\_12c & 00:45:00.4 & -24:34:47 & QSO & 0.807$^a$ & 109.0 & 110.6 & 0.13 & 6.0 & $<$5.1 \\ %& lobe \\
  MRC 0042-248[d] & SB13291\_component\_12d & 00:45:04.5 & -24:34:15 & QSO & 0.807$^a$ & 68.9 & 114.3 & 0.13 & 5.7 & $<$8.2 \\ %& lobe \\
  &&\\
  
  %MRC 0055-258[a] & SB13291\_component\_110a & 00:57:43.3 & -25:33:22 & G & 0.414$^b$ & 61.4 & 132.3 & 0.10 & 4.0 & $<$4.8 & lobe \\
  %MRC 0055-258[b] & SB13291\_component\_110b & 00:57:44.2 & -25:33:29 & G & 0.414$^b$ & 49.9 & 60.1 & 0.10 & 3.8 & $<$6.3 & lobe \\
  %&&\\ % Removed from sample

  MRC 0058-229[a] & SB13291\_component\_48a & 01:00:42.0 & -22:39:43 & QSO & 0.706 & 190.2 & 423.4 & 0.09 & 5.6 & $<$2.7 \\ %& lobe \\
  MRC 0058-229[b] & SB13291\_component\_48b & 01:00:41.9 & -22:40:28 & QSO & 0.706 & 99.3 & 292.6 & 0.09 & 5.7 & $<$5.5 \\ %& lobe \\
  &&\\

  MRC 0123-226[a] & SB37449\_component\_7a & 01:26:14.8 & -22:22:31 & QSO & 0.720 & 767.6 & 804.9 & 0.19 & 7.6 & $<$0.9 \\ %& core+lobe \\
  MRC 0123-226[b] & SB37449\_component\_7b & 01:26:15.0 & -22:22:36 & QSO & 0.720 & 121.9 & 107.9 & 0.19 & 7.4 & $<$5.5 \\ %& core+lobe \\
  &&\\

  MRC 0135-247[a] & SB37449\_component\_9a & 01:37:38.2 & -24:30:53 & QSO & 0.838 & 590.8 & 645.2 & 0.17 & 18.1 & $<$2.9\\
  MRC 0135-247[b] & SB37449\_component\_9b & 01:37:37.1 & -24:30:54 & QSO & 0.838 & 455.7 & 550.7 & 0.17 & 17.7 & $<$3.4\\
  MRC 0135-247[c] & SB37449\_component\_9c & 01:37:39.4 & -24:30:57 & QSO & 0.838 & 91.7 & 108.5 & 0.17 & 18.0 & $<$16.7\\
  &&\\
  
  MRC 0143-246[a] & SB13281\_component\_15a & 01:45:21.3 & -24:23:54 & G & 0.716 & 387.0 & 433.3 & 0.09 & 4.5 & $<$1.1 \\  %& lobe \\
  MRC 0143-246[b] & SB13281\_component\_15b & 01:45:21.4 & -24:23:06 & G & 0.716 & 204.6 & 306.1 & 0.09 & 4.3 & $<$1.9 \\ %& lobe \\
  MRC 0143-246[c] & SB13281\_component\_15c & 01:45:21.4 & -24:23:46 & G & 0.716 & 72.1 & 139.7 & 0.09 & 4.4 & $<$5.4 \\ %& core+lobe\\
  &&\\

  MRC 0149-299[a] & SB41226\_component\_3a & 01:52:01.3 & -29:41:00 & G & 0.603 & 504.3 & 493.9 & 0.12 & 6.1 & $<$1.1\\
  MRC 0149-299[b] & SB41226\_component\_3b & 01:52:01.2 & -29:41:03 & G & 0.603 & 384.2 & 438.6 & 0.12 & 6.1 & $<$1.5\\
  MRC 0149-299[c] & SB41226\_component\_3c & 01:52:00.9 & -29:41:00 & G & 0.603 & 305.4 & 401.4 & 0.12 & 6.2 & $<$1.9\\
  MRC 0149-299[d] & SB41226\_component\_3c & 01:52:00.3 & -29:40:51 & G & 0.603 & 264.7 & 235.3 & 0.12 & 6.2 & $<$2.2\\
  MRC 0149-299[e] & SB41226\_component\_3c & 01:52:00.4 & -29:40:58 & G & 0.603 & 79.8 & 81.9 & 0.12 & 6.2 & $<$7.3\\
  &&\\

  MRC 0205-229[a] & SB13268\_component\_20a & 02:07:40.0 & -22:44:50 & G & 0.680 & 351.5 & 579.9 & 0.22 & 6.0 & $<$1.8\\
  MRC 0205-229[b] & SB13268\_component\_20b & 02:07:38.6 & -22:45:05 & G & 0.680 & 324.9 & 558.6 & 0.22 & 6.0 & $<$2.0\\
  &&\\

  MRC 0209-282[a] & SB13268\_component\_11a & 02:12:10.3 & -28:00:17 & G & 0.600 & 431.1 & 454.2 & 0.13 & 5.0 & $<$1.1\\
  MRC 0209-282[b] & SB13268\_component\_11b & 02:12:10.3 & -28:00:07 & G & 0.600 & 236.9 & 256.3 & 0.13 & 5.2 & $<$1.9\\
  MRC 0209-282[c] & SB13268\_component\_11c & 02:12:10.1 & -28:00:02 & G & 0.600 & 155.9 & 140.7 & 0.13 & 5.2 & $<$2.8\\
  &&\\

  MRC 0223-245[a] & SB13268\_component\_7a & 02:25:29.5 & -24:21:15 & G & 0.634 & 577.1 & 569.1 & 0.11 & 4.7 & $<$0.7\\
  MRC 0223-245[b] & SB13268\_component\_7b & 02:25:29.7 & -24:21:19 & G & 0.634 & 150.9 & 130.7 & 0.11 & 4.9 & $<$3.0\\
  &&\\

  MRC 0231-235[a] & SB15212\_component\_8a & 02:33:23.9 & -23:21:14 & G & 0.810 & 807.9 & 1391.2 & 0.43 & 7.5 & $<$1.0\\
  MRC 0231-235[b] & SB15212\_component\_8b & 02:33:25.8 & -23:20:51 & G & 0.810 & 708.6 & 1021.8 & 0.43 & 7.7 & $<$1.4\\
  &&\\

\hline
\hline
\end{tabular}

\end{table}
 
\end{landscape} 

%Table 3 (contd) 
\begin {landscape}
\begin{table}
 \setcounter{table}{4}
\ContinuedFloat
\caption{continued..}

\begin{tabular}{|l|l|l|l|l|l|l|l|l|l|l|}
\hline
\hline
  \multicolumn{1}{|c|}{Source name} &
  \multicolumn{1}{c|}{Component ID} &
  \multicolumn{1}{c|}{RA} &
  \multicolumn{1}{c|}{DEC} &
  \multicolumn{1}{l|}{Source$^1$} &
  \multicolumn{1}{c|}{Redshift$^2$} &
  \multicolumn{1}{c|}{$\rm F_{peak}$} &
  \multicolumn{1}{c|}{$\rm F_{int}$} &
  \multicolumn{1}{c|}{$\sigma _ {image}$} &
  \multicolumn{1}{c|}{$\sigma _ {channel}$} &
  \multicolumn{1}{c|}{$\tau$} \\
 
\multicolumn{1}{|c|}{ } &
  \multicolumn{1}{c|}{[FLASH]} &
  \multicolumn{1}{c|}{ } &
  \multicolumn{1}{c|}{ } &
  \multicolumn{1}{l|}{type } &
  \multicolumn{1}{c|}{[z] } &
  \multicolumn{1}{c|}{ [mJy] } &
  \multicolumn{1}{c|}{[mJy] } &
  \multicolumn{1}{c|}{ [mJy] } &
  \multicolumn{1}{c|}{ [mJy] } &
  \multicolumn{1}{c|}{ [$\rm km \ s^{-1}$]} \\
  
      \multicolumn{1}{|c|}{(1) } &
  \multicolumn{1}{c|}{(2)} &
  \multicolumn{1}{c|}{(3) } &
  \multicolumn{1}{c|}{(4) } &
  \multicolumn{1}{c|}{(5) } &
  \multicolumn{1}{c|}{(6) } &
  \multicolumn{1}{c|}{ (7) } &
  \multicolumn{1}{c|}{(8) } &
  \multicolumn{1}{c|}{ (9) } &
  \multicolumn{1}{c|}{ (10) } &
  \multicolumn{1}{c|}{ (11)} \\
  
  \hline
\hline

  MRC 0254-274[a] & SB13269\_component\_104a & 02:56:51.5 & -27:18:03 & G & 0.480 & 76.1 & 153.0 & 0.11 & 6.1 & $<$6.5\\
  MRC 0254-274[b] & SB13269\_component\_104b & 02:56:51.1 & -27:17:48 & G & 0.480 & 63.5 & 103.1 & 0.11 & 6.2 & $<$7.7\\
  &&\\

  MRC 0254-236[a] & SB13269\_component\_1a & 02:56:15.5 & -23:24:40 & G & 0.509 & 845.6 & 1406.2 & 0.19 & 6.4 & $<$0.7\\
  MRC 0254-236[b] & SB13269\_component\_1b & 02:56:14.3 & -23:25:06 & G & 0.509 & 729.6 & 1161.5 & 0.19 & 6.7 & $<$1.0\\
  &&\\

  MRC 0320-267[a] & SB13269\_component\_3a & 03:23:00.8 & -26:36:14 & G & 0.890 & 681.8 & 724.9 & 0.14 & 7.7 & $<$1.1\\
  MRC 0320-267[b] & SB13269\_component\_3b & 03:23:01.3 & -26:36:11 & G & 0.890 & 58.4 & 50.4 & 0.14 & 7.8 & $<$13.4\\ 
  &&\\

  MRC 0325-260[a] & SB37432\_component\_57a & 03:27:41.0 & -25:51:28 & G & 0.638 & 168.3 & 182.8 & 0.20 & 6.8 & $<$3.7\\
  MRC 0325-260[b] & SB37432\_component\_57b & 03:27:41.5 & -25:52:24 & G & 0.638 & 86.3 & 112.5 & 0.20 & 6.6 & $<$6.3\\
  MRC 0325-260[c] & SB37432\_component\_57c & 03:27:41.4 & -25:51:40 & G & 0.638 & 43.7 & 93.2 & 0.20 & 6.6 & $<$13.9\\
  MRC 0325-260[d] & SB37432\_component\_57d & 03:27:41.5 & -25:52:10 & G & 0.638 & 40.3 & 97.2 & 0.20 & 6.9 & $<$15.8\\
  &&\\

  MRC 0327-241[a] & SB37432\_component\_6a & 03:29:54.0 & -23:57:08 & QSO & 0.895 & 709.2 & 804.3 & 0.48 & 8.3 & $<$1.0\\
  MRC 0327-241[b] & SB37432\_component\_6b & 03:29:53.6 & -23:57:10 & QSO & 0.895 & 147.2 & 161.8 & 0.48 & 8.2 & $<$4.6\\  
  &&\\

  MRC 0428-281[a] & SB37797\_component\_8a & 04:30:15.9 & -28:00:48 & G & 0.650 & 609.6 & 634.8 & 0.28 & 7.6 & $<$1.1\\
  MRC 0428-281[b] & SB37797\_component\_8b & 04:30:20.4 & -28:00:39 & G & 0.650 & 503.8 & 542.3 & 0.28 & 7.8 & $<$1.2\\
  MRC 0428-281[c] & SB37797\_component\_8c & 04:30:16.8 & -28:00:48 & G & 0.650 & 138.8 & 205.7 & 0.28 & 7.5 & $<$4.7\\
  MRC 0428-281[d] & SB37797\_component\_8d & 04:30:19.4 & -28:00:41 & G & 0.650 & 101.8 & 154.2 & 0.28 & 7.8 & $<$6.1\\
  &&\\

  MRC 0428-271[a] & SB37797\_component\_13a & 04:30:20.3 & -27:00:01 & G & 0.840 & 503.4 & 546.5 & 0.20 & 8.2 & $<$1.7\\
  MRC 0428-271[b] & SB37797\_component\_13b & 04:30:17.3 & -26:59:49 & G & 0.840 & 356.4 & 389.5 & 0.20 & 8.0 & $<$2.2\\
  MRC 0428-271[c] & SB37797\_component\_13c & 04:30:18.7 & -26:59:54 & G & 0.840 & 63.9 & 105.6 & 0.20 & 8.1 & $<$12.9\\  
  &&\\

  MRC 0430-235[a] & SB37797\_component\_67a & 04:32:55.4 & -23:25:20 & G & 0.820 & 129.5 & 152.4 & 0.16 & 7.5 & $<$5.3\\
  MRC 0430-235[b] & SB37797\_component\_67b & 04:32:51.8 & -23:24:08 & G & 0.820 & 117.7 & 132.0 & 0.16 & 7.6 & $<$5.8\\
  MRC 0430-235[c] & SB37797\_component\_67c & 04:32:52.0 & -23:24:17 & G & 0.820 & 93.6 & 142.6 & 0.16 & 7.8 & $<$7.6\\
  MRC 0430-235[d] & SB37797\_component\_67d & 04:32:51.7 & -23:24:26 & G & 0.820 & 79.5 & 116.9 & 0.16 & 7.8 & $<$9.1\\
  &&\\
  
  MRC 0437-244[a] & SB37797\_component\_25a & 04:39:08.3 & -24:21:22 & QSO & 0.834 & 273.2 & 341.5 & 0.11 & 8.3 & $<$3.0\\
  MRC 0437-244[b] & SB37797\_component\_25b & 04:39:10.8 & -24:23:18 & QSO & 0.834 & 149.4 & 248.4 & 0.11 & 8.1 & $<$4.5\\
  MRC 0437-244[c] & SB37797\_component\_25c & 04:39:08.8 & -24:21:37 & QSO & 0.834 & 58.2 & 199.4 & 0.11 & 8.2 & $<$13.2\\
  &&\\

  MRC 0454-220[a] & SB34547\_component\_5a & 04:56:12.2 & -21:59:38 & QSO & 0.533 & 1207.9 & 1334.9 & 0.19 & 9.9 & $<$1.2\\
  MRC 0454-220[b] & SB34547\_component\_5b & 04:56:07.2 & -21:58:56 & QSO & 0.533 & 860.4 & 1018.6 & 0.19 & 9.2 & $<$1.3\\
  MRC 0454-220[c] & SB34547\_component\_5c & 04:56:12.0 & -21:59:35 & QSO & 0.533 & 161.9 & 218.1 & 0.19 & 10.2 & $<$9.5\\
  MRC 0454-220[d] & SB34547\_component\_5d & 04:56:08.9 & -21:59:11 & QSO & 0.533 & 128.2 & 112.3 & 0.19 & 8.9 & $<$8.0\\
  MRC 0454-220[e] & SB34547\_component\_5e & 04:56:09.3 & -21:59:15 & QSO & 0.533 & 90.3 & 374.8 & 0.19 & 9.2 & $<$11.6\\
  &&\\
  
  MRC 0522-239[a] & SB41061\_component\_20a & 05:24:54.0 & -23:52:20 & G & 0.500 & 206.5 & 279.9 & 0.15 & 5.4 & $<$2.2\\
  MRC 0522-239[b] & SB41061\_component\_20b & 05:24:54.6 & -23:52:15 & G & 0.500 & 184.2 & 218.2 & 0.15 & 5.4 & $<$2.4\\
  MRC 0522-239[c] & SB41061\_component\_20c & 05:24:53.5 & -23:52:25 & G & 0.500 & 150.3 & 158.7 & 0.15 & 5.3 & $<$3.0\\
  &&\\

\hline
\hline
\end{tabular}

\end{table}
 
\end{landscape} 

%Table 3 (contd)
\begin {landscape}
\begin{table}
 \setcounter{table}{4}
\ContinuedFloat
\caption{continued..}

\begin{tabular}{|l|l|l|l|l|l|l|l|l|l|l|}
\hline
\hline
  \multicolumn{1}{|c|}{Source name} &
  \multicolumn{1}{c|}{Component ID} &
  \multicolumn{1}{c|}{RA} &
  \multicolumn{1}{c|}{DEC} &
  \multicolumn{1}{l|}{Source$^1$} &
  \multicolumn{1}{c|}{Redshift$^2$} &
  \multicolumn{1}{c|}{$\rm F_{peak}$} &
  \multicolumn{1}{c|}{$\rm F_{int}$} &
  \multicolumn{1}{c|}{$\sigma _ {image}$} &
  \multicolumn{1}{c|}{$\sigma _ {channel}$} &
  \multicolumn{1}{c|}{$\tau$} \\
 
\multicolumn{1}{|c|}{ } &
  \multicolumn{1}{c|}{[FLASH]} &
  \multicolumn{1}{c|}{ } &
  \multicolumn{1}{c|}{ } &
  \multicolumn{1}{l|}{type } &
  \multicolumn{1}{c|}{[z] } &
  \multicolumn{1}{c|}{ [mJy] } &
  \multicolumn{1}{c|}{[mJy] } &
  \multicolumn{1}{c|}{ [mJy] } &
  \multicolumn{1}{c|}{ [mJy] } &
  \multicolumn{1}{c|}{ [$\rm km \ s^{-1}$]} \\
  
      \multicolumn{1}{|c|}{(1) } &
  \multicolumn{1}{c|}{(2)} &
  \multicolumn{1}{c|}{(3) } &
  \multicolumn{1}{c|}{(4) } &
  \multicolumn{1}{c|}{(5) } &
  \multicolumn{1}{c|}{(6) } &
  \multicolumn{1}{c|}{ (7) } &
  \multicolumn{1}{c|}{(8) } &
  \multicolumn{1}{c|}{ (9) } &
  \multicolumn{1}{c|}{ (10) } &
  \multicolumn{1}{c|}{ (11)} \\
  
  \hline
\hline

  MRC 0941-200[a] & SB34571\_component\_24a & 09:43:50.4 & -20:19:41 & QSO & 0.715 & 287.3 & 437.6 & 0.09 & 7.9 & $<$3.1\\
  MRC 0941-200[b] & SB34571\_component\_24b & 09:43:50.3 & -20:19:01 & QSO & 0.715 & 86.9 & 216.7 & 0.09 & 7.6 & $<$9.7\\
  &&\\

  MRC 1002-215[a] & SB34561\_component\_6a & 10:05:11.4 & -21:45:09 & G & 0.589 & 1043.7 & 1241.2 & 0.38 & 64.1 & $<$5.1\\
  MRC 1002-215[b] & SB34561\_component\_6b & 10:05:11.6 & -21:44:41 & G & 0.589 & 1008.8 & 1278.9 & 0.38 & 71.5 & $<$5.9\\
  &&\\

  MRC 2201-272[a] & SB10849\_component\_7a & 22:04:35.1 & -27:01:17 & G & 0.930 & 865.6 & 929.8 & 0.31 & 7.3 & $<$0.8\\
  MRC 2201-272[b] & SB10849\_component\_7b & 22:04:34.8 & -27:01:24 & G & 0.930 & 58.8 & 45.3 & 0.31 & 7.3 & $<$10.5\\
  &&\\
   
  MRC 2213-283[a] & SB10849\_component\_10a & 22:16:03.0 & -28:03:31 & QSO & 0.948 & 499.2 & 552.7 & 0.25 & 7.9 & $<$1.5\\
  MRC 2213-283[b] & SB10849\_component\_10b & 22:15:58.9 & -28:03:30 & QSO & 0.948 & 476.4 & 489.6 & 0.25 & 8.4 & $<$1.7\\
  MRC 2213-283[c] & SB10849\_component\_10c & 22:15:59.6 & -28:03:30 & QSO & 0.948 & 124.4 & 170.9 & 0.25 & 8.2 & $<$6.6\\
  MRC 2213-283[d] & SB10849\_component\_10d & 22:16:02.2 & -28:03:29 & QSO & 0.948 & 93.6 & 192.8 & 0.25 & 8.3 & $<$8.7\\
  &&\\

  MRC 2229-228[a] & SB11051\_component\_5a & 22:32:05.5 & -22:33:41 & G & 0.542 & 343.1 & 305.8 & 0.60 & 5.5 & $<$1.3\\
  MRC 2229-228[b] & SB11051\_component\_5b & 22:32:05.9 & -22:33:43 & G & 0.542 & 211.5 & 195.5 & 0.60 & 5.4 & $<$2.1\\
  MRC 2229-228[c] & SB11051\_component\_5c & 22:32:05.7 & -22:33:37 & G & 0.542 & 178.1 & 189.7 & 0.60 & 5.4 & $<$2.5\\
  MRC 2229-228[d] & SB11051\_component\_5d & 22:32:05.1 & -22:33:43 & G & 0.542 & 70.8 & 41.7 & 0.60 & 5.5 & $<$6.5\\
  &&\\

  MRC 2232-232[a] & SB11051\_component\_3a & 22:35:20.5 & -22:59:42 & G & 0.870 & 914.7 & 1001.4 & 0.18 & 6.8 & $<$0.7\\
  MRC 2232-232[b] & SB11051\_component\_3b & 22:35:19.6 & -22:59:53 & G & 0.870 & 315.4 & 318.1 & 0.18 & 6.9 & $<$2.3\\
  &&\\

  MRC 2236-264[a] & SB11051\_component\_27a & 22:39:20.3 & -26:10:18 & G & 0.430 & 171.9 & 270.9 & 0.21 & 4.2 & $<$2.0\\
  MRC 2236-264[b] & SB11051\_component\_27b & 22:39:19.1 & -26:09:57 & G & 0.430 & 88.9 & 204.0 & 0.21 & 4.3 & $<$4.0\\
  MRC 2236-264[c] & SB11051\_component\_27c & 22:39:19.9 & -26:10:08 & G & 0.430 & 57.8 & 146.3 & 0.21 & 4.3 & $<$6.1\\
  MRC 2236-264[d] & SB11051\_component\_27d & 22:39:20.5 & -26:10:16 & G & 0.430 & 56.9 & 62.3 & 0.21 & 4.2 & $<$6.1\\
  &&\\

  MRC 2240-260[a] & SB11051\_component\_2a & 22:43:26.3 & -25:44:30 & QSO & 0.774 & 731.4 & 775.0 & 0.22 & 5.2 & $<$0.7\\
  MRC 2240-260[b] & SB11051\_component\_2b & 22:43:26.5 & -25:44:28 & QSO & 0.774 & 374.9 & 437.7 & 0.22 & 5.3 & $<$1.3\\
  MRC 2240-260[c] & SB11051\_component\_2c & 22:43:26.2 & -25:44:30 & QSO & 0.774 & 51.6 & 121.1 & 0.22 & 5.2 & $<$10.0\\
  &&\\

  MRC 2254-248[a] & SB11052\_component\_2a & 22:57:40.2 & -24:37:33 & G & 0.540 & 949.6 & 973.0 & 0.17 & 4.3 & $<$0.4\\
  MRC 2254-248[b] & SB11052\_component\_2b & 22:57:40.2 & -24:37:28 & G & 0.540 & 195.4 & 179.5 & 0.17 & 4.3 & $<$2.0\\
  MRC 2254-248[c] & SB11052\_component\_2c & 22:57:41.7 & -24:37:41 & G & 0.540 & 49.5 & 67.4 & 0.17 & 4.4 & $<$8.0\\
  &&\\

  MRC 2311-222[a] & SB11052\_component\_4a & 23:14:39.0 & -21:55:54 & G & 0.434 & 893.2 & 942.0 & 0.25 & 7.7 & $<$0.7\\
  MRC 2311-222[b] & SB11052\_component\_4b & 23:14:33.6 & -21:55:55 & G & 0.434 & 136.8 & 218.6 & 0.25 & 7.8 & $<$4.4\\
  MRC 2311-222[c] & SB11052\_component\_4c & 23:14:37.6 & -21:55:54 & G & 0.434 & 52.0 & 184.5 & 0.25 & 7.7 & $<$11.6\\
  &&\\

  MRC 2313-277[a] & SB11052\_component\_11a & 23:16:18.9 & -27:29:51 & G & 0.614 & 507.6 & 590.4 & 0.16 & 6.6 & $<$1.1\\
  MRC 2313-277[b] & SB11052\_component\_11b & 23:16:21.9 & -27:30:02 & G & 0.614 & 231.0 & 243.8 & 0.16 & 6.6 & $<$2.3\\
  MRC 2313-277[c] & SB11052\_component\_11c & 23:16:20.8 & -27:30:00 & G & 0.614 & 129.9 & 262.8 & 0.16 & 6.4 & $<$3.6\\
  MRC 2313-277[d] & SB11052\_component\_11d & 23:16:19.7 & -27:29:53 & G & 0.614 & 80.1 & 92.8 & 0.16 & 6.7 & $<$6.8\\
  &&\\

\hline
\hline
\end{tabular}

\end{table}
 
\end{landscape}

%Table 3 (contd)
\begin {landscape}
\begin{table}
 \setcounter{table}{4}
\ContinuedFloat
\caption{continued..}

\begin{tabular}{|l|l|l|l|l|l|l|l|l|l|l|}
\hline
\hline
  \multicolumn{1}{|c|}{Source name} &
  \multicolumn{1}{c|}{Component ID} &
  \multicolumn{1}{c|}{RA} &
  \multicolumn{1}{c|}{DEC} &
  \multicolumn{1}{l|}{Source$^1$} &
  \multicolumn{1}{c|}{Redshift$^2$} &
  \multicolumn{1}{c|}{$\rm F_{peak}$} &
  \multicolumn{1}{c|}{$\rm F_{int}$} &
  \multicolumn{1}{c|}{$\sigma _ {image}$} &
  \multicolumn{1}{c|}{$\sigma _ {channel}$} &
  \multicolumn{1}{c|}{$\tau$} \\
 
\multicolumn{1}{|c|}{ } &
  \multicolumn{1}{c|}{[FLASH]} &
  \multicolumn{1}{c|}{ } &
  \multicolumn{1}{c|}{ } &
  \multicolumn{1}{l|}{type } &
  \multicolumn{1}{c|}{[z] } &
  \multicolumn{1}{c|}{ [mJy] } &
  \multicolumn{1}{c|}{[mJy] } &
  \multicolumn{1}{c|}{ [mJy] } &
  \multicolumn{1}{c|}{ [mJy] } &
  \multicolumn{1}{c|}{ [$\rm km \ s^{-1}$]} \\
  
      \multicolumn{1}{|c|}{(1) } &
  \multicolumn{1}{c|}{(2)} &
  \multicolumn{1}{c|}{(3) } &
  \multicolumn{1}{c|}{(4) } &
  \multicolumn{1}{c|}{(5) } &
  \multicolumn{1}{c|}{(6) } &
  \multicolumn{1}{c|}{ (7) } &
  \multicolumn{1}{c|}{(8) } &
  \multicolumn{1}{c|}{ (9) } &
  \multicolumn{1}{c|}{ (10) } &
  \multicolumn{1}{c|}{ (11)} \\
  
  \hline
\hline

  MRC 2338-290[a] & SB13279\_component\_56a & 23:40:51.8 & -28:48:03 & QSO & 0.449 & 124.4 & 166.8 & 0.20 & 16.3 & $<$10.6\\
  MRC 2338-290[b] & SB13279\_component\_56b & 23:40:49.7 & -28:49:00 & QSO & 0.449 & 74.7 & 180.9 & 0.20 & 16.9 & $<$20.1\\
  MRC 2338-290[c] & SB13279\_component\_56c & 23:40:48.9 & -28:49:05 & QSO & 0.449 & 65.2 & 92.6 & 0.20 & 16.7 & $<$22.5\\
  MRC 2338-290[d] & SB13279\_component\_56d & 23:40:50.8 & -28:48:48 & QSO & 0.449 & 49.6 & 188.4 & 0.20 & 16.8 & $<$30.1\\
  &&\\

  MRC 2338-233[a] & SB13279\_component\_14a & 23:40:45.6 & -23:02:53 & QSO & 0.715 & 386.0 & 451.9 & 0.12 & 6.3 & $<$1.6\\
  MRC 2338-233[b] & SB13279\_component\_14b & 23:40:45.3 & -23:02:27 & QSO & 0.715 & 196.0 & 230.7 & 0.12 & 6.5 & $<$3.2\\
  &&\\

  MRC 2340-219[a] & SB13279\_component\_16a & 23:43:24.2 & -21:41:46 & G & 0.766 & 344.3 & 394.4 & 0.23 & 12.1 & $<$3.7\\
  MRC 2340-219[b] & SB13279\_component\_16b & 23:43:24.1 & -21:41:21 & G & 0.766 & 263.6 & 290.6 & 0.23 & 12.0 & $<$4.1\\
  MRC 2340-219[c] & SB13279\_component\_16c & 23:43:24.2 & -21:41:31 & G & 0.766 & 159.6 & 278.2 & 0.23 & 12.2 & $<$8.2\\
  &&\\

  MRC 2343-243[a] & SB10850\_component\_13a & 23:45:46.8 & -24:02:15 & G & 0.600 & 351.1 & 366.4 & 0.24 & 7.7 & $<$2.2\\
  MRC 2343-243[b] & SB10850\_component\_13b & 23:45:43.8 & -24:02:41 & G & 0.600 & 274.1 & 307.5 & 0.24 & 7.7 & $<$2.7\\
  MRC 2343-243[c] & SB10850\_component\_13c & 23:45:46.0 & -24:02:17 & G & 0.600 & 109.9 & 165.1 & 0.24 & 7.9 & $<$7.2\\
  MRC 2343-243[d] & SB10850\_component\_13d & 23:45:44.9 & -24:02:34 & G & 0.600 & 108.6 & 205.8 & 0.24 & 8.3 & $<$6.7\\
  &&\\

  MRC 2348-235[a] & SB10850\_component\_14a & 23:51:28.3 & -23:17:33 & G & 0.952 & 365.9 & 554.5 & 0.22 & 8.1 & $<$2.2\\
  MRC 2348-235[b] & SB10850\_component\_14b & 23:51:28.0 & -23:16:37 & G & 0.952 & 296.7 & 373.0 & 0.22 & 7.7 & $<$2.3\\

\hline
\hline
\end{tabular}
%Note: $^1$: The Gaussian components in various sources are labelled with alphabets [a], [b], [c] ..etc, appended with the MRC source name.

Notes $^1$ $^2$: References for source classification: \citet[][]{mccarthy1996, kapahi1998a, kapahi1998b}. References for redshift: \citet[][]{mccarthy1996, kapahi1998b}.
Notes$^{a,}$$^b$: for MRC 0042-248 and MRC 0055-258 the redshift references are \citet[][]{pocock84} and \citet[][]{brown2001} respectively.
\end{table}
 
\end{landscape}

\newpage

%New Table 4
\begin{table*}
         \caption{MRC 1-Jy sources observed in the FLASH pilot surveys for which measurements of interplanetary scintillation (IPS) are also available.  }\label{table_ips}
%    \centering
  \begin{tabular}{llcclcrrrrl}
    \hline
    \hline
        GLEAM name & S$_{162}$ & MRC name & Type & $z$ & FLASH  & NSI\_fit & NSI\_err & LAS$_{\rm 5GHz}$\\  
         & (Jy) &  & & & components & IPS & IPS & (arcsec) \\
         (1) & (2) & (3) & (4) & (5) & (6) & (7) & (8) & (9) \\
\hline
\hline
J003244-214414 & 1.79 & MRC 0030-220 & Q & 0.806 & 1 & 0.204 & 0.014 & 3.9 \\
J003824-225256 & 2.33 & MRC 0035-231 & G & 0.685 & 1 & 0.371 & 0.025 & 2.3 \\ 
J004503-243417 & 2.64 & MRC 0042-248 & Q & 0.807 & 4 & 0.067 & 0.010 & 73  \\
J005242-215540 & 2.45 & MRC 0050-222 & G & 0.654 & 1 & 0.559 & 0.037 & $<2$   \\ 
J010041-223955 & 2.61 & MRC 0058-229 & Q & 0.706 & 2 & $<0.030$ & .. & 63 \\
&& \\
J010902-230728 & 1.94 & MRC 0106-233 & Q & 0.818 & 1 & 0.435 & 0.031 & 2.5 \\
J012031-270125 & 1.87 & MRC 0118-272 & Q & $>0.557$ & 1 & 0.294 & 0.039 & $<2$ \\ %'unresolved' \\
J012614-222227 & 2.57 & MRC 0123-226 & Q & 0.720 & 2 & 0.169 & 0.013 & 5.1 \\
J013737-243048 & 3.76 & MRC 0135-247 & Q & 0.838 & 3 & 0.215 & 0.016 & $<2$  \\ %'unresolved' \\
J014521-242333 & 3.12 & MRC 0143-246 & G & 0.716 & 3 & 0.217 & 0.018 & 52.5 \\
&& \\
J014709-223234 & 1.98 & MRC 0144-227 & G & 0.600 & 1 & 0.618 & 0.013 & $<2$ \\
J020332-211342 & 1.02 & MRC 0201-214 & G & 0.915 & 1 & 0.886 & 0.059 & $<2$ \\   
J020739-224454 & 3.25 & MRC 0205-229 & G & 0.680 & 2 & 0.056 & 0.010 & 34.4 \\
J021128-232819 & 2.59 & MRC 0209-237 & Q & 0.680 & 1 & 0.087 & 0.012 & 18 \\ %marginal
J021209-280009 & 2.68 & MRC 0209-282 & G & 0.600 & 3 & $<$0.107 & .. & 13.3 \\ %upper_limit
&& \\
J022529-242113 & 1.62 & MRC 0223-245 & G & 0.634 & 2 & 0.925 & 0.066 & $<1$ \\ %detected
J022733-235450 & 4.05 & MRC 0225-241 & G & 0.520 & 1 & 0.076 & 0.007 & 5.2 \\     %detected
J023230-242201 & 4.05 & MRC 0230-245 & G & 0.880 & 1 & 0.148 & 0.011 & 11.3 \\     %detected
J023324-232059 & 8.56 & MRC 0231-235 & G & 0.810 & 2 & 0.067 & 0.005 & 38.4 \\     %detected
J023556-285044 & 4.01 & MRC 0233-290 & G & 0.725 & 1 & 0.245 & 0.026 & $<5$ \\     %detected
&& \\
J025614-232447 & 12.95 & MRC 0254-236 & G & 0.509 & 2 & 0.032 & 0.002 & 33.4 \\     %detected
J025651-271756 & 2.54 & MRC 0254-274 & G & 0.480 & 2 & 0.086 & 0.023 & 37.4 \\    %marginal
J032300-263613 & 1.54 & MRC 0320-267 & G & 0.890 & 2 & 0.302 & 0.037 & 4.3 \\     %detected
J032741-255151 & 1.45 & MRC 0325-260 & G & 0.638 & 4 & $<$0.062 & ..  & 59 \\ %upper_limit
J032954-235707 & 2.25 & MRC 0327-241 & Q & 0.895 & 2 & 0.104 & 0.019 & $<2$ \\     %marginal
&& \\
J034039-254734 & 2.01 & MRC 0338-259 & Q & 0.439 & 1 & $<$0.092 & .. & 19.7 \\  %upper_limit
J043252-232432 & 1.92 & MRC 0430-235 & G & 0.820 & 4 & $<$0.118 & .. & 91 \\ %upper_limit
J043909-242155 & 2.64 & MRC 0437-244 & Q & 0.834 & 3 & $<$0.063 & .. & 126 \\  %upper_limit
J045244-220117 & 6.55 & MRC 0450-221 & Q & 0.898 & 1 & 0.091 & 0.011 & 14.3 \\  %detected
J045610-215922 & 8.00 & MRC 0454-220 & Q & 0.533 & 5 & 0.054 & 0.008 & 84 \\    %marginal
&& \\
J052454-235214 & 1.81 & MRC 0522-239 & G & 0.500 & 3 & $<$0.154 & .. & 23.4 \\ %upper_limit
J093245-201750 & 6.28 & MRC 0930-200 & G & 0.769 & 1 & 0.136 & 0.009 & 20.2 \\    %detected
J094350-201929 & 2.12 & MRC 0941-200 & Q & 0.715 & 2 & 0.054 & 0.013 & 47.7 \\ %marginal
J100507-215221 & 2.52 & MRC 1002-216 & G & 0.490 & 1 & 0.784 & 0.058 & $<2$ \\    %detected
J100511-214451 & 16.12 & MRC 1002-215 & G & 0.589 & 2 & 0.034 & 0.003 & 29.1 \\    %detected
&& \\
J202704-213616 & 3.97 & MRC 2024-217 & Q & 0.459 & 1 & 0.076 & 0.005 & 31 \\  %detected
J215924-241751 & 1.95 & MRC 2156-245 & Q & 0.862 & 1 & 0.795 & 0.051 & $<2$ \\   %detected
J220435-270119 & 2.73 & MRC 2201-272 & G & 0.930 & 2 & 0.319 & 0.024 & 5.8 \\    %detected
J221600-280329 & 5.82 & MRC 2213-283 & Q & 0.948 & 4 & 0.038 & 0.004 & 58  \\ %detected 
J221942-275626 & 9.46 & MRC 2216-281 & G & 0.657 & 1 & 0.823 & 0.072 & $<2$ \\  
& \\
J223205-223339 & 1.90 & MRC 2229-228 & G & 0.542 & 4 & 0.594 & 0.046 & 6.6 \\    %detected
J223520-225946 & 3.27 & MRC 2232-232 & G & 0.870 & 2 & 0.266 & 0.020 & 15 \\ %detected 
J223919-261008 & 1.91 & MRC 2236-264 & G & 0.430 & 4 & $<$0.026 & .. & 25.7 \\  %upper_limit
J224326-254429 & 2.38 & MRC 2240-260 & Q & 0.774 & 3 & 0.283 & 0.020 & 11 \\ %detected 
J225740-243728 & 2.56 & MRC 2254-248 & G & 0.549 & 3 & 0.380 & 0.027 & 25 \\  %detected
&& \\ 
J225805-275818 & 1.29 & MRC 2255-282 & Q & 0.926 & 1 & 0.204 & 0.018 & $<1$ \\  %detected
J230627-250653 & 5.70 & MRC 2303-253 & G & 0.740 & 1 & 0.112 & 0.009 & 18.6 \\  %detected
J231438-215552 & 3.59 & MRC 2311-222 & G & 0.434 & 3 & 0.218 & 0.017 & 80 \\    %detected
J231620-272953 & 3.87 & MRC 2313-277 & G & 0.614 & 4 & 0.073 & 0.007 & 44.2 \\   %detected
J234045-230243 & 1.53 & MRC 2338-233 & Q & 0.715 & 2 & 0.096 & 0.011  & 19.5 \\  %detected
        \hline
        \hline
    \end{tabular}
\end{table*}

\newpage

\begin{table*}
 \setcounter{table}{5}
\ContinuedFloat
\caption{continued..}

%    \centering
  \begin{tabular}{llcclcrrrrl}
    \hline
    \hline
        GLEAM name & S$_{162}$ & MRC name & Type & $z$ & FLASH  & NSI\_fit & NSI\_err & LAS$_{\rm 5GHz}$\\  
         & (Jy) &  & & & components & IPS & IPS & (arcsec) \\
         (1) & (2) & (3) & (4) & (5) & (6) & (7) & (8) & (9) \\
\hline
\hline

J234045-230243 & 1.53 & MRC 2338-290 & Q & 0.449 & 4 & $<$0.013 & .. & 73 \\ %upper_limit  
J234324-214129 & 4.52 & MRC 2340-219 & G & 0.766 & 3 & 0.052 & 0.004 & 27.1 \\    %detected
J234412-240744 & 2.20 & MRC 2341-244 & G & 0.590 & 1 & 0.466 & 0.035 & 2.1 \\  %detected
J234545-240232 & 3.66 & MRC 2343-243 & G & 0.600 & 4 & 0.074 & 0.006  & 48.3 \\    %detected
J235128-231708 & 2.82 & MRC 2348-235 & G & 0.952 & 2 & $<$0.035 & .. & 68.7 \\ %upper_limit
        \hline
        \hline
    \end{tabular}
\end{table*}

%Table 5
\begin{table*}
         \caption{Peaked-spectrum and compact steep-spectrum sources from the MRC 1-Jy sample observed in the FLASH pilot survey }\label{table_peak}
%    \centering
    \begin{tabular}{crcrrlrl}
    \hline
    \hline
    MRC name & $z$ & Type & FLASH  & $\nu_{\rm peak, obs.}$ & Notes  & \\  
      &   &  & components & (MHz) & &  \\
         (1) & (2) & (3) & (4) & (5) & (6)  \\
\hline
\hline
\multicolumn{5}{l}{(i) Peaked-spectrum sources } \\
MRC 0201-214 & 0.915 & G & 1 & 267.6 &  \\ 
MRC 0223-245 & 0.634 & G & 2 & 213.7 &  \\
MRC 0531-237 & 0.851 & G & 1 & 610.6  &  \hi absorption detected \\
MRC 2216-281 & 0.657 & G & 1 & 161.9 & \hi absorption detected \\
\multicolumn{5}{l}{(ii) Compact steep-spectrum (CSS) sources } \\
MRC 0035-231 & 0.685 & G & 1 & .. & \\
MRC 0050-222 & 0.654 & G & 1 & .. & \\
MRC 0144-227 & 0.600 & G & 1 & .. & \\ 
MRC 0233-290 & 0.725 & G & 1 & .. & \\
MRC 0319-298 & 0.583 & G & 1 & .. & \\
MRC 0418-288 & 0.850 & QSO & 1 & .. & \\
MRC 1002-216 & 0.490 & G & 1 & .. & \\
MRC 2156-245 & 0.862 & QSO & 1 & .. &  \hi absorption detected \\
MRC 2341-244 & 0.590 & G  & 1 & .. & \\
%\multicolumn{5}{l}{(iii) Quasars with  } \\
    \hline
    \hline
    \end{tabular}
\end{table*}

\newpage

%Table 6
\begin {landscape}
\begin{table}
\caption{Luminosities and WISE colours for the sample. The columns are: (1) the source name, (2) the FLASH component ID, (3) the source right ascension, (4) the source declination, (5) the source redshift, z, (6) the logarithm of rest-frame 1.4 GHz luminosity, $\rm L^{'}_{rad}$ = $\rm log_{10}({L_{1.4 \ GHz} / W \ Hz^{-1}})$, (7) the logarithm UV luminosity at 1216 \AA, $\rm L^{'}_{UV}$ = $\rm log_{10}({L_{UV} / W \ Hz^{-1}})$  (8) the logarithm of UV ionising photon rate, $\rm Q^{'}_{UV}$ = $\rm Q^{'}$ = $\rm log_{10}{Q / s^{-1}}$,  (9) \& (10), (11) \& (12) and (13) \& (14)  the WISE fluxes in mag., in the W1, W2 and W3 bands, and their uncertainties, eW1, eW2 and eW3, respectively.
}\label{rad_uv_lum} 

\begin{tabular}{|l|l|l|l|c|r|r|r|r|r|r|r|r|r|r|}
\hline
\hline

  \multicolumn{1}{|c|}{Source name} &
    \multicolumn{1}{|c|}{Component ID} &
  \multicolumn{1}{c|}{RA} &
  \multicolumn{1}{c|}{DEC} &
  \multicolumn{1}{c|}{Redshift} &
  \multicolumn{1}{c|}{$\rm L^{'}_{rad}$} &
  \multicolumn{1}{c|}{$\rm L^{'}_{UV}$} &
  \multicolumn{1}{c|}{$\rm Q^{'}_{UV}$} &
  \multicolumn{1}{c|}{$\rm W1$} &
    \multicolumn{1}{c|}{$\rm eW1$} &
  \multicolumn{1}{c|}{$\rm W2$}  & 
    \multicolumn{1}{c|}{$\rm eW2$} &  
     \multicolumn{1}{c|}{$\rm W3$} &
       \multicolumn{1}{c|}{$\rm eW3$} \\

\multicolumn{1}{|c|}{ } &
\multicolumn{1}{|c|}{[FLASH] } &
  \multicolumn{1}{c|}{ } &
  \multicolumn{1}{c|}{ } &
  \multicolumn{1}{c|}{[z] } &
  \multicolumn{1}{c|}{  } &
  \multicolumn{1}{c|}{ } &
  \multicolumn{1}{c|}{  } &
  \multicolumn{1}{c|}{ mag } &
  \multicolumn{1}{c|}{ mag } &
    \multicolumn{1}{c|}{ mag } &
      \multicolumn{1}{c|}{ mag } &
        \multicolumn{1}{c|}{ mag } &
  \multicolumn{1}{c|}{ mag} \\

        \multicolumn{1}{|c|}{(1) } &
  \multicolumn{1}{c|}{(2)} &
  \multicolumn{1}{c|}{(3) } &
  \multicolumn{1}{c|}{(4) } &
  \multicolumn{1}{c|}{(5) } &
  \multicolumn{1}{c|}{(6) } &
  \multicolumn{1}{c|}{ (7) } &
  \multicolumn{1}{c|}{(8) } &
  \multicolumn{1}{c|}{ (9) } &
  \multicolumn{1}{c|}{ (10) } &
    \multicolumn{1}{c|}{ (11) } &
      \multicolumn{1}{c|}{ (12) } &
    \multicolumn{1}{c|}{ (13) } &
        \multicolumn{1}{c|}{ (14) } \\ 
 
  \hline
\hline

   MRC 0030-220 & SB11053\_component\_10a & 00:32:44.6 & -21:44:21 & 0.806 & 26.90 & -- & -- & 15.612 & 0.049 & 14.886 & 0.075 & 11.955 & 0.298\\
  MRC 0035-231 & SB11053\_component\_6a & 00:38:24.9 & -22:53:03 & 0.685 & 26.92 & $<$22.40 & $<$55.53 & 15.470 & 0.043 & 15.317 & 0.093 & $>$12.526 & -- \\
  MRC 0042-248 & SB13291\_component\_12d & 00:45:05.0 & -24:34:10 & 0.807 & 26.07 & -- & -- & 16.855 & 0.104 & 17.097 & 0.504 & $>$12.440 & -- \\
  MRC 0050-222 & SB13291\_component\_7a & 00:52:42.8 & -21:55:47 & 0.654 & 26.87 & 22.12 & 56.95 & 15.635 & 0.045 & 15.455 & 0.105 & 11.046 & 0.127\\
  MRC 0058-229 & SB13291\_component\_48a & 01:00:42.0 & -22:39:43 & 0.706 & 26.39 & -- & -- & 15.658 & 0.046 & 15.477 & 0.104 & 12.594 & 0.476\\
  MRC 0106-233 & SB13291\_component\_9a & 01:09:02.9 & -23:07:30 & 0.818 & 26.98 & 20.97 & 53.58 & 14.920 & 0.035 & 14.227 & 0.045 & 10.879 & 0.112\\
  MRC 0118-272 & SB37449\_component\_5a & 01:20:31.5 & -27:01:23 & 0.559 & 26.99 & 23.19 & 56.24 & 11.316 & 0.022 & 10.385 & 0.020 & 7.761 & 0.020\\
  MRC 0123-226 & SB37449\_component\_7b & 01:26:14.8 & -22:22:31 & 0.720 & 26.22 & 23.08 & 56.36 & 13.877 & 0.026 & 12.844 & 0.028 & 10.201 & 0.083\\
  MRC 0135-247 & SB13281\_component\_5a & 01:37:38.2 & -24:30:53 & 0.838 & 27.10 & 23.75 & 57.39 & 12.920 & 0.024 & 11.744 & 0.022 & 8.650 & 0.022\\
  MRC 0143-246 & SB13281\_component\_15c & 01:45:21.3 & -24:23:54 & 0.716 & 25.98 & 21.67 & 54.56 & 16.842 & 0.101 & 16.897 & 0.384 & $>$12.633 & -- \\
  MRC 0144-227 & SB37475\_component\_6a & 01:47:09.2 & -22:32:41 & 0.600 & 26.81 & -- & -- & 15.733 & 0.045 & 15.327 & 0.094 & $>$12.185 & -- \\
  MRC 0149-299 & SB35939\_component\_2b & 01:52:01.2 & -29:41:02 & 0.603 & 26.53 & -- & -- & 15.137 & 0.033 & 14.468 & 0.045 & 11.855 & 0.174\\
  MRC 0201-214 & SB13268\_component\_15a & 02:03:32.1 & -21:13:46 & 0.915 & 26.97 & -- & -- & 16.012 & 0.049 & 15.558 & 0.094 & 11.832 & 0.184\\
  MRC 0205-229 & SB13268\_component\_20a & 02:07:40.0 & -22:44:50 & 0.680 & 26.63 & $<$23.15 & $<$56.68 & 16.473 & 0.140 & 16.247 & 0.264 & $>$12.198 & --\\
  MRC 0209-237 & SB13268\_component\_12a & 02:11:28.3 & -23:28:23 & 0.680 & 26.69 & -- & -- & 14.364 & 0.028 & 13.509 & 0.031 & 10.951 & 0.104\\
  MRC 0209-282 & SB34584\_component\_18a & 02:12:10.3 & -28:00:17 & 0.600 & 26.69 & -- & -- & 15.705 & 0.041 & 15.691 & 0.111 & $>$12.889 & -- \\
  MRC 0223-245 & SB13268\_component\_7a & 02:25:29.5 & -24:21:15 & 0.634 & 26.78 & -- & -- & 15.898 & 0.049 & 16.025 & 0.170 & $>$11.981 & -- \\
  MRC 0225-241 & SB15212\_component\_4a & 02:27:33.5 & -23:54:55 & 0.520 & 26.88 & 21.27 & 54.15 & 14.813 & 0.029 & 14.215 & 0.042 & 11.926 & 0.201\\
  MRC 0230-245 & SB15212\_component\_5a & 02:32:30.4 & -24:22:05 & 0.880 & 27.27 & -- & -- & 15.943 & 0.043 & 15.465 & 0.081 & 12.058 & 0.22\\
  MRC 0231-235 & SB15212\_component\_8a & 02:33:23.9 & -23:21:14 & 0.810 & 27.14 & $<$21.67 & $<$54.50 & 16.372 & 0.054 & 15.881 & 0.103 & 12.758 & 0.417\\
  MRC 0233-290 & SB37451\_component\_1a & 02:35:56.2 & -28:50:46 & 0.725 & 27.14 & -- & -- & 15.811 & 0.044 & 15.466 & 0.092 & 12.259 & 0.299\\
  MRC 0254-236 & SB15212\_component\_6d & 02:56:15.5 & -23:24:40 & 0.509 & 26.16 & -- & -- & 16.593 & 0.067 & 16.361 & 0.182 & $>$12.975 & -- \\
  MRC 0254-274 & SB13269\_component\_104a & 02:56:51.5 & -27:18:03 & 0.480 & 25.65 & $<$20.93 & $<$53.90 & 15.7 & 0.102 & 15.540 & 0.149 & $>$12.258 & --\\
  MRC 0319-298 & SB37452\_component\_1a & 03:21:28.8 & -29:40:45 & 0.583 & 27.40 & -- & -- & 13.970 & 0.026 & 12.851 & 0.025 & 9.566 & 0.033\\
  MRC 0320-267 & SB13269\_component\_3a & 03:23:00.8 & -26:36:14 & 0.890 & 27.15 & -- & -- & 16.644 & 0.064 & 16.582 & 0.193 & $>$12.614 & -- \\
  MRC 0325-260 & SB37432\_component\_57c & 03:27:41.0 & -25:51:28 & 0.638 & 25.66 & -- & -- & 13.303 & 0.024 & 13.206 & 0.026 & $>$12.600 & -- \\
  MRC 0327-241 & SB37432\_component\_6a & 03:29:54.0 & -23:57:08 & 0.895 & 27.17 & 22.16 & 54.80 & 14.869 & 0.030 & 13.840 & 0.032 & 10.910 & 0.080\\
  MRC 0338-259 & SB37432\_component\_25a & 03:40:39.4 & -25:47:32 & 0.439 & 26.26 & -- & -- & 16.378 & 0.057 & 15.537 & 0.086 & 12.647 & 0.520\\
  MRC 0418-288 & SB37797\_component\_12a & 04:20:36.4 & -28:41:15 & 0.850 & 27.02 & -- & -- & 14.507 & 0.030 & 13.436 & 0.031 & 10.595 & 0.073\\
  MRC 0424-268 & SB37797\_component\_3a & 04:26:40.6 & -26:43:45 & 0.469 & 26.93 & -- & -- & 14.896 & 0.035 & 14.754 & 0.058 & $>$12.616 & --\\
  MRC 0428-271 & SB37797\_component\_13c & 04:30:20.3 & -27:00:01 & 0.840 & 26.07 & -- & -- & 15.493 & 0.040 & 15.376 & 0.084 & $>$12.374 & -- \\
  MRC 0428-281 & SB37797\_component\_8d & 04:30:15.9 & -28:00:48 & 0.650 & 26.05 & -- & -- & 14.990 & 0.032 & 14.628 & 0.051 & 10.634 & 0.083\\
  MRC 0430-235 & SB37797\_component\_67d & 04:32:55.4 & -23:25:20 & 0.820 & 26.15 & -- & -- & 16.157 & 0.055 & $<$16.721 & -- & $>$12.635 & -- \\
  MRC 0437-244 & SB37797\_component\_25c & 04:39:08.3 & -24:21:22 & 0.834 & 26.03 & -- & -- & 15.358 & 0.038 & 15.369 & 0.083 & $>$12.244 & -- \\
  MRC 0450-221 & SB34547\_component\_6a & 04:52:44.7 & -22:01:18 & 0.898 & 27.31 & $<$22.65 & $<$55.68 & 13.784 & 0.025 & 12.639 & 0.023 & 9.664 & 0.036\\
  MRC 0454-220 & SB34547\_component\_5d & 04:56:12.2 & -21:59:38 & 0.533 & 25.97 & 23.91 & 57.69 & 15.653 & 0.034 & 15.636 & 0.079 & $>$12.312 & -- \\
  MRC 0522-239 & SB34570\_component\_21a & 05:24:53.8 & -23:52:21 & 0.500 & 26.17 & -- & -- & 15.605 & 0.041 & 15.217 & 0.076 & $>$12.680 & -- \\
  MRC 0531-237 & SB34570\_component\_1a & 05:33:54.5 & -23:44:31 & 0.851 & 27.54 & -- & -- & 15.556 & 0.042 & 15.151 & 0.074 & $>$12.259 & -- \\
  MRC 0930-200 & SB34571\_component\_1a & 09:32:45.9 & -20:17:53 & 0.769 & 27.13 & -- & -- & 14.930 & 0.033 & 14.193 & 0.042 & 11.658 & 0.212\\
  MRC 0941-200 & SB34571\_component\_24a & 09:43:50.4 & -20:19:41 & 0.715 & 26.58 & 22.53 & 55.25 & 15.397 & 0.060 & 15.020 & 0.097 & $>$11.382 & -- \\

\hline
\hline
\end{tabular}

\end{table}
 
\end{landscape}

\begin {landscape}
\begin{table}
 \setcounter{table}{6}
\caption{continued..}\label{rad_uv_lum}  

\begin{tabular}{|l|l|l|l|c|r|r|r|r|r|r|r|r|r|r|}
\hline
\hline
  \multicolumn{1}{|c|}{Source name} &
    \multicolumn{1}{|c|}{Component ID} &
  \multicolumn{1}{c|}{RA} &
  \multicolumn{1}{c|}{DEC} &
  \multicolumn{1}{c|}{Redshift} &
  \multicolumn{1}{c|}{$\rm L^{'}_{rad}$} &
  \multicolumn{1}{c|}{$\rm L^{'}_{UV}$} &
  \multicolumn{1}{c|}{$\rm Q^{'}_{UV}$} &
  \multicolumn{1}{c|}{$\rm W1$} &
    \multicolumn{1}{c|}{$\rm eW1$} &
  \multicolumn{1}{c|}{$\rm W2$}  & 
    \multicolumn{1}{c|}{$\rm eW2$} &  
     \multicolumn{1}{c|}{$\rm W3$} &
       \multicolumn{1}{c|}{$\rm eW3$} \\
 
\multicolumn{1}{|c|}{ } &
\multicolumn{1}{|c|}{[FLASH] } &
  \multicolumn{1}{c|}{ } &
  \multicolumn{1}{c|}{ } &
  \multicolumn{1}{c|}{[z] } &
  \multicolumn{1}{c|}{  } &
  \multicolumn{1}{c|}{ } &
  \multicolumn{1}{c|}{  } &
  \multicolumn{1}{c|}{ mag } &
  \multicolumn{1}{c|}{ mag } &
    \multicolumn{1}{c|}{ mag } &
      \multicolumn{1}{c|}{ mag } &
        \multicolumn{1}{c|}{ mag } &
  \multicolumn{1}{c|}{ mag} \\
  
      \multicolumn{1}{|c|}{(1) } &
  \multicolumn{1}{c|}{(2)} &
  \multicolumn{1}{c|}{(3) } &
  \multicolumn{1}{c|}{(4) } &
  \multicolumn{1}{c|}{(5) } &
  \multicolumn{1}{c|}{(6) } &
  \multicolumn{1}{c|}{ (7) } &
  \multicolumn{1}{c|}{(8) } &
  \multicolumn{1}{c|}{ (9) } &
  \multicolumn{1}{c|}{ (10) } &
    \multicolumn{1}{c|}{ (11) } &
      \multicolumn{1}{c|}{ (12) } &
    \multicolumn{1}{c|}{ (13) } &
        \multicolumn{1}{c|}{ (14) } \\ 
  \hline
\hline

MRC 1002-215 & SB34561\_component\_6b & 10:05:11.4 & -21:45:09 & 0.589 & 26.96 & -- & -- & 15.468 & 0.041 & 15.329 & 0.087 & $>$12.208 & -- \\
  MRC 1002-216 & SB34561\_component\_1a & 10:05:08.3 & -21:52:25 & 0.490 & 26.92 & -- & -- & 15.475 & 0.042 & 15.168 & 0.088 & $>$12.603 & -- \\
  MRC 2024-217 & SB13372\_component\_4a & 20:27:04.2 & -21:36:19 & 0.459 & 26.43 & 21.30 & 54.25 & 14.411 & 0.031 & 13.376 & 0.033 & 10.890 & 0.140\\
  MRC 2156-245 & SB10849\_component\_8a & 21:59:24.9 & -24:17:52 & 0.862 & 27.20 & -- & -- & 14.549 & 0.031 & 13.486 & 0.033 & 10.283 & 0.078\\
  MRC 2201-272 & SB10849\_component\_7a & 22:04:35.1 & -27:01:17 & 0.930 & 27.29 & -- & -- & 15.954 & 0.053 & 16.293 & 0.226 & $>$12.301 & -- \\
  MRC 2213-283 & SB10849\_component\_10c & 22:16:03.0 & -28:03:31 & 0.948 & 26.47 & $<$23.56 & $<$56.97 & 14.301 & 0.029 & 14.036 & 0.042 & 11.634 & 0.240\\
  MRC 2216-281 & SB10849\_component\_1a & 22:19:42.6 & -27:56:27 & 0.657 & 27.61 & -- & -- & 15.432 & 0.041 & 15.297 & 0.098 & $>$11.750 & -- \\
  MRC 2229-228 & SB11051\_component\_5b & 22:32:05.5 & -22:33:41 & 0.542 & 26.20 & -- & -- & 15.720 & 0.050 & 14.906 & 0.070 & 10.758 & 0.101\\
  MRC 2232-232 & SB11051\_component\_3a & 22:35:20.5 & -22:59:42 & 0.870 & 27.26 & -- & -- & 15.503 & 0.047 & 13.913 & 0.042 & 9.626 & 0.045\\
  MRC 2236-264 & SB11051\_component\_27c & 22:39:20.3 & -26:10:18 & 0.430 & 25.43 & $<$20.26 & $<$52.95 & 14.110 & 0.029 & 13.438 & 0.034 & 11.041 & 0.126\\
  MRC 2240-260 & SB11051\_component\_2a & 22:43:26.3 & -25:44:30 & 0.774 & 27.06 & 22.41 & 55.50 & 12.054 & 0.022 & 11.087 & 0.021 & 8.574 & 0.027\\
  MRC 2254-248 & SB11052\_component\_2a & 22:57:40.2 & -24:37:33 & 0.540 & 26.85 & -- & -- & 14.696 & 0.032 & 13.667 & 0.038 & 10.646 & 0.093\\
  MRC 2255-282 & SB11052\_component\_3a & 22:58:06.0 & -27:58:21 & 0.926 & 27.34 & $<$23.73 & $<$56.92 & 11.963 & 0.023 & 10.846 & 0.021 & 7.864 & 0.021\\
  MRC 2303-253 & SB11052\_component\_9a & 23:06:26.9 & -25:06:54 & 0.740 & 26.94 & -- & -- & 15.710 & 0.049 & 15.157 & 0.089 & 11.742 & 0.243\\
  MRC 2311-222 & SB11052\_component\_4c & 23:14:39.0 & -21:55:54 & 0.434 & 25.39 & 21.31 & 54.11 & 16.450 & 0.075 & 16.663 & 0.331 & $>$12.317 & --\\
  MRC 2313-277 & SB11052\_component\_11c & 23:16:18.9 & -27:29:51 & 0.614 & 26.10 & $<$22.80 & $<$55.64 & 16.116 & 0.059 & 16.349 & 0.229 & 12.252 & 0.351\\
  MRC 2338-233 & SB13279\_component\_14b & 23:40:45.6 & -23:02:53 & 0.715 & 26.42 & $<$23.20 & $<$56.63 & 13.749 & 0.026 & 12.630 & 0.024 & 10.044 & 0.056\\
  MRC 2338-290 & SB13279\_component\_56d & 23:40:49.7 & -28:49:00 & 0.449 & 25.41 & 21.83 & 55.23 & 17.464 & 0.155 & 16.790 & 0.338 & $>$12.479 & --\\
  MRC 2340-219 & SB13279\_component\_16c & 23:43:24.2 & -21:41:46 & 0.766 & 26.39 & -- & -- & 15.097 & 0.036 & 15.305 & 0.092 & $>$12.143 & -- \\
  MRC 2341-244 & SB13279\_component\_3a & 23:44:12.2 & -24:07:41 & 0.590 & 26.95 & $<$21.49 & $<$54.35 & 15.209 & 0.038 & 14.973 & 0.078 & $>$11.946 & --\\
  MRC 2343-243 & SB13279\_component\_17d & 23:45:46.8 & -24:02:15 & 0.600 & 25.80 & $<$22.29 & $<$55.29 & 16.087 & 0.061 & 16.076 & 0.185 & $>$12.196 & -- \\
  MRC 2348-235 & SB10850\_component\_14a & 23:51:28.3 & -23:17:33 & 0.952 & 26.94 & -- & -- & 17.644 & 0.202 & $<$16.551 & -- & $>$12.149 & --\\

\hline
\hline
\end{tabular}

\end{table}
 
\end{landscape}

\newpage
\appendix

\section{Notes on individual sources}

\noindent
{\bf MRC 0030-220} \\
A 4.8\,GHz continuum image from \cite{kapahi1998b} shows that this $z=0.806$ quasar has two lobes separated by about 4\,arcsec. The central core is relatively weak (9.5\,mJy at 8.4\,GHz).
MRC\,0030-220 was observed with HST by \cite{baker2002}, who detected associated C\,IV absorption. The HST spectrum was re-analysed by \cite{neeleman16} to search for intervening Damped Lyman-$\alpha$ (DLA) absorption lines in the redshift range 0.533 to 0.788, but no Ly-$\alpha$\ absorption was seen. 

MRC 0030-220 has been detected as an X-ray source by {\it Chandra}/ACIS \citep[2CXO J003244.6$-$214420;][]{evans2019}, with a rest-frame X-ray luminosity L(2--10 keV) of 6 ($\pm2 $) $\times 10^{44}$ erg\,s$^{-1}$. \\

\noindent
{\bf MRC 0042-248} \\
The host of this radio source was classified as a galaxy by \cite{mccarthy1996}, but an optical objective-prism spectrum published by \cite{pocock84} shows broad emission lines of C\,III] and Mg\,II at $z=0.807$ - leading these authors to classify MRC 0042-248 as a quasar.  Based on this, we have reclassified this object as a QSO rather than a galaxy. \\

\noindent
{\bf MRC 0050-222} \\
This source is not well-studied in the literature. It appears to be compact (angular size $<0.5$\,arcsec) and core-dominated even at frequencies as low as 160\,MHz, based on the scintillation index listed in Table \ref{table_ips}. This implies a linear size smaller than about 3.5\,kpc at the published redshift of $z=0.654$, so we classify MRC 0050-222 as a compact steep-spectrum (CSS) source.  \\

\noindent
{\bf MRC 0106-233} \\
\cite{kapahi1998b} present a 4.8\,GHz continuum image of this source, showing two lobes separated by about 3\,arcsec. They note that the core emission in MRC\,0106-233 is relatively weak (5\,mJy at 8.4\,GHz). \\

\noindent
{\bf MRC 0118-272} \\
The radio spectrum of this BL Lac object shows long-term variability (and a possible spectral peak) at frequencies above a few GHz. While the radio source is classified by \cite{kapahi1998b} as unresolved at 5\,GHz, IPS data (Table \ref{table_ips}) suggest the presence of some extended emission at 162\,MHz. 
The optical spectrum published by \cite{falomo91} is largely featureless but shows a MgII absorption doublet at $z=0.557$. Since this may be an intervening line, \cite{kapahi1998b} adopted a lower limit of $z>0.557$ for the redshift of MRC\,0118-227. Several later papers quote the redshift as $z=0.557$\ without acknowledging that this is only a lower limit. 

MRC 0118-272 has been detected as an X-ray source by {\it Swift}/XRT  \citep[2SXPS J012031.6$-$270123;][]{evans2020}, with an average rest-frame X-ray luminosity L(2--10 keV) of $>1.36 \times 10^{45}$ erg\,s$^{-1}$ (this is a lower limit because we have only a lower limit to the redshift of this source).  \\

\noindent
{\bf MRC 0144-227} \\
IPS measurements \citep[][; see table \ref{table_ips}]{chhetri2018a} imply that this source is smaller than 1\,arcsec in size at low radio frequencies. Based on its radio SED, we classify MRC 0144-227 as a CSS source. \\

\noindent
{\bf MRC 0201-214} \\
This object was identified as a peaked-spectrum radio source by \cite{callingham2017}. \\

\noindent
{\bf MRC 0209-237} \\
The 4.8\,GHz continuum image published by \cite{kapahi1998b} shows that this source is roughly 20\,arcsec in angular extent, with a core and two prominent hotspots. 

MRC 0209-237 was detected as an X-ray source by {\it ROSAT}/PSPC \citep[2RXS J021129.4$-$232819;][]{boller2016}. Converting to flux using the prescriptions of \citep{dwelly2017}, we estimate a rest-frame X-ray luminosity L(2--10 keV) of 1.0 ($\pm0.3 $) $\times 10^{45}$ erg\,s$^{-1}$. \\

\noindent
{\bf MRC 0223-245} \\
\cite{kapahi1998a} classify this as an unresolved source with an angular size $<1$\,arcsec, and it is identified as a peaked-spectrum source by \cite{callingham2017}. The CASDA catalogue lists two continuum components for this source. This might indicate the presence of extended emission at lower frequencies, but the source also shows strong interplanetary scintillation suggesting that the weaker CASDA component could be an artefact.    \\

\noindent
{\bf MRC 0230-245} \\
The 5\,GHz image from \citet{kapahi1998a} shows a pair of lobes separated by about 10\,arcsec with no prominent radio core.  \\

\noindent
{\bf MRC 0233-290} \\
No MWA IPS measurements are available for this object, but we tentatively classify it as a CSS source based on its compactness at 5\,GHz \citep{kapahi1998a}. \\

\noindent
{\bf MRC 0418-288} \\
The provenance for the redshift of z=0.85 listed by \cite{baker1999} is given as `McCarthy 1994, private communication' and it appears that no published spectrum is available. We classify this as a CSS source based on its radio SED and compactness at 5\,GHz \citep{kapahi1998a}. \\

\noindent
{\bf MRC 0531-237} \\
This object was identified as a peaked-spectrum radio source by \cite{callingham2017}. \\

\noindent
{\bf MRC 1002-216} \\
\cite{kapahi1998a} classify this as an unresolved source with an angular size $<2$\,arcsec, and it is classified as a peaked-spectrum source by \cite{callingham2017}. \\

\noindent
{\bf MRC 2156-245} \\
\cite{baker1999} show an optical spectrum of this quasar, noting that it has weak emission lines and that ``Mg II may be heavily absorbed''.  As noted in the text, the optical redshift of z=0.862 listed by \cite{baker1999} is based on measurements of several emission lines and appears secure, so there is a significant velocity offset from the \hi redshift of z=0.8679. We classify this object as a CSS source. \\

\noindent
{\bf MRC 2216-281} \\
This object was identified as a peaked-spectrum radio source by \cite{callingham2017}. \\

\begin{figure}

\includegraphics[width = 6.5cm]{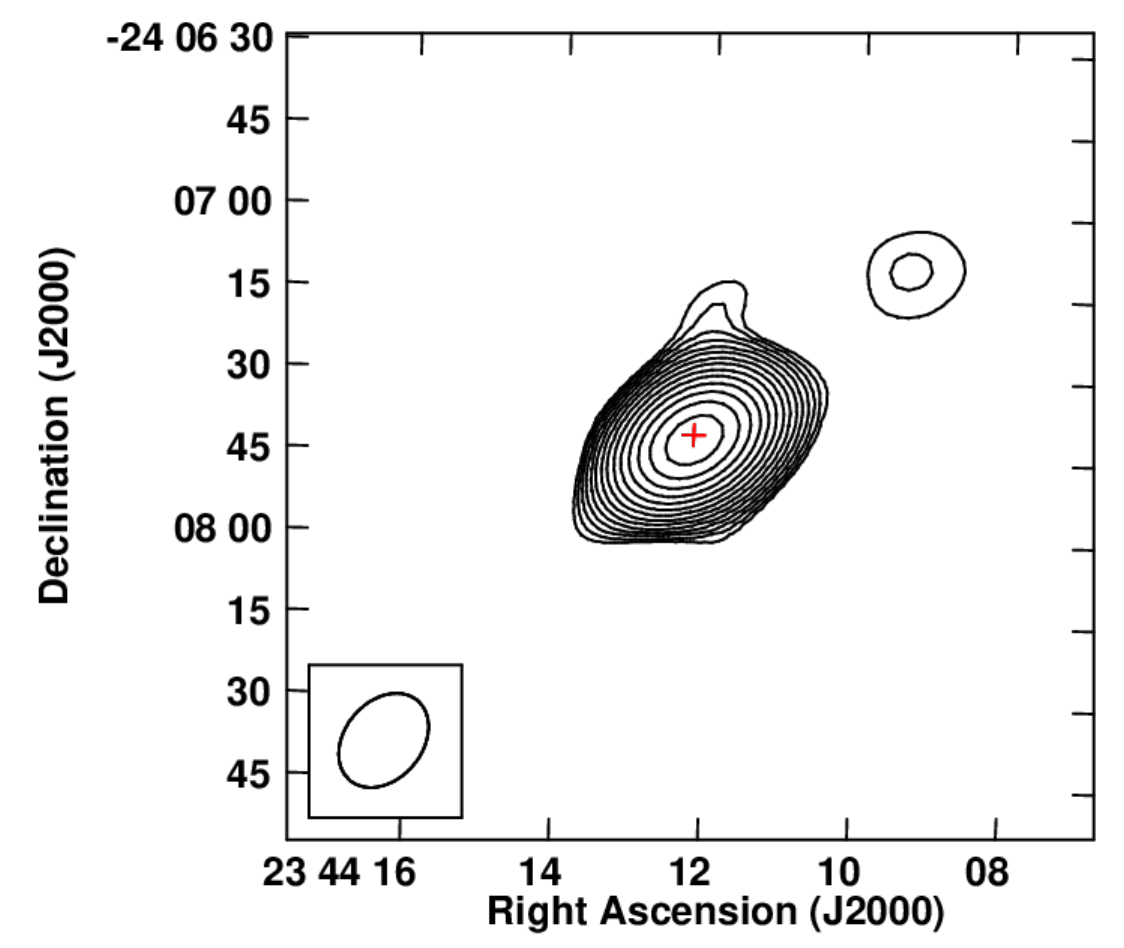} 
\\

\caption[]{ASKAP continuum image of MRC 2341-244. The red marker shows the continuum peak.}\label{mrc2341-244}

\end{figure}

\noindent
{\bf MRC 2341-244} \\
This source is relatively compact at 5\,GHz \citep{kapahi1998a}, but has a low NSI value of only 0.05 at 160\,MHz - implying that extended emission contributes a significant fraction of the low-frequency flux density. The FLASH continuum image at 856\,MHz (Figure~\ref{mrc2341-244}) shows two weaker sources within 1\,acmin of the MRC position that may account for the low NSI value.  
We have tentatively listed MRC 2341-244 among the CSS sources in Table 5, but caution that it may be an extended triple source rather than a genuine CSS object. \\ 
\\

\label{lastpage}

\end{document}